\documentstyle[12pt]{article}

  \def\pp{{\mathchoice
          {
              \kern 1pt%
              \raise 1pt
              \vbox{\hrule width5pt height0.4pt depth0pt
                    \kern -2pt
                    \hbox{\kern 2.3pt
                          \vrule width0.4pt height6pt depth0pt
                          }
                    \kern -2pt
                    \hrule width5pt height0.4pt depth0pt}%
                    \kern 1pt
           }
            {
              \kern 1pt%
              \raise 1pt
              \vbox{\hrule width4.3pt height0.4pt depth0pt
                    \kern -1.8pt
                    \hbox{\kern 1.95pt
                          \vrule width0.4pt height5.4pt depth0pt
                          }
                    \kern -1.8pt
                    \hrule width4.3pt height0.4pt depth0pt}%
                    \kern 1pt
            }
            {
              \kern 0.5pt%
              \raise 1pt
              \vbox{\hrule width4.0pt height0.3pt depth0pt
                    \kern -1.9pt  
                    \hbox{\kern 1.85pt
                          \vrule width0.3pt height5.7pt depth0pt
                          }
                    \kern -1.9pt
                    \hrule width4.0pt height0.3pt depth0pt}%
                    \kern 0.5pt
            }
            {
              \kern 0.5pt%
              \raise 1pt
              \vbox{\hrule width3.6pt height0.3pt depth0pt
                    \kern -1.5pt
                    \hbox{\kern 1.65pt
                          \vrule width0.3pt height4.5pt depth0pt
                          }
                    \kern -1.5pt
                    \hrule width3.6pt height0.3pt depth0pt}%
                    \kern 0.5pt
            }
        }}

  \def\mm{{\mathchoice
                       {
                             \kern 1pt
               \raise 1pt    \vbox{\hrule width5pt height0.4pt depth0pt
                                  \kern 2pt
                                  \hrule width5pt height0.4pt depth0pt}
                             \kern 1pt}
                       {
                            \kern 1pt
               \raise 1pt \vbox{\hrule width4.3pt height0.4pt depth0pt
                                  \kern 1.8pt
                                  \hrule width4.3pt height0.4pt depth0pt}
                             \kern 1pt}
                       {
                            \kern 0.5pt
               \raise 1pt
                            \vbox{\hrule width4.0pt height0.3pt depth0pt
                                  \kern 1.9pt
                                  \hrule width4.0pt height0.3pt depth0pt}
                            \kern 1pt}
                       {
                           \kern 0.5pt
             \raise 1pt  \vbox{\hrule width3.6pt height0.3pt depth0pt
                                  \kern 1.5pt
                                  \hrule width3.6pt height0.3pt depth0pt}
                           \kern 0.5pt}
                       }}

\catcode`@=11
\def\un#1{\relax\ifmmode\@@underline#1\else
        $\@@underline{\hbox{#1}}$\relax\fi}
\catcode`@=12

\let\du=\du                     

\def\a{\alpha}
\def\b{\beta}
\def\c{\chi}
\def\d{\delta}

\def\f{\phi}
\def\g{\gamma}
\def\h{\eta}

\def\j{\psi}

\def\l{\lambda}
\def\m{\mu}
\def\n{\nu}
\def\o{\omega}
\def\p{\pi}
\def\q{\theta}
\def\r{\rho}
\def\s{\sigma}
\def\t{\tau}

\def\x{\xi}

\def\D{\Delta}
\def\F{\Phi}
\def\G{\Gamma}
\def\J{\Psi}
\def\L{\Lambda}
\def\O{\Omega}
\def\P{\Pi}
\def\Q{\Theta}
\def\S{\Sigma}

\def\ve{\varepsilon}
\def\vf{\varphi}

\def\vq{\vartheta}

\def\cf{{\cal F}}
\def\cg{{\cal G}}

\def\cl{{\cal L}}
\def\cm{{\cal M}}
\def\cn{{\cal N}}

\def\bo{{\raise-.5ex\hbox{\large$\Box$}}}               
\def\pa{\partial}                                       
\def\de{\nabla}                                         
\def\TH{{\raise.2ex\hbox{$\displaystyle \bigodot$}\mskip-4.7mu \llap H \;}}
\def\face{{\raise.2ex\hbox{$\displaystyle \bigodot$}\mskip-2.2mu \llap {$\ddot
        \smile$}}}                                      
\def\dg{\sp\dagger}                                     

\def\sp#1{{}^{#1}}                              
\def\ket#1{\left| #1\right\rangle}              
\def\VEV#1{\left\langle #1\right\rangle}        
\def\abs#1{\left| #1\right|}                    
\def\leftrightarrowfill{$\mathsurround=0pt \mathord\leftarrow \mkern-6mu
        \cleaders\hbox{$\mkern-2mu \mathord- \mkern-2mu$}\hfill
        \mkern-6mu \mathord\rightarrow$}
\def\dvec#1{\vbox{\ialign{##\crcr
        \leftrightarrowfill\crcr\noalign{\kern-1pt\nointerlineskip}
        $\hfil\displaystyle{#1}\hfil$\crcr}}}           
\def\dt#1{{\buildrel {\hbox{\LARGE .}} \over {#1}}}     

\def\frac#1#2{{\textstyle{#1\over\vphantom2\smash{\raise.20ex
        \hbox{$\scriptstyle{#2}$}}}}}                   
\def\half{\frac12}                                        
\def\sfrac#1#2{{\vphantom1\smash{\lower.5ex\hbox{\small$#1$}}\over
        \vphantom1\smash{\raise.4ex\hbox{\small$#2$}}}} 
\def\bfrac#1#2{{\vphantom1\smash{\lower.5ex\hbox{$#1$}}\over
        \vphantom1\smash{\raise.3ex\hbox{$#2$}}}}       
\def\afrac#1#2{{\vphantom1\smash{\lower.5ex\hbox{$#1$}}\over#2}}    

\def\[{\lfloor{\hskip 0.35pt}\!\!\!\lceil}
\def\]{\rfloor{\hskip 0.35pt}\!\!\!\rceil}
\def\Lag{{\cal L}}
\def\du#1#2{_{#1}{}^{#2}}
\def\ud#1#2{^{#1}{}_{#2}}

\def\fracm#1#2{\hbox{\large{${\frac{{#1}}{{#2}}}$}}}
\def\ha{{\fracmm12}}
\def\tr{{\rm tr}}

\def\ul{\underline}
\def\un{\underline}
\def\fracmm#1#2{{{#1}\over{#2}}}

\def\low#1{{\raise -3pt\hbox{${\hskip 0.75pt}\!_{#1}$}}}

\def\Dot#1{\buildrel{_{_{\hskip 0.01in}\bullet}}\over{#1}}
\def\dt#1{\Dot{#1}}

\newskip\humongous \humongous=0pt plus 1000pt minus 1000pt
\def\caja{\mathsurround=0pt}
\def\eqalign#1{\,\vcenter{\openup2\jot \caja
        \ialign{\strut \hfil$\displaystyle{##}$&$
        \displaystyle{{}##}$\hfil\crcr#1\crcr}}\,}
\newif\ifdtup

\def\ref#1{$\sp{#1)}$}

\parskip=\medskipamount                 
\thicklines                         

\begin{document}


\thispagestyle{empty}               

\def\border{                                            
        \setlength{\unitlength}{1mm}
        \newcount\xco
        \newcount\yco
        \xco=-24
        \yco=12
        \begin{picture}(140,0)
        \put(-20,11){\tiny Institut f\"ur Theoretische Physik Universit\"at
Hannover~~ Institut f\"ur Theoretische Physik Universit\"at Hannover~~
Institut f\"ur Theoretische Physik Hannover}
        \put(-20,-241.5){\tiny Institut f\"ur Theoretische Physik Universit\"at
Hannover~~ Institut f\"ur Theoretische Physik Universit\"at Hannover~~
Institut f\"ur Theoretische Physik Hannover}
        \end{picture}
        \par\vskip-8mm}

\def\headpic{                                           
        \indent
        \setlength{\unitlength}{.8mm}
        \thinlines
        \par
        \begin{picture}(29,16)
        \put(75,16){\line(1,0){4}}
        \put(80,16){\line(1,0){4}}
        \put(85,16){\line(1,0){4}}
        \put(92,16){\line(1,0){4}}

        \put(85,0){\line(1,0){4}}
        \put(89,8){\line(1,0){3}}
        \put(92,0){\line(1,0){4}}

        \put(85,0){\line(0,1){16}}
        \put(96,0){\line(0,1){16}}
        \put(92,16){\line(1,0){4}}

        \put(85,0){\line(1,0){4}}
        \put(89,8){\line(1,0){3}}
        \put(92,0){\line(1,0){4}}

        \put(85,0){\line(0,1){16}}
        \put(96,0){\line(0,1){16}}
        \put(79,0){\line(0,1){16}}
        \put(80,0){\line(0,1){16}}
        \put(89,0){\line(0,1){16}}
        \put(92,0){\line(0,1){16}}
        \put(79,16){\oval(8,32)[bl]}
        \put(80,16){\oval(8,32)[br]}

        \end{picture}
        \par\vskip-6.5mm
        \thicklines}

\border\headpic {\hbox to\hsize{
\vbox{\noindent DESY 96 -- 244 \hfill November 1996 \\
ITP--UH--23/96 \hfill hep-th/9611209 }}}

\noindent
\vskip1.3cm
\begin{center}

{\Large\bf SOLITONS, MONOPOLES, AND DUALITY~: 
\vglue.1in
           from Sine--Gordon to Seiberg--Witten~\footnote{Supported in part 
by the `Deutsche Forschungsgemeinschaft' and the `Volkswagen Stiftung'}}\\
\vglue.3in

Sergei V. Ketov \footnote{
On leave of absence from:
High Current Electronics Institute of the Russian Academy of Sciences,
\newline ${~~~~~}$ Siberian Branch, Akademichesky~4, Tomsk 634055, Russia}

{\it Institut f\"ur Theoretische Physik, Universit\"at Hannover}\\
{\it Appelstra\ss{}e 2, 30167 Hannover, Germany}\\
{\sl ketov@itp.uni-hannover.de}
\end{center}
\vglue.2in
\begin{center}
{\Large\bf Abstract}
\end{center}

An elementary introduction into the Seiberg-Witten theory is given. Many 
efforts are made to get it as pedagogical as possible, and within a reasonable 
size. The selection of the relevant material is heavily oriented towards 
graduate students. The basic ideas about solitons, monopoles, supersymmetry 
and duality are reviewed from first principles, and they are illustrated on 
the simplest examples. The exact Seiberg-Witten solution to the low-energy 
effective action of the four-dimensional N=2 supersymmetric pure Yang-Mills 
theory with the gauge group SU(2) is the main subject of the review. Other 
gauge groups are also discussed. Some related issues (like adding matter, 
confinement, string dualities) are outlined. 

\newpage

{\Large\bf INTRODUCTION}
\vglue.2in

The recent years gave several remarkable achievements in theoretical high energy
physics, which constitute a significant progress in understanding the {\it 
strongly coupled} supersymmetric gauge theories and their superstring 
generalizations. It shed new light on some `old' but still `hot' problems, such 
as confinement, spontaneous symmetry breaking and the role of supersymmetry. The 
key concepts behind the new developments are (i) {\it supersymmetry}, 
(ii) {\it holomorphicity}, (iii) {\it duality}, and (iv) {\it integrability}. 

The significance of that new results is so important, that it already changed many
traditional ways of thinking about quantum field theory and string theory.
At the same time, the precise content of many recent results about duality was 
not enough appreciated outside of the relatively small community of scientists 
working in string theory or in a few related areas. Despite of the appearance of 
several reviews during the recent two years (see, for example, 
\cite{olive,alv,gomez,lerche,bilal,har,vecchia}, it is still rather difficult for
a non-expert to understand the Seiberg-Witten results~\cite{sw1}. It happens, in
particular, because some of the available reviews are not elementary enough, while
the other do not contain the pre-requisite information. Also, the information 
required is usually scattered over many sources. It is the purpose of this paper 
to provide the information which is really needed. I hope that it may be useful 
for those students who are willing to understand the logic in the beautiful 
papers of Seiberg and Witten~\cite{sw1}. The uniqueness of their solution was
recently proved from first principles~\cite{padova,dublin}. A solid understanding
 of the well-established facts about the strong-weak coupling duality in the 
four-dimensional supersymmetric gauge theories may help to enter the more 
fascinating world of superstring dualities. 

The standard example of duality is provided the two-dimensional
{\it Ising model}. It is defined by taking a set of spins $\s_i$,
whose values are restricted to $\pm 1$, which live on a square two-dimensional
lattice with nearest neighbourhood interactions of strength J. The partition
function reads
$$ Z(K)=\sum_{\s}\exp\left(K\sum_{(ij)}\s_i\s_j\right)~,$$
where the sum $(ij)$ goes over all the nearest neighbours, while the sum on
$\s$ goes over all spin configurations, and $K=J/k_BT$. The Ising model
is exactly solvable (Onsager), and it exhibits a first-order phase transition
to a ferromagnetic state at a critical temperature $T_c\,$. It is also known
(Kramers and Wannier) that the Ising partition function can be represented in
two different ways as a sum over plaquettes of a lattice. In the first form,
the sum goes over plaquettes of the original lattice with the coupling constant
$K$. In the second form, the sum goes over plaquettes of the {\it dual} square
lattice whose vertices are the centers of the faces of the original lattice,
with another coupling $K^*$, where ${\rm sinh}\,2K^*=({\rm \sinh}\,2K)^{-1}$. 
Both formulations are equivalent, but have different coupling constants. There
exists a {\it duality} symmetry which exchanges high temperature or weak
coupling $(K\ll 1)$ with low temperature or strong coupling $(K^* \gg 1)$.
Remarkably, the sole existence of the duality symmetry allows one to exactly
determine the critical temperature which must occur as the self-dual point where
$K=K^*$ or ${\rm sinh}(2J/k_BT_c)=1$. One may view the very existence of the phase
transition as a consequence of the fact that the dual weak and strong coupling 
regimes can be consistently `patched' together. In addition, one can learn about 
the strong coupling in the Ising model by considering its weakly-coupled dual
formulation. Similarly, in the Seiberg-Witten theory, the leading terms in the
quantum effective action at any coupling can be obtained from duality, by
using the known weak-coupling behaviour together with some additional
information, provided by extended supersymmetry, as regards patching together
the different regimes. Duality is not the property of the weak-coupling
(perturbative) expansion of the quantum theory, but it is the property
of the full (exact) theory.

It is usually very difficult to make any exact dynamical statements about 
non-perturbative phenomena in the `realistic' Standard Model of elementary 
particles, which is based on the gauge quantum field theory, even if its $(N=1)$ 
supersymmetric version is considered. It is nevertheless possible to extract some
partial information about its non-perturbative behaviour, whose origin can be 
most clearly seen in the gauge theories with {\it extended} $(N>1)$ supersymmetry.
 We are going to start in Part I with the basic facts about monopoles and 
instantons, which are the main attributes of non-perturbative physics in field 
theory. Then, we introduce supersymmetry in Part II. The information collected in
Parts I and II is necessary for understanding the Seiberg-Witten results, as well
as some of their generalizations in the main Part III.

The material appearing in this review is based on my notes collected for the 
student seminars at the Institute of Theoretical Physics in Hannover during the 
Spring and Summer 1996. The notes were used for preparing some of my seminar 
talks at DESY (Hamburg), JINR (Dubna) and Tomsk State University in Russia, all
intended for students and based on the already existing literature. The list of 
references is very far from being complete. Its only purpose is to help the 
reader to find his own way through the literature (see e.g., refs.~[1--10] and 
references therein for more).
\vglue.2in

\newpage

\begin{center}
{\bf TABLE~ of~ CONTENTS}
\end{center}

\noindent
{\bf Part I: Basic Examples of Duality in Field Theory} 
. . . . . . . . . . . . . . . \hfill 5
\begin{enumerate}
\item Sine-Gordon solitons and Thirring model  
. . . . . . . . . . . . . . . . . . . . . \hfill 5
\item Dirac monopole and electro-magnetic duality 
 . . . . . . . . . . . . . . . . . . . \hfill 9
\item t`Hooft--Polyakov monopole  
 . . . . . . . . . . . . . . . . . . . . . . . . . . . . . \hfill 13
\item Bogomol'nyi-Prasad-Sommerfield limit  
 . . . . . . . . . . . . . . . . . . . . . . \hfill 16
\item Witten effect and S duality  
. . . . . . . . . . . . . . . . . . . . . . . . . . . . . \hfill 21
\end{enumerate}
{\bf Part II: Introducing Supersymmetry} 
. . . . . . . . . . . . . . . . . . . . . . . . \hfill 25
\begin{enumerate}
\item Supersymmetry algebras and their representations  
  . . . . . . . . . . . . . . . \hfill 25
\item $N=1$ field theories and superspace  
. . . . . . . . . . . . . . . . . . . . . . . . \hfill 28
\item $N=2$ super-Yang-Mills theory  
. . . . . . . . . . . . . . . . . . . . . . . . . . . \hfill 31
\item $N=4$ super-Yang-Mills theory  
. . . . . . . . . . . . . . . . . . . . . . . . . . . \hfill 35
\item Moduli space of the $N=2$~ SYM~ theory  
. . . . . . . . . . . . . . . . . . . . . \hfill 37   
\item $N=2$~ SYM~ low-energy effective action and renormalization group 
 . . . . . \hfill 40
\end{enumerate}
{\bf Part III: Seiberg-Witten theory}   
. . . . . . . . . . . . . . .  . .. . . . . . . . . . . \hfill 45
\begin{enumerate}
\item Quantum moduli space in the $SU(2)$ pure $N=2$~ SYM~ theory 
. . . . . . . \hfill 45
\item Duality transformations  
 . . . . . . . . . . . . . . . . . . . . . . . . . . . . . . . \hfill 50
\item Seiberg-Witten elliptic curve  
 . . . . . . . . . . . . . . . . . . . . . . . . . . . . \hfill 52
\item Solution to the low-energy effective action  
 . . . . . . . . . . . . . . . . . . . . \hfill 55
\item Other groups, and adding ~$N=2$~ matter
. . . . . . . . . . . . . . . . . . . . \hfill 57
\item  Seiberg-Witten version of confinement 
. . . . . . . . . . . . . . . . . . . . . . . \hfill 65
\item Conclusion 
 . . . . . . . . . . . . . . . . . . . . . . . . . . . . . . . . . . . . . . . 
\hfill 66
\end{enumerate}
{\bf References} 
. . . . . . . . . . . . . . . . . . . . . . . . . . . . . . . . . . . . . . . . .
. \hfill 68

\newpage

{\Large\bf PART I~: ~~BASIC EXAMPLES OF DUALITY 
\vglue.1in ${~~~~~~~~~~~~~~~}$  IN ~FIElD ~THEORY}
\vglue.2in

In this introductory part, various aspects of duality in field theory are 
discussed. We start with the basic explicit example provided by the 
Sine--Gordon/Thirring models in two spacetime dimensions. Next, the Dirac 
quantization condition and the t'Hooft--Polyakov monopole in four spacetime 
dimensions are derived from the first principles. Taken all together, it provides
the necessary background for understanding further developments such as the 
Bogomo'lnyi bound, the BPS states, the Witten effect and S-duality. 
\vglue.2in

\section{Sine-Gordon solitons and Thirring model}

Let us consider the two-dimensional relativistic field theory characterized by 
the action
$$ I_{\rm SG}[\f] =\int d^2x\,\left[ \frac{1}{2} \pa_{\m}\f\pa^{\m}\f 
+\fracmm{\a}{\b^2}(\cos\b\f-1)\right]~,\eqno(1.1)$$
where $\a$ and $\b$ are constants, $\a>0$. By expanding the potential, one finds 
that the $\sqrt{\a}\equiv m$ plays the role of the mass parameter for the 
perturbative `{\it meson}' excitations (after second quantization), whereas 
$\b^2\equiv \l/m^2$ acts as the coupling constant. By changing the variables
to $\tilde{\f}=\b\f$ and $\tilde{x}^{\m}=mx^{\m}=m(t,x)$, one can put the 
equation of motion into the form
$$ \bo\tilde{\f} +\sin\tilde{\f}=0~,\eqno(1.2)$$
known as the {\it sine-Gordon} equation. This equation enjoys the discrete 
symmetries $\tilde{\f}\to -\tilde{\f}$ and $\tilde{\f}\to \tilde{\f}+2\p$, and it
has constant solutions of vanishing energy,
$$ \tilde{\f}_N =2\p N~, \qquad N\in {\bf Z}~.\eqno(1.3)$$

These solutions are not the only classical solutions of finite energy 
(generically called {\it solitons}) for the sine-Gordon equation. Since all such 
solutions at the spacial infinity must approach $\tilde{\f}_N$, one can associate
with each of them the {\it topological charge} 
$$ Q=\fracmm{1}{2\p}\int^{+\infty}_{-\infty} dx \fracmm{\pa\tilde{\f}}{\pa x
}=N_1-N_2~.\eqno(1.4)$$
The corresponding {\it topological} (i.e. non-Noether) current is given by
$$ J^{\m}=\fracmm{1}{2\p}\ve^{\m\n}\pa_{\n}\tilde{\f}~.\eqno(1.5)$$
which is conserved without using any equations of motion. Note that $Q$ does 
not contain canonical momenta.
 
The simplest one-soliton $(Q=\pm 1)$ solution can be obtained by a Lorentz boost 
of a static solution with finite energy, the latter being obtained by solving 
the one-dimensional classical mechanics problem for the potential 
$-U(\tilde{\f})=\cos\tilde{\f}-1$: 
$$ \fracm{1}{2}(\tilde{\f}')^2=U(\tilde{\f})~,\quad {\rm or}\quad
\tilde{x}-\tilde{x}_0 =\pm \int^{\tilde{\f}(\tilde{x})}_{\tilde{\f}(\tilde{x}_0)}
\fracmm{d\f}{2\sin(\f/2)}~.\eqno(1.6)$$
The solution moving with velocity $u$ reads
$$ \tilde{\f}_{\rm S,A}=\pm 4\tan^{-1}\left[ \exp\left( \fracmm{
\tilde{x}-\tilde{x}_0-u\tilde{t}}{\sqrt{1-u^2}}\right)\right]~,\eqno(1.7)$$
where $\pm$ stands for a soliton or an anti-soliton, with $Q_{\rm S,A}=\pm 1$,
respectively. More complicated multi-soliton solutions for any $Q\in {\bf Z}$
comprising any number of solitons and anti-solitons under collision are also 
known to exist, and each of them is reducible at $t\to\pm\infty$ to the sum of 
the well-separated solitons and anti-solitons up to certain time delays, 
with the velocities and energy profiles being {\it unchanged}~\cite{raja}. 
It is also clear that a multi-soliton solution with a given $Q$ cannot `decay' 
into solitons with a different $Q$  because of the topological charge 
conservation (the {\it superselection} rule).

The fact that solitons maintain their shape despite their collisions and have 
finite classical mass (defined by the static energy) suggests their physical 
interpretation as classical particles. Therefore, we have two apparently 
different `sorts' of particles in the sine-Gordon theory: the perturbative 
`mesons' as the small excitations of the second-quantized field with the mass
$m$, and the non-perturbative solitons of mass $M=8m/\b^2=8m^3/\l$, which are 
the extended classical solutions of the non-linear field equations. The solitons
interpolate between different minima of the potential, and they are absent in the 
perturbative spectrum. In the weak coupling limit, $\l\to 0$ or $\b\to 0$, the  
`meson' mass $m$ is constant or small, while the soliton mass $M$ is large.

In fact, these two `sorts' of particles can be considered on equal footing in 
the full quantum theory~\cite{skyrme}. The whole point is the known {\it quantum 
equivalence} of the sine-Gordon model to the {\it Thirring} model to be defined 
by the action
$$ I_{\rm T}[\j]=\int d^2x\left[ \bar{\j}i\g^{\m}\pa_{\m}\j -m_{\rm F}\bar{\j}\j 
-\fracmm{g}{2}(\bar{\j}\g^{\m}\j)^2\right]~.\eqno(1.8)$$
The equivalence is established via bosonization~\cite{colem,mandel}:
$$ \j_{\pm}(x)=\exp\left\{ \fracmm{2\p}{i\b}\int^x_{-\infty}
\fracmm{\pa\f(x')}{\pa t}dx' \mp\fracmm{i\b}{2}\f(x)\right\}~,\eqno(1.9)$$
where the two spinor components are distinguished by $\pm$. One can show that
the $\j_{\pm}$ satisfies the Thirring equations of motion provided the $\f$ 
satisfies the sine-Gordon equation and vice versa. The {\it vertex operator 
construction} of eq.~(1.9) establishes the equivalence of the correlation 
functions in both theories, while the correspondence between their coupling 
constants turns out to be~\cite{colem,mandel}
$$ \fracmm{\b^2}{4\p}=\fracmm{1}{1+g/\p}~.\eqno(1.10)$$
The strong coupling in the T-theory (large $g$) is thus mapped to the weak 
coupling (small $\b$) in the {\it dual} SG-theory and vice versa. It allows one 
to identify the particles corresponding to the fluctuations of $\j_{\pm}$ with 
solitons and anti-solitons. One can show that the meson SG states correspond to 
the fermion-antifermion {\it bound} states in the Thirring theory~\cite{raja}. 

Actually, we were rather sloppy above, since we ignored the effects of
renormalization in quantum field theory. Fortunately, the renormalization effects
in the SG- and T-theories are under control, and they can be fully taken
into account by normal ordering, in terms of bare parameters $m^2_0$ and 
$m_{\rm F}$, and the fermionic field renormalization parameters $C_{\pm}$. One
uses the Baker-Hausdorff identity to show that~\cite{dashen}
$$\fracmm{m^4}{\l}:\left[\cos\left(\fracmm{\sqrt{\l}}{m}\f\right)-1\right]:
=m_0^2\fracmm{m^2}{\l}\left[\cos\left(\fracmm{\sqrt{\l}}{m}\f\right)-1\right]~.
\eqno(1.11)$$
The action-angle variables, in which the classical SG hamiltonian  reduces to a 
free particle form, are known~\cite{fad}, which implies that the SG model is 
exactly solvable both as a classical theory and as a quantum one (semiclassical 
quantization is exact in this case). Accordingly, the quantum 
renormalization in the SG theory amounts to replacing the naive coupling 
constant $\b^2$ to a renormalized coupling constant $\g$,
$$\g=\fracmm{\b^2}{1-\b^2/8\p}=\fracmm{8\p}{1+2g/\p}~.\eqno(1.12)$$

The quantum bosonization rules are given by the normal-ordered equation 
(1.9): 
$$\j_a(x)=C_a:\exp\[A_a(x)]:~,\quad A_{\pm}(x)=\fracmm{2\p m}{i\sqrt{\l}}
\left(\int^x_{-\infty}\Dot{\f}(x')dx'\right)
\mp \fracmm{i\sqrt{\l}}{2m}\f(x)~,\eqno(1.13)$$
where $a=\{\pm\}=(1,2)$. In particular, 
it implies the relations~\cite{colem,mandel}
$$ \eqalign{
m^2_0(m^2/\l)\cos\left[ (\sqrt{\l}/m)\f\right] =~&~-m_{\rm F}\bar{\j}\j~,\cr
-(\sqrt{\l}/2\p m)\ve^{\m\n}\pa_{\n}\f=\bar{\j}\g^{\m}\j~,\cr}\eqno(1.14)$$
while the fermionic charge can be identified with the topological charge. 

It is not difficult to show that the fermions defined by eq.~(1.13) do satisfy
the local Fermi rules~\cite{mandel}. The canonical equal-time commutation 
relations
$$
\[ \f(x),\f(y) \]_- = \[ \Dot{\f}(x),\Dot{\f}(y) \]_- =0~,\quad
\[ \f(x),\Dot{\f}(y) \]_- =i\d(x-y)~,\eqno(1.15)$$
imply that $ \[ A_a(x),A_b(y) \]$ is either $i\p$ or $-i\p$ when $x\neq y$, which
leads to~\footnote{The renormalization coefficients $C_{\pm}$ are to be adjusted 
in the coincidence limit.}
$$\{ \j_a(x),\j_b(y)\}_+ =0~,\quad {\rm and}\quad \{ \j_a(x),\j^{\dg}_b(y)\}_+ 
=Z\d(x-y)\d_{ab}~,\eqno(1.16)$$
where $Z$ is another renormalization constant. In addition one finds 
that~\cite{raja}
$$ \[ \f(y),\j(x)\]=(2\p/\b)\q(x-y)\j(x)~,\qquad x\neq y~.\eqno(1.17)$$
Being applied to the soliton state with $\f(\infty)-\f(-\infty)=2\p/\b$, the 
operator $\j$ thus reduces it to a state in the vacuum sector with  
$\f(\infty)-\f(-\infty)=0$. Because of eq.~(1.17), $\j(x)$ alters a field $\f$ 
by
a step function which can be considered as a `point soliton' (obviously, a local 
operator cannot create an extended object). The physical (extended) soliton then 
arises via interactions. The $\j$ and $\j^{\dg}$ can therefore be interpreted as 
the destruction and creation operators for bare solitons.

One learns from the explicit duality between the T- and SG-models that 
\begin{itemize}
\item duality is a quantum correspondence which relates the strong coupling
in one theory with the weak coupling in another theory;
\item duality interchanges `fundamental' quanta with solitons, and thus 
establishes a `democracy' between them;
\item in addition, duality exchanges Noether currents with topological currents.
\end{itemize}

In other words, the full physical spectrum does not only contain the particles
corresponding to the fields present in the classical Lagrangian, but it also
contains other particles which correspond to the soliton solutions and which 
are required by duality.

It is highly non-trivial to generalize that ideas to {\it four} dimensions. 
In particular, the naive generalization of the two-dimensional sine-Gordon 
theory to a scalar field theory in higher dimensions does not work.~\footnote{The
absence of non-trivial static solutions for a very general class of scalar 
potentials in more \newline ${~~~~~}$ than two dimensions is known as the 
{\it Derrick} theorem~\cite{derrick}.} Hence, the need for some additional 
{\it gauge} fields becomes apparent. Moreover, we need a gauge theory in which 
the semiclassical properties are not destroyed by quantum corrections. It is the 
(extended) {\it supersymmetric} gauge theories that enjoy such a behaviour. In
what follows, both ideas will be discussed in some detail.
\vglue.2in

\section{Dirac monopole and electro-magnetic duality}

The Maxwell equations for the electromagnetism in 1+3 dimensions can be written 
down in the relativistic form as
$$ \pa_{\n}F^{\m\n}=-j_e^{\m}~,\qquad \pa_{\n}{}^*F^{\m\n}=0~,\eqno(2.1)$$
or in the vector form as
$$\eqalign{
 {\rm div} \vec{E}=\r_e~,\quad & \quad {\rm rot}\vec{E} 
+\fracmm{\pa\vec{B}}{\pa t}=0~,\cr
{\rm div} \vec{B}=0~,\quad & \quad {\rm rot}\vec{B} -
\fracmm{\pa\vec{E}}{\pa t}=\vec{J}_e~,\cr}\eqno(2.2)$$
where we use the notation $\m=(0,i)=(0,1,2,3)$, $\h_{\m\n}={\rm diag}(-,+,+,+,)$,
$\ve^{0123}=1$, $j^{\m}_e=(\r_e,\vec{J}_e)$, and define~\footnote{We 
normally take $c=1$ and $\hbar=1$, but sometimes reintroduce one or both of them,
in order to \newline ${~~~~~}$ emphasize the relativistic and/or quantum nature 
of some equations.}   
$$ F^{0i}=-E^i~,\quad F^{ij}=-\ve^{ijk}B^k~,\quad {\rm and}\quad 
{}^*F^{\m\n}=\frac{1}{2}\ve^{\m\n\l\r}F_{\l\r}~.\eqno(2.3)$$

In vacuum, where $\r_e=\vec{J}_e=0$, eq.~(2.2) can be rewritten to the form
$$ \vec{\de}\cdot (\vec{E}+i\vec{B})=0~,\qquad \vec{\de}\wedge(\vec{E}+i\vec{B})=
i\fracmm{\pa}{\pa t}(\vec{E}+i\vec{B})~,\eqno(2.4)$$
which is invariant under the duality rotations~\footnote{The Maxwell equations in
vacuum are also known to be invariant under Lorentz and conformal 
\newline ${~~~~~}$ transformations.} 
$$ \vec{E}+i\vec{B} \to e^{-i\q}(\vec{E}+i\vec{B})~,\eqno(2.5)$$
parametrized by an arbitrary angle $\q$. In particular, when $\q=\p/2$, 
there is a discrete symmetry
$$D:\qquad \vec{E}\to +\vec{B}~, \quad \vec{B}\to -\vec{E}~,\eqno(2.6)$$
whose square $D^2:(\vec{E},\vec{B})\to (-\vec{E},-\vec{B})$ is just the
charge conjugation $C$. Eq.~(2.6) is obviously equivalent to 
$$ F^{\m\n}\to {}^*F^{\m\n}~, \qquad {}^*F^{\m\n}\to -F^{\m\n}~,\eqno(2.7)$$
and it can only be valid in 1+3 dimensions because of the identity 
$(*)^2=-{\bf 1}$.~\footnote{Only in 1+3 dimensions do the electric and magnetic 
fields both constitute vectors.}

The energy and momentum density of the electro-magnetic field,
$$ \ha \abs{ \vec{E} + i\vec{B} }^2 = \ha\left( \vec{E}^2 + \vec{B}^2 \right)~,
\eqno(2.8)$$
and
$$\fracmm{1}{2i} \left( \vec{E} +i\vec{B} \right)^* \wedge 
\left( \vec{E} +i\vec{B} \right) = \vec{E} \wedge \vec{B}~,\eqno(2.9)$$
respectively, are invariant under the duality (2.5). As far as the Lagrangian and
the topological charge density are concerned, they are given by the real and 
imaginary part of 
$$ \ha \left( \vec{E} + i\vec{B} \right)^2 = \ha \left( \vec{E}^2 - \vec{B}^2 
\right) + i\vec{E}\cdot\vec{B}~,\eqno(2.10)$$
respectively, and, hence, they transform as a doublet 
under the duality~\cite{olive}.

The duality symmetry is lost if an electric current $j^{\m}$ 
enters the Maxwell equations. Therefore, if we want to keep the 
electro-magnetic duality in the presence of matter, we have to add magnetic 
source terms into the Maxwell equations as well, so that
$$ \pa_{\n}{}^*F^{\m\n}=-k^{\m}\neq 0~.\eqno(2.11)$$
For example, the discrete duality transformations (2.7) are to be appended by
$$j^{\m}\to k^{\m}~,\quad {\rm and}\quad k^{\m}\to -j^{\m}~.\eqno(2.12)$$

If the duality makes sense, it has also to be consistent with quantum mechanics
and non-abelian gauge theories (see also the next section). Consider a 
charged quantum particle with momentum $\vec{p}$, whose interaction with the
electromagnetic field via the standard substitution 
$\vec{p}=-i\vec{\de}\to -i(\vec{\de}-ie\vec{A})$ is governed by a
 potential $A^{\m}=(A_0,\vec{A})$ to be defined from the field strength
$$ F_{\m\n}=\pa_{\m}A_{\n}-\pa_{\n}A_{\m}~.\eqno(2.13)$$
The Schr\"odinger equation for the quantum particle,
$$ i\fracmm{\pa\j}{\pa t}=-\fracmm{1}{2m}(\vec{\de}-ie\vec{A})^2\j+V\j~,
\eqno(2.14)$$
is invariant under the gauge transformations
$$\j\to e^{-ie\c}\j~,\qquad \vec{A}\to\vec{A}-\vec{\de}\c=\vec{A}
-\fracm{i}{e}e^{ie\c}\vec{\de}e^{-ie\c}~,\eqno(2.15)$$
where the gauge parameter $\c$ enters via the $U(1)$ group element
$e^{ie\c}$, which must be single-valued and continuous. Hovever, it is the
potential $A^{\m}$ itself that gives a problem since its definition
(2.13) apparently implies that $\pa_{\n}{}^*F^{\m\n}=-k^{\m}=0$. Therefore,
the electromagnetic potential of a magnetic charge (called {\it monopole}), 
if exists, has to be singular inside the monopole.~\footnote{Since
we do not expect the electrodynamics to be a correct theory at very small 
distances, the \newline ${~~~~~}$ existence of singularity at the location of 
a monopole does not pose a serious problem.} 
The consistent solution outside the monopole of 
magnetic charge $g$, resulting in a magnetic field
$$ \vec{B}=\fracmm{g\vec{e}_r}{4\p r^2}~,\eqno(2.16)$$
 makes use of the ambiguity relating
the vector potential to the field strength~\cite{wy}: one can use different
potentials in different regions if their differences in the overlapping regions
are given by gauge transformations. It is the physically measurable field
strength $F^{\m\n}$ that has to be continuous and unambiguous. The simplest way
out is to divide a sphere $S^2$ surrounding the monopole into a
northern (N) and southern (S) hemispheres, corresponding to $0\leq \q\leq\p/2$ 
and $\p/2\leq \q \leq\p$, respectively, the equator (E) with $\q=\p/2$ being
the overlap region. A non-singular solution to the vector potential on the 
hemispheres reads~\footnote{A general solution can be understood in more
abstract terms (see below).}
$$\eqalign{
\vec{A}_{\rm N}=~&~+\fracmm{g}{4\p r}\fracmm{1-\cos\q}{\sin\q}\vec{e}_{\f}~,\cr
\vec{A}_{\rm S}=~&~-\fracmm{g}{4\p r}\fracmm{1+\cos\q}{\sin\q}\vec{e}_{\f}~,\cr}
\eqno(2.17)$$
so that $\vec{B}=\vec{\de}\times\vec{A}$ just yields eq.~(2.16). This
construction makes sense, since the difference of the vector potentials at 
$\q=\p/2$,
$$ \vec{A}_{\rm N}-\vec{A}_{\rm S}=-\vec{\de}\c~,\qquad 
\c=-\fracmm{g}{2\p}\f~,\eqno(2.18)$$
is a gauge transformation indeed, while the enclosed magnetic charge is given by
$$g=\int_{S^2}\vec{B}\cdot d\vec{S}=
\int_{N}\vec{B}\cdot d\vec{S}+\int_{S}\vec{B}\cdot d\vec{S}=
\int_{E}(\vec{A}_{\rm N}-\vec{A}_{\rm S})\cdot d\vec{l}=\c(0)-\c(2\p)\neq 0~,
\eqno(2.19)$$
as required. The gauge transformation parameter $\c$ in eq.~(2.18) is not a 
continuous function, but it is the function $e^{-ie\c}$ that has to be continuous
so that $\exp(-ieg)=1$. Reintroducing $\hbar$ and $c$, one can represent it
in the form
$$ eg=2\p n \hbar c~,\qquad n\in {\bf Z}~,\eqno(2.20)$$
known as the celebrated {\it Dirac quantization condition}~\cite{dirac}.

In mathematical terms, the sphere $S^2$ surrounding the monopole is just
the base space of a non-trivial $U(1)$ principal fibre bundle. The resulting 
structure is a manifold when the fibers are patched together in a globally 
consistent way, with gauge transformations as the transition functions. Because 
of eqs.~(2.18) and (2.19), the magnetic charge of the monopole can be directly
interpreted as the {\it winding number} of the gauge transformation, defining
a map from the overlap region (equator) $S^1$ to the gauge group $U(1)\sim S^1$.
These maps are classified by the first {\it homotopy} group $\p_1(U(1)) \sim
{\bf Z}$, whose elements can be identified with the integers $n$ appearing in
the Dirac quantization condition (2.20).~\footnote{In eq.~(2.17) above, the
case of $n=1$ was considered.} The same integer is given by the {\it first 
Chern class} $c_1$ of the bundle, which is defined by an integration of the
two-form $\fracm{1}{2\p}F_{\m\n}dx^{\m}\wedge dx^{\n}$ over $S^2$.

It is clear from eq.~(2.20) that just assuming the existence of a 
monopole~\footnote{No monopoles were observed in the experiments, which implies
that, if they nevertheless exist, \newline ${~~~~~}$ their masses are to be high 
enough.} is
sufficient for explaining the quantization of the electric charge $e$, as well
as another well-known experimental fact that the absolute values of the electron 
and proton electric charges are {\it exactly} equal. It is also clear from 
eqs.~(2.5) and (2.12) that the electro-magnetic duality requires the rotation of 
electric and magnetic charges of point particles representing  matter, in order 
to keep the Maxwell equations invariant,
$$ e+ig \to e^{-i\q}(e+ig)~.\eqno(2.21)$$

It should be noticed that the Dirac quantization condition (2.20) does {\it not} 
respect the symmetry (2.21). It is related to the (unjustified) hidden assumption
that the Dirac monopole does not carry an electric charge. In order to generalize
eq.~(2.20) to the form which is consistent with the electromagnetic duality, one
first notices that eq.~(2.20) can be obtained in many different ways. For example,
 when computing the orbital angular momentum
$$ \vec{L}=\int d^3r\, \vec{r}\times(\vec{E}\times\vec{B}) \eqno(2.22)$$
of a point particle with an electric charge $e$ in the field of the magnetic 
monopole with a magnetic charge $g$, just demanding the $\vec{L}$ be quantized in 
units of $\hbar/2$ also yields eq.~(2.20). Eq.~(2.22) can be easily generalized 
to the case of two {\it dyons}, having both electric and magnetic charges, 
$(q_1,g_1)$ and $(q_2,g_2)$. The momentum quantization then gives rise to the 
so-called {\it Dirac-Zwanziger-Schwinger} (DZS) quantization 
condition~\cite{dirac,zwan,schw},
$$ q_1g_2-q_2g_1=2\p n, \qquad n\in {\bf Z}~,\eqno(2.23)$$
which is invariant under the electromagnetic duality (2.21). The DZS
condition implies that the allowed electric and magnetic charges of a dyon are 
quantized, and they should lie on a two-dimensional lattice~\cite{vecchia}.

Similarly to the SG--T duality considered in the preceding section, the 
interchange of electricity and magnetism by exchanging the coupling constants 
leads to the interchange of weak and strong coupling. Like solitons in the SG 
theory, the Dirac monopole does not exist in the spectrum of standard quantum
electrodynamics, and no local theory exists which could accomodate both 
electrons and Dirac monopoles.

One learns from the electromagnetic duality that
\begin{itemize}
\item it requires magnetic monopoles,
\item the existence of monopoles in a gauge theory is closely related to the
existence of a compact $U(1)$ gauge group,
\item the magnetic charge is given by the topological quantity -- the winding
number --  which belongs to the first homotopy group of $U(1)$,
\item electro-magnetic duality implies $C$-invariance,
\item the electric and magnetic charges of dyons lie on a two-dimensional lattice.
\end{itemize}

The derivation of the Dirac quantization condition above considers a monopole 
from a distance, so it directly applies to an electron which is not confined 
unlike the quarks. It is also very general, since no particular underlying
theory was used for describing monopoles. However, in order to probe the monopole 
inside, one needs a deeper gauge theory, which contains both electrically and 
magnetically charged particles. The so-called {\it Georgi-Glashow model} is 
such a theory, as was independently found by t'Hooft and 
Polyakov~\cite{tho,pol}. This model is considered in the next section.
\vglue.2in

\section{t'Hooft-Polyakov monopole}

The basic idea is to embed the $U(1)$ generator $Q$ of electric charge into a 
larger compact gauge group, say, $SU(2)$ or $SO(3)$ for simplicity, i.e. to
switch to a non-abelian gauge theory. The standard Higgs mechanism can then be 
used to select the direction of $Q$ amongst the $SO(3)$ generators. The 
situation is very much analoguous to the SG theory (sect.1) having the discrete
vacuum symmetry (1.3) which is now replaced by the continuous gauge symmetry.
 
The Georgi-Glashow model consists of an $SO(3)$ gauge field
$A^a_{\m}$ and a Higgs triplet field $\F^a$, with the Lagrangian
$$ \cl_{\rm GG} 
= -\fracm{1}{4}F^a_{\m\n}F^{a\m\n}+\frac{1}{2}D^{\m}\F^aD_{\m}\F^a
-V(\F)~,\eqno(3.1)$$
where the Yang-Mills field strength
$$F^a_{\m\n}=\pa_{\m}A_{\n}^a -\pa_{\n}A_{\m}^a +e\ve^{abc}A_{\m}^bA_{\n}^c~,
\eqno(3.2)$$
the covariant derivative
$$ D_{\m}\F^a=\pa_{\m}\F^a +e\ve^{abc}A_{\m}^b\F^c~,\eqno(3.3)$$
and the Higgs potential
$$ V(\F)=\fracm{\l}{4}(\F^a\F^a-v^2)^2~,\eqno(3.4)$$
have been introduced. The corresponding equations of motion read
$$ D_{\m}F^{a\m\n}=e\ve^{abc}\F^bD^{\n}\F^c~,\quad (D^{\m}D_{\m}\F)^a=-\l
\F^a(\F^b\F^b-v^2)~,\eqno(3.5)$$
and the Bianchi identity is
$$D_{\m}{}^*F^{a\m\n}=0~.\eqno(3.6)$$

Like in the SG theory, our strategy is to find static classical solutions of
the equations of motion with a finite energy. 
The {\it improved}~\footnote{The improved stress-energy tensor is symmetric 
and gauge-invariant by definition.} stress-energy tensor is given by
$$\vq^{\m\n}=\fracmm{2}{\sqrt{-g}}\fracmm{\d I_{\rm GG}}{\d g_{\m\n}}=
-F^{a\m\r}F_{\r}^{a\n}+D^{\m}\F^aD^{\n}\F^a-\h^{\m\n}\cl_{\rm GG}~,
\eqno(3.7)$$
and it is classically conformally invariant, $\vq^{\m}_{\m}=0$, if
$\l=0$. It still makes sense to choose $V(\F)=0$ while maintaining 
$\VEV{\F}\neq 0$ which spontaneously breaks both the gauge and scale 
invariances, and it is going to be used later, in the next section. Because
of eq.~(3.7), the energy density reads
$$ \vq_{00} = \fracm{1}{2}\left(\vec{E}^a\vec{E}^a 
+\vec{B}^a\vec{B}^a+\P^a\P^a+
\vec{D}\F^a\cdot\vec{D}\F^a\right) +V(\F)~,\eqno(3.8)$$
where we have introduced the momenta $\P^a\equiv D_0\F^a$~, and 
defined $E^{ai}=-F^{a0i}$, $B^{ai}=-\frac{1}{2}\ve^{ijk}F^{a}_{jk}$. 
Obviously, we have
$\vq_{00}\geq 0$, while $\vq_{00}=0$ if and only if 
$F^{a\m\n}=D^{\m}\F^a=V(\F)=0$. The Higgs vacuum $\cm_{\rm H}$ is therefore 
given by the vanishing gauge field and a constant Higgs field, $\F^a\F^a=v^2$,
i.e. $\cm_{\rm H}=S^2$. The perturbative spectrum consists of a massless
`photon', massive spin-one gauge bosons $W^{\pm}$ of mass $\abs{e}v$ and a Higgs 
field whose mass is $v\sqrt{2\l}$. 

The finite energy solutions must lie in $\cm_{\rm H}$ at the spacial infinity,
whereas the Higgs field overthere provides a map from $S^2_{\infty}$ to
$\cm_{\rm H}=S^2$. Such maps are topologically classified by the integer 
winding number which is an element of the second 
homotopy group of $S^2$,
$$ \p_2(S^2)={\bf Z}~.\eqno(3.9)$$

It is easy to check that finite-energy field configurations with a non-trivial
winding number require a non-vanishing gauge field. Indeed, it follows from 
the relations
$$ \vq_{00}\geq \int d^3x\,\fracm{1}{2}\vec{\de}\F^a\vec{\de}\F^a~,\eqno(3.10)
$$
and
$$ (\vec{\de}\F^a)^2=(\fracmm{\pa\F^a}{\pa r})^2 +(\vec{e}_r\times\vec{\de}
\F^a)^2~,\eqno(3.11)$$
that at $A^{\m}=0$ one arrives at a linearly divergent integral,
$$\vq_{00}\geq \int^{\infty}_0 \fracmm{r^2dr}{r^2}~,\eqno(3.12)$$
since the non-trivial winding number implies non-vanishing angular derivatives
 of $\F^a$ at spacial infinity, and their contribution alone in eq.~(3.11)
leads to the divergence (3.12) in eq.~(3.10). Simultaneously, this argument 
shows that, in order to achieve a finite-energy solution, there should be a 
cancellation between the angular part of the vector potential (which must fall
as $1/r$) and the angular derivative of $\F$, such that the {\it covariant} 
derivative of the Higgs field vanishes at spacial infinity.

It is not difficult to see that the $1/r$ falloff in the angular component of
the gauge field $A_{\m}^a$ gives rise to a non-vanishing magnetic field at 
spacial infinity, i.e. it gives a monopole~!  When taking into account only 
the leading $1/r$-terms, the general solution to the equation
$$ D_{\m}\F^a=\pa_{\m}\F^a +e\ve^{abc}A_{\m}^b\F^c\sim 0\eqno(3.13)$$
reads
$$ A^a_{\m}\sim -\,\fracmm{1}{ev^2}\ve^{abc}\F^b\pa_{\m}\F^c
+\fracmm{1}{v}\F^aA_{\m}~,\eqno(3.14)$$
where $A_{\m}$ is an arbitrary field. Accordingly, the field strength takes 
the form
$$ F^{a\m\n}=\fracmm{1}{v}\F^aF^{\m\n}~,\quad {\rm where}\quad
F_{\m\n}=-\,\fracmm{1}{ev^3}\ve^{abc}\F^a\pa_{\m}\F^b\pa_{\n}\F^c
+\pa_{\m}A_{\n}-\pa_{\n}A_{\m}~.\eqno(3.15)$$

The equations of motion (3.5) together with the Bianchi identity (3.6) 
imply in addition that
$$ \pa_{\m}F^{\m\n}=\pa_{\m}{}^*F^{\m\n}=0~,\eqno(3.16)$$
outside of the core of the monopole. It is therefore the Higgs field that 
is solely responsible for the non-vanishing magnetic charge of the gauge 
field configuration (3.15):
$$g=\int_{S^2_{\infty}}\vec{B}\cdot d\vec{S}
=\fracmm{1}{2ev^3}\int_{S^2_{\infty}}
dS^i\,\ve^{ijk}\ve^{abc}\F^a\pa^j\F^b\pa^k\F^c
=\fracmm{4\p}{e}n~,\qquad n\in {\bf Z}~.\eqno(3.17)$$
The just found quantization condition, 
$$eg=4\p n~,\eqno(3.18)$$
 differs by a factor of $2$ from the Dirac quantization condition (2.20).
It is related to the fact that we could add into our theory more fields in 
the fundamental representation $\ul{2}$ of $SU(2)$ whose quanta carry an
electric charge $\pm e/2$. It is the Dirac quantization condition with 
respect to the electric charge $\pm e/2$ that yields eq.~(3.18).
 
The main lesson one learns from this section is that there exists a deep 
connection between the Dirac monopoles and the Higgs mechanism~\cite{gol},
namely,
\begin{itemize}
\item finite-energy solutions with non-vanishing topological charge in the 
Georgi-Glashow model are necessarily magnetic monopoles which satisfy the
Dirac quantization condition.
\end{itemize}
In mathematical terms, on the one hand, given a gauge (simply connected) group
$G$ broken down to a subgroup $H$ by the non-vanishing Higgs field vacuum 
expectation value, the topology of the Higgs 
vacuum is classified by $\p_2(G/H)$. On the other hand, the general Dirac 
monopole configurations to be constructed by patching together $H$-gauge 
fields along the equator are classified by $\p_1(H)$. It is just the 
topology theorem that
$$\p_2(G/H)=\p_1(H)~.\eqno(3.19)$$ 

The exact solution to the Georgi-Glashow model in the limit of vanishing 
potential $(V=0)$ is discussed in the next section.
\vglue.2in

\section{Bogomol'nyi-Prasad-Sommerfield limit}

An exact monopole solution  with a non-vanishing topological charge, $n\neq 0$,
 cannot be 
invariant under the rotational subgroup $SO(3)_{\rm R}$ of the Lorentz group, 
because the Higgs fields must vary at spacial infinity. The solution cannot be
invariant under the global gauge transformations $SO(3)_{\rm G}$ either since,
otherwise, the Higgs fields must vanish. However, the lowest-energy monopole 
solution may still be invariant under the {\it diagonal} subgroup $SO(3)$ of 
the $SO(3)_{\rm R}\otimes SO(3)_{\rm G}$. When imposing this symmetry, one is 
left with the unique {\it Ansatz}~\footnote{Strictly speaking, the additional 
discrete symmetry which is a combination of parity and a sign \newline
${~~~~~}$ change of $\F$ has to be imposed too.}
$$ \F^a=\fracmm{e^a_r}{er}H(evr)~,\quad  A^a_0=0~,\quad
A^a_i=-\ve\ud{a}{ij}\fracmm{e^j_r}{er}\left[ 1-K(evr)\right]~,\eqno(4.1)$$
in terms of two radial real functions, $H$ and $K$, subject to the boundary
conditions (sect.~3)
$$\eqalign{
K(\x)\to 1~,\quad H(\x)\to 0~,\quad {\rm at}\quad ~&~ \x\to 0~,\cr
K(\x)\to 0~,\quad \fracmm{H(\x)}{\x}\to 1~,\quad {\rm at}\quad ~&~ \x\to 
\infty~,\cr}\eqno(4.2)$$
where the dimensionless parameter $\x=evr$ has been introduced. The mass of this
static field configuration is determined by eq.~(3.8),
$$\eqalign{
 M_{\rm M}=~&~\fracmm{4\p v}{e}\int_0^{\infty} \fracmm{d\x}{\x^2}\left[
\x^2\left(\fracmm{dK}{d\x}\right)^2 + K^2H^2 
+\ha\left(\x\fracmm{dH}{d\x}-H\right)^2 \right. \cr
~&~ \left. +\ha\left(K^2-1\right)^2 
+\fracmm{\l}{4e^2}\left(H^2-\x^2\right)^2\right]~,\cr}\eqno(4.3)$$
whereas the equations of motion (3.5) take the form
$$\eqalign{ \x^2 \fracmm{d^2K}{d\x^2}=&~KH^2 +K(K^2-1)~,\cr
 \x^2 \fracmm{d^2H}{d\x^2}=&~2K^2H+\fracmm{\l}{e^2}H(H^2-\x^2)~.\cr}
\eqno(4.4)$$

The system of non-linear differential equations (4.4) for the unknown radial 
functions $H$ and $K$ admits a finite energy solution, and it can be explicitly
integrated in a certain limit~\cite{ps}. In order to 
understand the nature of this limit, let us discuss first the
so-called {\it Bogomol'nyi bound}~\cite{bog}. This bound can be obtained by
considering the mass $M_{\rm M}$ of a static configuration with vanishing 
electric field,
$$\eqalign{
M_{\rm M}~&~=\int d^3r\,\left[ \fracm{1}{2}(\vec{B}^a\cdot\vec{B}^a
+\vec{D}\F^a\cdot\vec{D}\F^a)+V(\F) \right] \geq 
\int d^3r\, \fracm{1}{2}(\vec{B}^a\cdot\vec{B}^a
+\vec{D}\F^a\cdot\vec{D}\F^a)\cr
~&~=\fracm{1}{2}\int d^3r \,
(\vec{B}^a-\vec{D}\F^a)(\vec{B}^a-\vec{D}\F^a) +vg~,\cr}\eqno(4.5)$$
where we have used the expression for the magnetic charge $g$ in the form
$$ g=\int_{S^2_{\infty}}\vec{B}\cdot d\vec{S}=\fracmm{1}{v}\int_{S^2_{\infty}}
\F^a\vec{B}^a\cdot d\vec{S}=\fracmm{1}{v}\int \vec{B}^a(\vec{D}\F)^ad^3r~,
\eqno(4.6)$$
because of eq.~(3.15) and the Bianchi identity $\vec{D}\cdot \vec{B}^a=0$.
Eq.~(4.5) yields the famous {\it Bogomol'nyi bound}~\cite{bog}:
$$ M_{\rm M}\geq \abs{vg}~.\eqno(4.7)$$
This bound is saturated if and only if $V(\F)=0$ (and, of course, $\vec{E}=0$)
and the Bogomol'nyi equation
$$ \vec{B}^a=\vec{D}\F^a \eqno(4.8)$$
is satisfied.~\footnote{Writing $\F\equiv A_4$, the Bogomol'nyi equation in 
$R^3$ can be rewritten as the {\it self-dual} Yang-Mills \newline ${~~~~~}$
equation in Euclidean space  $R^4$: $F_{ab}={}^*F_{ab}$, where $a,b=1,2,3,4$, 
and all the fields are suppo-\newline ${~~~~~}$ sed to be independent 
upon $x_4$.} It should be noticed that the {\it first-order} Bogomol'nyi equation
implies the second-order equations of motion. The corresponding limit is known as 
the {\it Bogomol'nyi-Prasad-Sommerfield} (BPS) limit~\cite{ps,bog}:
$$\vec{E}^a=0~,\quad D_0\F^a=0~,\quad \vec{B}^a=\pm\vec{D}\F^a~.\eqno(4.9)$$
In quantum theory, where even the vanishing scalar potential may have radiative
corrections, it is therefore important to protect flat directions of the
potential, in order to achieve the Bogomol'nyi bound (see Part II).

After a substitution of the Ansatz (4.1) into the Bogomol'nyi equation (4.8),
one finds
$$ \x \fracmm{dK}{d\x}=-KH~,\qquad \x\fracmm{dH}{d\x}=H-(K^2-1)~,\eqno(4.10)$$
whose solution is~\cite{ps}
$$ H(\x)=\fracmm{\x}{\tanh\x}-1~,\qquad  K(\x)=\fracmm{\x}{\sinh\x}~.\eqno(4.11)$$

When inserting this solution into eq.~(4.3), one finds that the energy density is
concentrated in the small region around the origin (i.e. in the core of a 
monopole). At distances greater than a Compton wavelength 
$(ve)^{-1}\sim M_W^{-1}$, where $M_W$ is the mass of the $W^{\pm}$ gauge 
particles resulting from the spontaneous
symmetry breaking, the function $K$ exponentially vanishes. 
In physical terms, it means that there is a cloud of $W^{\pm}$ fields around 
the monopole while, well outside the monopole core, the magnetic field falls 
like $r^{-2}$, thus leaving that field configuration to be indistinguishable from 
the Dirac monopole. The Higgs fields also exponentially decay at spacial infinity,
but they also have a long-range piece falling as $r^{-1}$:
$$ \F^a\stackrel{r\to\infty}{\longrightarrow} ve_r^a-\fracmm{e_r^a}{er}~.
\eqno(4.12)$$
The presence of the last term follows from the Nambu-Goldstone theorem which 
predicts a massless `dilaton' field $D$ associated to the spontaneous breakdown
 of scale invariance. The field $D$ can be introduced as~\cite{har}
$$ \F^a=ve_r^ae^D~,\eqno(4.13)$$
while its dimensionless $D$-charge is given by
$$Q_D\equiv v\int_{S^2_{\infty}}\vec{\de}D\cdot d\vec{S}=\fracmm{4\p}{e}=g=
\fracmm{M_{\rm M}}{v}~.\eqno(4.14)$$

One can conveniently describe the field of a point monopole by a 
{\it dual} potential $\tilde{A}^{\m}=(\tilde{A}^0,\tilde{\vec{A}})$ defined by 
${}^*F=d\tilde{A}$. A coupling of another point monopole (of mass $M$) to this 
field is described by the action which is `dual' to the standard action for an 
electrically charged point particle in an electro-magnetic field,
$$ I_M=\int dt\left( -M\sqrt{1-\vec{v}^2}-g\tilde{A}^0+g\vec{v}\cdot
\tilde{\vec{A}}\right)~,\eqno(4.15)$$
where $\vec{v}$ is a velocity of the test monopole. A sum of the standard 
electromagnetic action, $I_{\rm e.-m.}=-\frac{1}{4}\int d^4x\,{}^*F{}^*F$, and
eq.~(4.15) defines the total action which gives rise to a Coulomb magnetic field 
for a monopole at rest in the origin, as well as to the standard Coulomb 
repulsion between like sign monopoles, as it should have been expected from the 
dual picture. This picture should however be corrected since the theory also 
includes the massless `dilaton' field $D$, as we already know, with the free 
action
 $I_D=\frac{v^2}{2}\int d^4x\,\pa_{\m}D\pa^{\m}D$. Accordingly, the full action 
must also include the coupling to the `dilaton' field, which is dictated by the 
fact that a shift of the `dilaton' field is equivalent to a shift in the mass of 
the monopole (scale invariance is spontaneously broken~!). The action (4.15) 
should therefore be modified as~\cite{man}
$$ I_{M,D}=\int dt\left( -[M+vD]\sqrt{1-\vec{v}^2}-g\tilde{A}^0+g\vec{v}\cdot
\tilde{\vec{A}}\right)~.\eqno(4.16)$$
The ultimate force between two stationary monopoles to be computed from that 
action turns out to be zero, which is consistent with the existence of 
multi-monopole static configurations~\cite{man}.~\footnote{There is, of course, a
non-zero interaction between a monopole and an anti-monopole.} The space of 
solutions to the Bogomol'nyi equation (4.8) is called the {\it moduli space}, 
and it has dimension $4m$~\cite{ah}. Amongst the $4m$ {\it moduli} parameterizing 
the moduli space, $3m$ are just the space coordinates of the monopole locations, 
whereas the rest $(m)$ corresponds to the monopole excitations of the 
electrically charged $W^{\pm}$ fields in the core of the monopole.

In quantum theory, the classical BPS solution corresponds to a new particle -- 
a {\it BPS state} -- which is not present in the perturbative spectrum
of the quantized Georgi-Glashow model, and whose mass is proportional to the 
{\it inverse} of the gauge coupling constant $e$, according to eq.~(4.3). The last
remark also explains why this BPS state cannot be seen in the weak coupling
limit -- simply because the mass of this state becomes very large when $e\to 0$.

The electro-magnetic duality (2.21) implies a generalization of the
Bogomol'nyi bound (4.7) to the form
$$ M_{\rm D}\geq v\sqrt{q^2+g^2}~,\eqno(4.17)$$
which applies to {\it dyons} having both a magnetic charge $g$ and an electric 
charge $q$. In order to verify eq.~(4.17), one has to construct a dyon solution. 
It was found by allowing a nonvanishing electric potential of the type~\cite{jz}
$$ A^a_0 = \fracmm{e^a_r}{er} J(r)~,\eqno(4.18)$$
in the Ansatz (4.1). The equations of motion are therefore modified, namely,
$$\eqalign{
\x^2\fracmm{d^2K}{d\x^2}~=&~K\left(K^2+H^2-J^2-1\right)~,\cr
\x^2\fracmm{d^2H}{d\x^2}~=&~2K^2H+\fracmm{\l}{e^2}H\left(H^2-\x^2\right)~,\cr
\x^2\fracmm{d^2J}{d\x^2}~=&~2K^2J~,\cr}\eqno(4.19)$$
whose solution {\it in the BPS limit} $\l\to 0$ reads~\cite{jz}
$$ \eqalign{
H(\x)~=&~\cosh\g\left(\fracmm{\x}{\tanh\x}-1\right)~,\cr
K(\x)~=&~\fracmm{\x}{\sinh\x}~,\cr
J(\x)~=&~\sinh\g\left(\fracmm{\x}{\tanh\x}-1\right)~,\cr}
\eqno(4.20)$$
where $\g$ is an arbitrary constant. The charges and the mass of this classical 
object are given by ({\it cf.} eq.~(4.6))
$$q\equiv\fracmm{1}{v}\int d^3x\,D_i\F^aE^a_i=\fracmm{4\p}{e}\sinh\g~,\qquad
g\equiv \fracmm{1}{v}\int d^3x\,D_i\F^aB^a_i=\fracmm{4\p}{e},\eqno(4.21)$$
and
$$ M=\fracmm{4\p}{e}v\cosh^2\g~.\eqno(4.22)$$
It is now easy to verify the bound in eq.~(4.17). It is remarkable that the BPS 
mass formula 
$$ M_{\rm BPS}=v\sqrt{q^2+g^2}~,\eqno(4.23)$$
 does not distinguish between the `fundamental' quantum particles and the 
monopoles, being applicable to all of them, like the meson-soliton democracy in 
the SG model. Semiclassical quantization of the dyon solution leads to the
electric charge quantization~\cite{osborn} (see also the next sect.~5),
$$ q=en_e~,\quad {\rm where}\quad n_e\in {\bf Z}.~\eqno(4.24)$$

Thus, we just learned in this section that 
\begin{itemize} 
\item the BPS limit implies the existence of new BPS states in the quantum 
theory, which are absent in the perturbative spectrum,
\item the BPS mass formula (4.23) is universal, and it is unvariant under 
the electro-magnetic duality,
\item the Coulomb repulsion between like sign static monopoles is exactly 
cancelled by the dilaton attraction.
\end{itemize}

\section{Witten effect and S duality}

The previous discussion of the electro-magnetic duality, although being supported
by the BPS mass formula and the moduli space structure, still leaves many 
quesitons to be unanswered within the framework of the Georgi-Glashow model. 
For instance, the quantum  Georgi-Glashow model cannot be duality invariant, since
there are quantum corrections to all masses, which are not under control in that 
model. Also, since the $W$-bosons have spin, the magnetic monopoles should also 
have spin, whose origin in the Georgi-Glashow model is unclear. One still needs 
an underlying theory for describing dyons. The necessary additional input is
provided by the extended supersymmetry~\footnote{The supersymmetry is discussed 
in the Part II.} and the so-called $\q${\it -term} (or {\it vacuum angle}), 
which can be added to the 
Yang-Mills Lagrangian without spoiling its renormalizability,
$$\cl_{\q}=-\,\fracmm{\q e^2}{32\p^2}F^a_{\m\n}{}^*F^{a\m\n}~.\eqno(5.1)$$
Being a total derivative, it does not affect the classical equations of 
motion, it violates $P$ and $CP$, but not $C$, which makes it as a good 
candidate for generalizing the long-range behaviour of the theory while 
maintaining duality.

As was first noticed by Witten~\cite{wi79}, the allowed values of electric
charge in the monopole sector of the theory become shifted by the $\q$-term. 
For instance, an electromagnteic field in the presence of a Dirac monopole
takes the form
$$ \vec{E}=\vec{\de}A_0~,\qquad \vec{B}=\vec{\de}\times\vec{A}
+\fracmm{g}{4\p}\fracmm{\vec{e}_r}{r^2}~.\eqno(5.2)$$
Its substitution into eq.~(5.1) yields
$$\eqalign{
L_{\q}=\int d^3r\,\cl_{\q}=&~\fracmm{\q e^2}{8\p^2}\int d^3r\,
\vec{\de}A_0\cdot
\left( \vec{\de}\times\vec{A} +\fracmm{g}{4\p}\fracmm{\vec{e}_r}{r^2}\right)
\cr
&~=-\,\fracmm{\q e^2g}{32\p^3}\int d^3r\,A_0\vec{\de}\fracmm{\vec{e}_r}{r^2}
=-\,\fracmm{\q e^2g}{8\p^2}\int d^3r\,A_0\d^3(\vec{r})~,\cr}\eqno(5.3)$$
which is just the coupling of the scalar potential $A_0$ to an {\it 
electric} charge of magnitude $-\q e^2g/(8\p^2)$ located at the origin. In 
other words, the magnetic monopole has aquired an electric charge~!

A more fundamental derivation of the same fact is based on the full 
spontaneously broken gauge theory with $\q$-term, whose total Lagrangian
$\cl_{\rm tot}$ is given by the sum of eqs.~(3.1) and (5.1) in the BPS 
limit~\cite{wi79}. By the full theory here one means the non-local theory where 
magnetically neutral particles occur as quantum excitations of the fields present
in the action, whereas magnetically charged particles (the BPS states) accur as 
solitons. Consider now a gauge transformation from the unbroken $U(1)$ subgroup 
(i.e. about the axis $e_{\F}^a\equiv\F^a/\abs{\F^a}$) with the gauge 
parameter approaching a constant at spacial infinity. In the infinitesimal 
form, it is given by
$$ \d A_{\m}^a=\fracmm{1}{ve}(D_{\m}\F)^a~,\eqno(5.4)$$
where $\F^a$ is the background monopole Higgs field. Let $\cn$ be the 
generator of that gauge transformation. Its explicit form can be easily
computed via the Noether method,
$$ \cn=\fracmm{\pa\cl_{\rm tot}}{\pa(\pa_0 A^a_{\m})}\d A^a_{\m}=
\fracmm{q}{e} +\fracmm{\q eg}{8\p^2}~,\eqno(5.5)$$
where we have used $\d A^a_{\m}$ as of eq.~(5.4), as well as the definitions 
of the total magnetic and electric charges in eq.~(4.21).
Since the rotation by $2\p$ about the axis $e_{\F}^a$ should 
be trivial on physical states, we must have
$$ \exp (2\p i\cn)=1~,\quad {\rm or,~~equivalently,} 
\quad \cn=n_{\rm e} \in {\bf Z}~.\eqno(5.6)$$
Together with the Dirac quantization condition, $eg=4\p n_{\rm m}$ where
$n_{\rm m} \in {\bf Z}$, eq.~(5.5) now implies
$$ q=en_{\rm e} -\fracmm{e\q}{2\p}n_{\rm m}~,\eqno(5.7)$$
which is a generalization of eq.~(4.24). The {\it Witten effect} described by 
eq.~(5.7) provides the physical meaning to the shift $\q\to\q+2\p$ which changes 
the induced electric charge of the BPS monopole. 

It was the original {\it Montonen-Olive conjecture}~\cite{mo} that the 
Georgi-Glashow model, i.e. the $SO(3)$ Yang-Mills-Higgs theory in the BPS limit 
(at $\q=0$) has an exact duality symmetry under the exchange of the fields, 
$\vec{E}\to\vec{B}$ and $\vec{B}\to -\vec{E}$, and the  exchange of the coupling 
constants,
$$e\to g=\pm\,\fracmm{4\p\hbar}{e}~.\eqno(5.8)$$
The dual or `magnetic' formulation of the theory will also be a spontaneously 
broken gauge theory with essentially the same Lagrangian, where the $W^{\pm}$ 
bosons would appear as solitons, while the BPS monopoles would be `fundamental'. 
It is clear that eq.~(5.8) represents a strong-weak coupling transformation, 
like that in eq.~(1.10) for the SG--T quantum equivalence. Unfortunately, the 
corresponding `vertex operator construction' connecting the two dual gauge theory
formulations is not known in four dimensions. 

The Montonen-Olive idea becomes extended when the $\q$-term is also taken into
account. First, both coupling constants $e$ and $\q$ can be united into one 
{\it complex} parameter $\t$,
$$ \t\equiv \fracmm{\q}{2\p} +\fracmm{4\p i}{e^2}~.\eqno(5.9)$$
Since the physics is periodic in $\q$ with period $2\p$, we have a duality
transformation
$$T:\qquad \t ~\to~ \t+1~,\eqno(5.10)$$
whereas the Montonen-Olive duality transformation (5.8) in terms of $\t$ takes
the form~\footnote{Note that it is the $\t$, not $e$, which is inverted.}
$$S:\qquad \t ~\to~ -\,\fracmm{1}{\t}~.\eqno(5.11)$$
It seems to be quite reasonable that the full duality symmetry is generated
by the two transformations (5.10) and (5.11). They generarate the group 
$SL(2,{\bf Z})$ of projective transformations
$$ \t ~\to~ \fracmm{a\t + b}{c\t + d}~,\quad {\rm where}\quad
a,b,c,d \in {\bf Z}~,\quad{\rm and}\quad ad-bc=1~.\eqno(5.12)$$

Since $e^2>0$, the parameter $\t$ naturally lives on the upper half plane,
${\rm Im}\,\t\geq 0$. Because of eqs.~(5.7) and (5.9), the transformation (5.10)
shifts the electric charge by $-1$ (for $n_{\rm m}=1$), while the 
transformation (5.11) exchanges electric and magnetic quantum numbers 
$n_{\rm e}$ and $n_{\rm m}$. Putting all together, the action of the 
$SL(2,{\bf Z})$ on the quantum numbers reads
$$\left(\begin{array}{c} n_{\rm e} \\ n_{\rm m} \end{array}\right)
~\to~ \left( \begin{array}{cc} a & -b \\ c & -d \end{array}\right) 
\left(\begin{array}{c} n_{\rm e} \\ n_{\rm m} \end{array}\right)~.\eqno(5.13)$$

When being rewritten in terms of $\t$, the BPS bound
$M^2\geq v^2(Q^2_{\rm e}+Q^2_{\rm m})$, where $Q_{\rm e}\equiv 
en_{\rm e}-\fracm{\q e}{2\p}n_{\rm m}$ and $Q_{\rm m}\equiv 
\fracm{4\p}{e}n_{\rm m}$, takes the form
$$M^2\geq 4\p v^2(n_{\rm e},n_{\rm m})\fracmm{1}{{\rm Im}\,\t}\left(
\begin{array}{cc} 1 & -{\rm Re}\,\t \\ -{\rm Re}\,\t & \abs{\t}^2 \end{array}
\right)\left(\begin{array}{c} n_{\rm e} \\ n_{\rm m}\end{array}\right)~,
\eqno(5.14)$$
which is $SL(2,{\bf Z})$ invariant~! Most of the key equations above can be 
conveniently represented in terms of new variables
$$ a\equiv ve~,\qquad {\rm and }\qquad a_{\rm D}\equiv \t a~.\eqno(5.15)$$
In particular, eqs.~(5.10) and (5.11) now read
$$\eqalign{
T=\left( \begin{array}{cc} 1 & 1 \\ 0 & 1 \end{array} \right):~&~ 
\left( \begin{array}{c} a_{\rm D} \\ a \end{array}\right) ~\to~
\left( \begin{array}{c} a+a_{\rm D} \\ a \end{array}\right)~,\cr
S=\left( \begin{array}{cc} 0 & -1 \\ 1 & 0 \end{array} \right):~&~ 
\left( \begin{array}{c} a_{\rm D} \\ a \end{array}\right) ~\to~
\left( \begin{array}{c} -a \\ a_{\rm D} \end{array}\right)~,\cr}
\eqno(5.16)$$
while the BPS mass spectrum is given by
$$ M_{\rm BPS}=\abs{an_{\rm e} +a_{\rm D}n_{\rm m}}~.\eqno(5.17)$$

Since the mass formula should be duality invariant, the charge vector 
$q_m=(n_{\rm m},n_{\rm e})$ also gets transformed under $M\in
SL(2,{\bf Z})$ to $q'=qM^{-1}$, where ${q'}_m=({n'}_{\rm m},{n'}_{\rm e})$ are 
also integers. The {\it stable} BPS states
are those for which $n_{\rm m}$ and $n_{\rm e}$ are relatively prime~\cite{sen}.

Therefore, we just learned that
\begin{itemize}
\item in the presence of the $\q$-term, the naive (Montonen-Olive) 
electro-magnetic duality becomes extended to the projective transformations 
$SL(2,{\bf Z})$.
\end{itemize}

The extension (5.12) of the Montonen-Olive duality is called S-{\it duality} 
\cite{filq}. The S-duality invariance is a very strong 
requirement in quantum field theory. In particular, it implies that the 
renormalization group trajectories (if the theory has a non-vanishing 
beta-function) must be confined in the fundamental region of $SL(2,{\bf Z})$
in the $\t$-plane. If the beta-function is vanishing, the S-duality
implies that the partition function of the theory is modular invariant (i.e.
it must be a modular form). The only known candidates for such a behaviour
are given by the {\it finite} gauge theories with N=2 or N=4 extended 
supersymmetry. It is the extended supersymmetry that also explains from the 
fundamental point of view the Bogomol'nyi bound, and provides an exact quantum 
status to the BPS states. Therefore, it order to proceed further in our studies 
of duality, we need to learn more about the extended supersymmetry in the next 
Part II of the review.
\vglue.2in

\newpage



{\Large\bf PART II: INTRODUCING SUPERSYMMETRY}
\vglue.2in

In this Part of the review, some aspects of supersymmetry, which are going to be
relevant in the Part III, are discussed. The emphasis is made on the superspace 
approach to the supersymmetric gauge theories with $N=1$ and $N=2$ supersymmetry.
The BPS bound is related to the central charges appearing in the $N=2$ extened 
supersymmetry algebra. The field content and the classical component action of 
the $N=4$ supersymmetric gauge theory, which is believed to be exactly self-dual
under the S-duality, is given. As a pre-requisite to the Seiberg-Witten results
to be discussed in Part III, the moduli space of the $N=2$ {\it supersymmetric
Yang-Mills} (SYM) theory, its renormalization and the {\it low-energy effective 
action} (LEEA) are introduced. 
\vglue.2in

\setcounter{section}{0}

\section{Supersymmetry algebras and their \\ representations}

The Lorentz group $SO(1,3)$ has the covering group $SL(2,{\bf C})$. Accordingly, 
a four-component (complex and anticommuting) Dirac spinor $\J_{\rm D}$ is a 
reducible representation. One can introduce the irreducible two-component
complex spinors $\j_{\a}$ and $\bar{\c}^{\Dot{\a}}=(\c_{\a})^*$ instead,
$$ \J_{\rm D}=\left( \begin{array}{c} \j_{\a} \\
\bar{\c}^{\Dot{\a}} \end{array}\right)~.\eqno(1.1)$$
The two-component spinor indices are raised and lowered with the
antisymmetric $\ve$-tensors, which represent the charge conjugation matrix,
$$\ve^{\a\b}=\ve^{\Dot{\a}\Dot{\b}}=\left( \begin{array}{cc} 0 & 1 \\
-1 & 0 \end{array}\right)~.\eqno(1.2)$$
We use the notaion $\j\c\equiv\j^{\a}\c_{\a}$ and
$\bar{\j}\bar{\c}\equiv\bar{\j}_{\Dot{\a}}\bar{\c}^{\Dot{\a}}$, so that
$(\j\c)^{\dg}=\bar{\j}\bar{\c}$. A convenient representation for the
$4\times 4$ Dirac matrices is given by
$$ \g^{\m}=\left( \begin{array}{cc} 0 & \s^{\m} \\
\bar{\s}^{\m} & 0 \end{array}\right)~,\eqno(1.3)$$
where the $2\times 2$~ $\s$-matrices are defined by
$$ (\s^{\m})_{\a\Dot{\a}}=({\bf 1},\vec{\s})_{\a\Dot{\a}}~,\qquad
(\bar{\s}^{\m})^{\Dot{\a}\a}=\ve^{\Dot{\a}\Dot{\b}}\ve^{\a\b}(\s^{\m})_{\b
\Dot{\b}}=({\bf 1},-\vec{\s})^{\Dot{\a}\a}~,\eqno(1.4)$$
and $\vec{\s}=(\s^1,\s^2,\s^3)$ are Pauli matrices.
The $\g_5=-i\g^0\g^1\g^2\g^3$ is diagonal in this representation, while
$\g_5=1$ for the upper
two components of $\J_{\rm D}$ and $\g_5=-1$ for the lower two components.
The two-component spinors can be identified with the Weyl (chiral)
spinors. A Majorana spinor is defined by
$$ \J_{\rm M}=\left( \begin{array}{c} \j_{\a} \\
\bar{\j}^{\Dot{\a}} \end{array}\right)~.\eqno(1.5)$$
One finds
$$\s^{\m}P_{\m}=\left( \begin{array}{cc}  P^0-P^3 & -P^1+iP^2\\
-P^1-iP^2& P^0+P^3 \end{array}\right)~,\quad{\rm so~~that}\quad
\det\s^{\m}P_{\m}=P^{\m}P_{\m}~.\eqno(1.6)$$
The Lorentz transformations for two-component spinors are generated by
$$\eqalign{
(\s^{\m\n})\low{\a} {}^{\b} = & \fracmm{1}{4} \left[ \s^{\m}_{\a\Dot{\b}}
\bar{\s}^{\n\Dot{\b}\b}-(\m\leftrightarrow\n)\right]~,\cr
(\bar{\s}^{\m\n}) {}^{\Dot{\a}} {}_{\Dot{\b}} = &
\fracmm{1}{4}\left[ \bar{\s}^{\m\Dot{\a}\b}\s^{\n}_{\b\Dot{\b}}
-(\m\leftrightarrow\n) \right]~. \cr}\eqno(1.7)$$

The $N$-{\it extended supersymmetry} (susy) algebra without central charges
reads
$$ \{ Q^I\low{\a},\bar{Q}_{\Dot{\a}J} \}=2\s^{\m}_{\a\Dot{\a}}P_{\m}\d\ud{I}{J}~,
\qquad  \{ Q^I_{\a},Q^J_{\b} \}= \{ \bar{Q}\low{I\Dot{\a}},\bar{Q}_{J\Dot{\b}} \}
=0~,\eqno(1.8)$$
where $I,J=1,2,\ldots,N$.~\footnote{It is assumed in what follows that $N$ is
either 1, 2 or 4.}  The {\it massive} susy irreducible representations (irreps)
 can be easily found by using Wigner's method of induced representations.
Defining $P^{\m}=(M,0,0,0)$ and rescaling the charges, one can represent
eq.~(1.8) as {\it two} Clifford algebras, each having the form
$$\{ a^I,(a^J)^{\dg}\}=\d^{IJ}~.\eqno(1.9)$$
Hence, without incorporating CPT invariance, the susy irrep over the spin-$j$
`vacuum' $\ket{\O}_j$ has dimension $(2j+1)2^{2N}$. There is always an equal
number of bosons and fermions, all having the same mass. The maximal helicity gap
amongst the states in the representation is $N$. For example, if $N=1$ and $j=0$,
one arrives at a {\it chiral} $N=1$ susy multiplet comprising a Majorana (or Weyl)
 spinor and a complex scalar (2 bosonic and 2 fermionic degrees of freedom). 
Similarly, if $N=1$ and $j=1/2$, one finds an $N=1$ {\it vector} multiplet which 
can be represented in field theory by a vector field, a Dirac fermion and a real 
scalar (4 bosonic and 4 fermionic degrees of freedom). The minimal massive $N=2$ 
multiplet has $2^4=16$ states, whereas in the $N=4$ case the minimal number of 
states increases to $2^8=256$ while the spin $2$ appears.

As far as the {\it massless} susy irreps are concerned, the situation is 
different. When choosing a frame where $P^{\m}=M(1,0,0,1)$, one easily finds 
from eq.~(1.8) that
$$ \{ Q^I\low{\a},\bar{Q}_{\Dot{\a}J} \}=\left( \begin{array}{cc} 0 & 0 \\
0 & 4M \end{array} \right)\d\ud{I}{J}~.\eqno(1.10)$$
Hence, one of the Clifford algebras can be trivially realized, which effectively 
reduces the number of creation and destruction operators by half. As a result, 
there are now only $(2j+1)2^N$ states in the multiplets. If the
vacuum has helicity $\l$, the highest helicity state has $\l+\fracm{1}{2}N$.
Accordingly, it yields
$$\eqalign{
N=1:\qquad & \qquad \ket{\l},~\quad \ket{\l-1/2}~;\cr
N=2:\qquad & \qquad \ket{\l},~\quad 2\ket{\l-1/2}~,\quad \ket{\l-1}~;\cr
N=4:\qquad & \qquad \ket{\l},~\quad 4\ket{\l-1/2}~,\quad 6\ket{\l-1}~,\quad
4\ket{\l-3/2}~,\quad \ket{\l-2}~.\cr}\eqno(1.11)$$
In a local field theory, which is $CPT$ invariant, one has to append the states 
(1.11) with their $CPT$ conjugates, unless they are already $CPT$ invariant.

The extended susy algebra can be modified~\cite{hls}:
$$ \{ Q^I\low{\a},\bar{Q}_{\Dot{\a}J} \}=2\s^{\m}_{\a\Dot{\a}}P_{\m}\d\ud{I}{J}~,
\qquad  \{ Q^I_{\a},Q^J_{\b} \}=\ve_{\a\b}Z^{IJ}~,\quad
 \{ \bar{Q}_{I\Dot{\a}},\bar{Q}_{J\Dot{\b}} \}=\ve_{\Dot{\a}\Dot{\b}}Z^*_{IJ}~,
\eqno(1.12)$$
where the {\it central charges} $Z^{IJ}=-Z^{JI}$ have been introduced. In the
$N=2$ case, they reduce to a single (complex) central charge, $Z^{IJ}=2\ve^{IJ}Z$,
while $Z$ can be fixed to be real by a chiral rotation. Defining
$$\eqalign{
 a_{\a}=& \fracmm{1}{\sqrt{2}} [Q^1_{\a} +\ve_{\a\b}(Q^2_{\b})^{\dg} ]~,\cr
b_{\a}=& \fracmm{1}{\sqrt{2}} [Q^1_{\a} -\ve_{\a\b}(Q^2_{\b})^{\dg}]~,\cr}
\eqno(1.13)$$
one finds for massive represenations that
$$ \{ a_{\a},a_{\b}^{\dg}\}=2(M+\abs{Z})\d_{\a\b}~,\qquad
 \{ b_{\a},b_{\b}^{\dg}\}=2(M-\abs{Z})\d_{\a\b}~,\eqno(1.14)$$
while all the other anticommutators vanish. Eq.~(1.14) leads to the bound
$$ M\geq \abs{Z}~.\eqno(1.15)$$
When this bound becomes saturated, $\abs{Z}=M$, the massive representation
becomes smaller, and one gets a {\it reduced} massive multiplet comprising the
BPS states (sect.~I.4). The reduction
mechanism is quite similar to that for the massless susy representations without 
central charges, and it results in the same number of states at given $N$. This
fact is important for a consistency of the Higgs mechanism in supersymmetric
gauge theories, which assumes an equal number of degrees of freedom before and
after spontaneous gauge symmetry breaking. For example, a {\it reduced} or
{\it short} massive $N=2$ multiplet can have only two bosonic and two fermionic 
degrees of freedom. Similarly, there are only $16$ states in the short massive 
representation of $N=4$ supersymmetry. 

The states that become massive by the Higgs mechanism must belong to {\it short}
supermultiplets, as they were before the spontaneous symmetry breaking, since the
Higgs mechanism cannot generate the extra massive states which appear in an
unreduced (`long') massive supermultiplet.

One concludes that the status of BPS states in a short representation of extended
supersymmetry is well defined in quantum theory, unless the supersymmetry is not 
broken, because 
\begin{itemize}
\item the BPS states are protected by extended supersymmetry. Hence, their 
existence does not depend on the underlying dynamics of quantum theory. 
In particular, it remains to be true at strong coupling,
\item The short massive supermultiplets of BPS states can be equally defined by 
requiring {\it a half} of the supersymmetry generators to vanish on them, 
in the presence of central charges.
\end{itemize}

\section{$N=1$~ field theories and superspace}

Since supersymmetry representations in field theory appear as miltiplets
comprising bosonic and fermionic fields, one needs their unified description. 
Such a description is provided by {\it superspace}~\cite{wb,ggrs,west,psbook}.
The basic idea of superspace is to extend spacetime by anticommuting
coordinates that are spacetime spinors and whose number is just equal to the
number of supersymmetry generators. The supersymmetry transformations can then
be realized as certain translations in superspace, while a tensor in
superspace automatically provides a supersymmetry representation. In the case
of the unextended ({\it simple}) $N=1$ supersymmetry, the $N=1$ superspace 
coordinates are
given by $z^M=(x^{\m},\q^{\a},\bar{\q}^{\Dot{\a}})$. The superspace
realization of the supersymmetry generators (without central charges) reads
$$ Q_{\a} =+\fracmm{\pa}{\pa\q^{\a}}-i\s^{\m}_{\a\Dot{\a}}\bar{\q}^{\Dot{\a}}
\pa_{\m}~,\qquad
\bar{Q}_{\Dot{\a}}=-\fracmm{\pa}{\pa\bar{\q}^{\Dot{\a}}}+i\s^{\m}_{\a\Dot{\a}}
\q^{\a}\pa_{\m}~.\eqno(2.1)$$
However, there is a problem since a general superfield provides a reducible
representation of supersymmetry. Hence, one needs to develop a covariant
calculus in superspace. The main tools are the spinorial covariant derivatives
$$ D_{\a} =+\fracmm{\pa}{\pa\q^{\a}}+i\s^{\m}_{\a\Dot{\a}}\bar{\q}^{\Dot{\a}}
\pa_{\m}~,\qquad
\bar{D}_{\Dot{\a}}=-\fracmm{\pa}{\pa\bar{\q}^{\Dot{\a}}}-i\s^{\m}_{\a\Dot{\a}}
\q^{\a}\pa_{\m}~,\eqno(2.2)$$
which anticommute with the supersymmetry generators (2.1), and satisfy a
similar algebra. The simplest chiral scalar multiplet is given by the {\it
chiral} scalar superfield $\F$ satisfying the superspace constraint
$$ \bar{D}_{\Dot{\a}}\F=0~.\eqno(2.3)$$
This constraint can be easily solved,~\footnote{We use the conventional
notation: $\q^2=\q^{\a}\q_{\a}$, $\q\s^{\m}\bar{\q}=\q^{\a}\s^{\m}_{\a\Dot{\a}}
\bar{\q}^{\Dot{\a}}$, etc.} $\F=\F(y,\q)=\f(y) +\sqrt{2}\q\j(y)+\q^2F(y)$,
where $y^{\m}=x^{\m}+i\q\s^{\m}\bar{\q}$, or, more explicitly,
$$ \F=\f(x) +\sqrt{2}\q\j(x) +\q^2F(x) +i\q\s^{\m}\bar{\q}\pa_{\m}\f(x)
-\fracmm{i}{\sqrt{2}}\q^2
(\pa_{\m}\j(x)\s^{\m}\bar{\q})-\fracm{1}{4}\q^2\bar{\q}^2\pa^2\f(x)~.
\eqno(2.4)$$
The {\it antichiral} superfield $\F^{\dg}$ satisfies $D_{\a}\F^{\dg}=0$,
whereas a supersymmetry invariant action is simply given by a full superspace
integral
$$\fracmm{1}{4} \int d^4 x d^2\q d^2\bar{\q}\,\F^{\dg}\F=\int d^4x\,
(\pa_{\m}\f\pa^{\m}\f^{\dg}-i\bar{\j}\bar{\s}^{\m}\pa_{\m}\j+F^{\dg}F)~.
\eqno(2.5)$$

Obviously, any function of chiral superfields is again a chiral superfield, so
that eq.~(2.5) can be easily generalized to include interactions, whose most
general form is
$$\cl=\int d^4\q\,K(\F,\bar{\F}) +\int d^2 \q\,W(\F)
+ \int d^2\bar{\q}\,\bar{W}(\bar{\F})~,\eqno(2.6)$$
where a {\it K\"ahler} potential $K$ and a holomorphic {\it superpotential} $W$
have been introduced.

A general real scalar superfield $V$ can be written down in the form~\cite{wb}
$$\eqalign{
V(x,\q,\bar{\q}) =  & C+i\q\c-i\bar{\q}\bar{\c}+\fracm{i}{2}\q^2(M+iN)
-\fracm{i}{2}\bar{\q}^2(M-iN) -\q\s^{\m}\bar{\q}A_{\m} \cr
& +i\q^2\bar{\q}(\bar{\l}+\frac{i}{2}\bar{\s}^{\m}\pa_{\m}\c)
-i\bar{\q}^2\q(\l+\frac{i}{2}\s^{\m}\pa_{\m}\bar{\c})+\fracm{1}{2}
\q^2\bar{\q}^2 (D-\frac{1}{2}\bo C)~.\cr}\eqno(2.7)$$
This superfield is a reducible representation of supersymmetry, since it 
contains the smaller chiral and antichiral superfields, $\L$ and $\L^{\dg}$. 
They can be effectively removed by imposing the gauge symmetry
$$ V\to V +i\L -i\L^{\dg}~.\eqno(2.8)$$
In the so-called {\it Wess-Zumino gauge}, one chooses $C=M=N=\c=0$, which
reduces eq.~(2.7) to
$$ V= -\q\s^{\m}\bar{\q}A_{\m} +i\q^2(\bar{\q}\bar{\l})
-i\bar{\q}^2(\q\l)+\fracm{1}{2}\q^2\bar{\q}^2 D~,\eqno(2.9)$$
thus defining a {\it vector} multiplet comprising a massless gauge field
$A_{\m}$, its superpartner ({\it gaugino}) $\l_{\a}$ and a real auxiliary
field $D$. Note that $V^3\equiv 0$ in the Wess-Zumino gauge.

The abelian gauge-invariant superfield strength is given by the chiral spinor
superfield $W_{\a}$,
$$W_{\a}= -\,\fracm{1}{4}\bar{D}^2D_{\a}V~,\qquad \bar{W}_{\Dot{\a}}=
-\,\fracm{1}{4}D^2\bar{D}_{\Dot{\a}}V~,\eqno(2.10)$$
satisfying the constraint $DW=\bar{D}\bar{W}$. In the Wess-Zumino gauge, it
reads
$$ W(y)=-i\l +\q D -i\s^{\m\n}\q(\pa_{\m}A_{\n}-\pa_{\n}A_{\m})
 +\q^2\s^{\m}\pa_{\m}\bar{\l}~.\eqno(2.11)$$

In the non-abelian theory, all the fields of the vector multiplet, as well as
the corresponding superfields. are to be assigned in the adjoint,
$A_{\m}=A^a_{\m}t^a$, $\[t^a,t^b\]=f^{abc}t^c$, $\tr(t^at^b)=2\d^{ab}$, etc.
The non-abelian version of eqs.~(2.8) and (2.10) actually follows from a
solution to the constraints on the gauge-covariant and super-covariant
spinorial derivatives, ${\cal D}_{\a}$ and $\bar{\cal D}_{\Dot{\a}}$, defining
the super-Yang-Mills theory in superspace~\cite{wb}. Instead of going into
detail, the form of the non-abelian solution can be anticipated from the
abelian eqs.~(2.8) and (2.10). For example, as regards the gauge
transformations with
a Lie-algebra valued chiral superfield parameter $\L$, one finds that
$$ e^{-2eV}\to e^{+i\L^{\dg}}e^{-2eV}e^{-i\L}~,\eqno(2.12)$$
whereas, as far as the non-abelian superfield strength is concerned, it reads
$$\eqalign{
W_{\a} &~\equiv\fracmm{1}{8e}\bar{D}^2\left(e^{2eV}D_{\a}e^{-2eV}\right)=
-\,\fracmm{1}{4}\bar{D}^2\left( D_{\a}V +e\[V,D_{\a}V\]\right)~,\cr
&~= -i\l +\q D -i\s^{\m\n}\q F_{\m\n} +\q^2\s^{\m}\de_{\m}\bar{\l}~,\cr}
\eqno(2.13)$$
where the Wess-Zumino gauge has been used, while
$F_{\m\n}=\pa_{\m}A_{\n}-\pa_{\n}A_{\m}-ie\[A_{\m},A_{\n}\]$ and
$\de_{\m}\l=\pa_{\m}\l-ie\[A_{\m},\l\]$ as usual. The gauge-invariant kinetic
term for the matter chiral superfields in some (for example, the adjoint)
representaion is given by
$$\eqalign{
I_{\rm matter} & =\fracmm{1}{4}\int d^4x d^2\q d^2\bar{\q}\, \tr
(\F^{\dg}e^{-2eV}\F) \cr
& = \fracmm{1}{4}\int d^4x d^2\q d^2\bar {\q} \,
\tr \left( \F^{\dg}\F -2e\F^{\dg}V\F +2e^2\F^{\dg}V^2\F\right)~,\cr
& = \int d^4x\,\tr\left( \abs{\de_{\m}\f}^2 -i\bar{\j}\bar{\s}^{\m}\de_{\m}\j
+F^{\dg}F \right. \cr
& \left. ~~~~~~-e\f^{\dg}\[D,\f\] -ie\sqrt{2}\f^{\dg}\{\l,\j \} 
+ie\sqrt{2}\bar{\j} \[ \bar{\l},\f \] \right)~,\cr} \eqno(2.14)$$
whereas the natural (complex) kinetic term for the gauge fields reads
$$ -\fracmm{1}{4}\int d^4x d^2\q\,\tr W^{\a}W_{\a}=
\int d^4x\,\tr\left[
-\fracm{1}{4} F_{\m\n}F^{\m\n}+\fracm{i}{4}F_{\m\n}{}^*F^{\m\n}
-i\l\s^{\m}\de_{\m}\bar{\l}+\fracm{1}{2}D^2\right]~.\eqno(2.15)$$
In addition to the standard kinetic term for the Yang-Mills field, eq.~(2.15)
also contains the $\q$-term, as required by supersymmetry. We are therefore
guided by supersymmetry to introduce the complex coupling constant $\t$ as in
eq.~(I.5.9), and then define the following {\it real} action:
$$\eqalign{
I_{\rm SYM} ~&~= \fracmm{1}{16\p}{\rm Im}\left[\t\int d^4x d^2\q\,\tr\,
W^{\a}W_{\a}\right] \cr
 ~&~=\fracmm{1}{e^2}\int d^4 x\,\tr\left[ -\fracm{1}{4} F_{\m\n}F^{\m\n}
-i\l\s^{\m}\de_{\m}\bar{\l}+\fracm{1}{2}D^2\right] +\fracmm{\q}{32\p^2}
\int d^4x\, F_{\m\n}{}^*F^{\m\n}~.\cr}\eqno(2.16)$$

It can be shown that the non-abelian superfield strength $W_{\a}$ is (i) a
covariantly chiral superfield, $\bar{\cal D}_{\Dot{\a}}W\low{\a}=
{\cal D}\low{\a}\bar{W}_{\Dot{\a}}=0$, and (ii) satisfies the constraint
$$ {\cal D}^{\a}W_{\a}=\bar{\cal D}_{\Dot{\a}}\bar{W}^{\Dot{\a}}~.\eqno(2.17)
$$
These two conditions actually {\it define} the $N=1$ super-Yang-Mills theory
in superspace, and determine the component content of the theory in the
Wess-Zumino gauge, as given above.
\vglue.2in

\section{N=2 super-Yang-Mills theory}

The most natural framework for the $N=2$ extended supersymmetry is provided
by $N=2$ superspace, whose coordinates $z^M=(x^{\m},\q^{\a}_i, \bar{\q}^{
\Dot{\a}}_i)$ contain two sets of the anticommuting spinor variables $(i=1,2)$
related to each other by internal symmetry rotations. The $N=2$
super-Yang-Mills (SYM) theory in $N=2$ superspace can be defined by
imposing appropriate constraints on the gauge-covariant and super-covariant
spinorial derivatives ${\cal D}\low{\a}^i$ and
$\bar{\cal D}_{\Dot{\a}}^i$~\cite{gsw}.
The constraints essentially amount to the existence of the $N=2$~ SYM field
strength -- a covariantly chiral scalar $N=2$ superfield $\J$ --
satisfying the reality condition
$$ {\cal D}^{\a}_i{\cal D}_{\a j}\J=
\bar{\cal D}_{\Dot{\a}i}\bar{\cal D}^{\Dot{\a}}_{j}\bar{\J}~,\eqno(3.1)$$
which is analogous to eq.~(2.17). However, unlike in the $N=1$ case, an $N=2$
supersymmetric solution to the $N=2$ non-abelian superspace constraints is not
known in an analytic form.~\footnote{The $N=2$ analogue of the $V$-superfield
is given by an unconstrained $N=2$ tensor superfield \newline ${~~~~~}$ $V_{ij}$
of dimension $-2$. An analytic relation between $\J$ and $V_{ij}$ is
not known in the non-abelian \newline ${~~~~~}$ case.} Therefore, instead of 
discussing the $N=2$ constraints and their solution in $N=2$ superspace, we are 
going to make a `short cut', and first construct the $N=2$~ SYM theory in terms 
of $N=1$ superfields.

Since the on-shell field content of an $N=2$ vector multiplet is given by a
sum of an $N=1$ vector multiplet and a chiral $N=1$ scalar multiplet, in the
Wess-Zumino gauge, where the super-gauge degrees of freedom are eliminated, we
should expect the gauge-covariant $N=2$~ SYM field strength $\J$ be expressible
in terms of the $N=1$ gauge-covariant superfields $\F$ and $W_{\a}$, all in the
adjoint representation of the gauge group. Expanding the $N=2$ covarianly
chiral superfield $\J$ in terms of a `half' of proper chiral anticommuting
coordinates,
$$ \J=\F + \sqrt{2}\Q^{\a}W_{\a} + \Q^2 G~,\eqno(3.2)$$
we can represent $\J$ in terms of three gauge-covariant $N=1$ chiral
superfields, $\F$, $W_{\a}$ and $G$. Using dimensional reasons, we can now
identify the $N=1$ superfields $\F$ and $W_{\a}$ with the superfields appearing
in eqs.~(2.4) and (2.13), respectively. The remaining $N=1$ superfield $G$ is
expected to be a (complicated) gauge-covariant chiral function of $\F$ and $V$,
 whose explicit form we do not need~\cite{bilal}.

As far as the action of the $N=2$~ SYM theory is concerned, it should be given
by a sum of eqs.~(2.14) and (2.16) with  proper relative
normalization.~\footnote{The relative normalization is easily fixed by
requiring all fermionic kinetic terms to have the \newline ${~~~~~}$ same 
coefficients.} Hence, the $N=2$~ SYM action in  $N=1$ superspace reads as
follows:
$$\eqalign{
I_{N=2~SYM}=&\int d^4x \left[ {\rm Im}\left( \fracmm{\t}{16\p}\int d^2\q\,\tr\,
W^{\a}W_{\a}\right)+\fracmm{1}{4e^2}\int d^2\q d^2\bar{\q}\,\tr\,\F^{\dg}
e^{-2eV}\F\right]~,\cr
&={\rm Im}\;\tr\,\int d^4x\,\fracmm{\t}{16\p}\left[ \int d^2\q \,W^{\a}W_{\a}
+\int d^2\q d^2\bar{\q}\, \F^{\dg}e^{-2eV}\F\right]~.\cr}\eqno(3.3)$$

When dealing with an $N=2$ theory in $N=1$ superspace, one does not take care
of the underlying off-shell $N=2$ supersymmetry structure of the $N=2$
theory, while the on-shell physics is of course the same. It is also possible to
write down the $N=2$~ SYM action in $N=2$ superspace. The $N=2$
action should have the form of a chiral integral (on dimensional reasons), and
the only gauge-invariant candidate is given by the trace of the $N=2$~ SYM
superfield strength $\J$ squared. The correct answer reads
$$ I_{N=2~SYM}={\rm Im}\left(\fracmm{\t}{16\p}\int d^4xd^4\q\,\fracm{1}{2}\tr\,
\J^2\right)~.\eqno(3.4)$$

The $N=2$~ SYM action in components can be easily recovered from eqs.~(2.14),
(2.16) and (3.3). In particular, the structure of auxiliary fields is governed
by the action
$$ I_{\rm aux} = \fracmm{1}{e^2} \int d^4x\,\tr\,\left[ \fracm{1}{2}D^2 -
e\f^{\dg}\[D,\f\]+F^{\dg}F\right]~.\eqno(3.5)$$
Eliminating the auxiliary fields $D$ and $F$ via their algebraic equations of 
motion yields
$$ I_{\rm aux} = -\ha \int d^4x\,\tr\,\left( \[ \f^{\dg},\f \]
\right)^2~.\eqno(3.6)$$
The potential $V(\f)\equiv\frac{1}{2}\tr(\[ \f^{\dg},\f \])^2$ is therefore
non-negative, but has {\it flat} directions. The non-trivial solutions to the
equation $V(\VEV{\f})=0$ follow from  
$$ \[ \VEV{\f^{\dg}},\VEV{\f}\]=0~,\qquad \VEV{\f}\neq 0~,\eqno(3.7)$$
or, equivalently,
$$\[ \VEV{S},\VEV{P} \]=0~,\eqno(3.8)$$
where the scalar $S$ and the pseudo-scalar $P$ have been introduced,
$\f\equiv \fracmm{1}{\sqrt{2}}(S+iP)$. The parity-conserving solution to 
eq.~(3.8) in the $SU(2)$ case is
$$ \VEV{S^a}=v\d^{a3}~,\qquad \VEV{P^a}=0~,\eqno(3.9)$$
where the value of the real parameter $v$ is arbitrary. The set of all solutions 
to eq.~(3.7) modulo gauge transformations is the classical moduli space of the 
theory, which is parametrized by the gauge-invariant parameter 
$\tr\VEV{\f^2}=\fracmm{1}{2}v^2$ (see sect.~5 for more).

The $N=2$~ SYM~ Lagrangian in components can be written down in the form
$$\eqalign{
\Lag_{N=2~SYM}=~&~\fracmm{1}{4\p}\,{\rm Im}\, \left\{ \left( \fracmm{\q}{2\p}
+i\fracmm{4\p}{e^2}\right) \tr\left[ -\fracmm{1}{4}\left(F_{\m\n}F^{\m\n}
-\ha\ve^{\m\n\r\l}F_{\m\n}F_{\r\l}\right)\right.\right. \cr
~&~\left.\left. 
+(D_{\m}\f)^{\dg}(D_{\m}\f) -\ha\left(\[\f,\f^{\dg}\]\right)^2 +\ldots
\right]\right\}~,\cr}\eqno(3.10)$$
where the scalar and spinor component fields have been rescaled, and the dots 
stand for fermionic terms. In the $SU(2)$ case, eq.~(3.10) has the structure 
which is very similar to that of the Georgi-Glashow model, except of the 
potential. The $N=2$~ SYM~ action is classically scale (and conformally) 
invariant, but this invariance is spontaneously broken, if $\VEV{\f}\neq 0$. 
Unbroken supersymmetry requires the vanishing vacuum expectation values for all 
the auxiliary fields and, hence, implies $V(\VEV{\f})=0$.

With $SU(2)$ as the SYM gauge group, eq.~(3.9) at $v\neq 0$ spontaneously breaks 
it down to $U(1)$. The BPS monopole solution (Part I) can be embedded into the 
$N=2$~ SYM theory, whose fields $S^a$ replace $\F^a$ overthere and 
satisfy the Bogomol'nyi bound $B^a_i=D_iS^a$. Unlike the Georgi-Glashow
model, the BPS limit in the $N=2$ SYM theory can be reached without sending the
potential coupling constant to zero.

One can check whether a charge-one monopole
solution has some supersymmetry. Since the fermionic fields have to vanish
initially, their supersymmetry variations have to vanish too. The $N=2$
supersymmetry variation of gaugino's in the BPS limit is governed by the operator
$$(\s^{\m\n}F_{\m\n}-\g^{\m}\de_{\m}S)\ve=\g^iB_i(1-\g_5)\ve=0~,\eqno(3.11)$$
which implies that a chiral {\it half} of the supersymmetry remains
unbroken.~\footnote{The non-vanishing (of opposite chirality) supersymmetry
variations of gaugino's are Dirac's \newline ${~~~~~}$ zero-modes in the
monopole background.}

As was shown in sect.~1, the $N=2$ extended supersymmetry algebra can be
modified by the inclusion of central charges. In an $N=2$ supersymmetric field 
theory, the supersymmetry  charges are expressed as space integrals of 
supersymmetry currents given by certain polynomials in fields and their 
derivatives. In the presence of monopoles carrying magnetic charges, the central 
terms in the $N=2$ supersymmetry algebra of the $N=2$~ SYM theory can therefore 
be explicitly calculated. It was done by Olive and Witten~\cite{ow} who found that
$$\eqalign{
{\rm Re}\,Z=~&~\int d^3x\,\pa_i\left[ S^aE^a_i +P^aB_i^a\right]=vQ_{\rm e}~,\cr
{\rm Im}\,Z=~&~\int d^3x\,\pa_i\left[ P^aE^a_i +S^aB_i^a\right]=vQ_{\rm m}~,\cr}
\eqno(3.12)$$
where eq.~(3.9) has been used, as well as the definitions of the total electric 
and magnetic charges (Part I). Hence, one gets  
$$Z=v(Q_{\rm e}+iQ_{\rm m})\quad {\rm or,~~equivalently}~,\quad
\abs{Z}^2=v^2(Q_{\rm e}^2+Q_{\rm m}^2)~,\eqno(3.13)$$
as well as the Bogomol'nyi bound $M\geq \abs{Z}=v\sqrt{Q_{\rm e}^2+Q_{\rm m}^2}$ 
as the direct consequences of extended supersymmetry~! Inverting the argument, 
the Bogomol'nyi equation follows by demanding the monopole solution be 
annihilated by half of the supersymmetries ( i.e. form a {\it short} 
representation of $N=2$ supersymmetry). Therefore, if $N=2$ supersymmetry is not 
dynamically broken in quantum theory, the Bogomol'nyi bound is not going to be 
modified by quantum corrections, and the BPS states with magnetic charges will 
occur in the full quantum $N=2$~ SYM theory as well. If the `fundamental' 
particles get their masses via the Higgs mechanism which does not change the 
number of physical degrees of freedom, they also fall into reduced (short) 
representations of $N=2$ supersymmetry, and they can therefore also be considered
as the BPS states.

Assuming that the $N=2$ supersymmetry of the classical $N=2$~ SYM theory is 
maintained in the full quantum theory,~\footnote{The Witten index does not vanish
for the $N=2$~ SYM theory, which means that the $N=2$ \newline ${~~~~~}$
supersymmetry cannot be dynamically broken in that theory~\cite{w1}.} 
it is possible to predict the form 
of its {\it low-energy effective action},~\footnote{The low-energy part of the 
full (non-local) effective action represents the component kinetic 
\newline ${~~~~~}$ 
terms with no more than two derivatives, and no more than four-fermion couplings.}
$$ I_{\cf}=\fracmm{1}{16\p}{\rm Im}\,\int d^4x d^4\q\,\cf(\J)~,\eqno(3.14)$$
where $\cf$ is a holomorphic function, called the $N=2$ {\it prepotential}.
The classical part of the $N=2$ prepotential is dictated by eq.~(3.10):
$$ \cf_{\rm class}(\J)=\half\,\tr\,\t_{\rm cl}\J^2~.\eqno(3.15)$$
where $\t_{\rm cl}$ is given by eq.~(I.5.9). The quadratic dependence in 
eq.~(3.15) is crucial for renormalisability. In $N=1$
superspace, the $N=2$~ SYM~ low-energy effective action (3.14) reads 
as follows~\cite{gates}:
$$ I_{\cf}=\fracmm{1}{16\p}{\rm Im}\,\int d^4x \left[ \int d^2\q\, \cf_{ab}(\F)
W^{a\a}W^b_{\a} +\int d^2\q d^2\bar{\q}\,\left(\F^{\dg}e^{-2eV}\right)^a
\cf_a(\F)\right]~,\eqno(3.16)$$
where we have used the notation
$$\cf_a(\F)\equiv \pa\cf(\F)/\pa\F^a~,\qquad
\cf_{ab}(\F)\equiv \pa^2\cf(\F)/\pa\F^a\pa\F^b~.\eqno(3.17)$$

One concludes that
\begin{itemize}
\item the BPS condition which was initially found at the classsical level 
(Part I) is maintained in the full quantum theory as well, because it is a
consequence of the extended supersymmetry,
\item the mass formula for the BPS states (see e.g., the right-hand-side of
eq.~(I.5.14)) is {\it exact}, i.e. it holds in the full 
quantum theory, and it is valid for all particles in the semiclassical spectrum,
\item the low-energy effective action of the $N=2$~ SYM theory is governed by a
holomorphic prepotential $\cf$.
\end{itemize}

The  holomorphic function $\cf$ is expected to receive both perturbative and
non-perturbative contributions after quantization. The tools to calculate the
$N=2$ prepotential exactly, by using a non-trivial interplay between 
holomorphicity, extended supersymmetry and duality, will be provided in Part III.
\vglue.2in

\section{N=4 super-Yang-Mills theory}

Though the $N=4$ super-Yang-Mills theory can be formulated on-shell in the 
conventional N=4 superspace, it is very difficult to construct its off-shell
$N=4$ supersymmetric formulation, if any. Therefore, we are going to confine
ourselves to its component formulation. The easiest way to construct the 
four-dimensional $N=4$~ SYM theory is provided by dimensional reduction of the 
{\it ten-dimensional} supersymmetric gauge theory down to four 
dimensions~\cite{bss}. 

The main point here is related to the dimension of a spinor representation in 
various space-time dimensions. The number of on-shell bosonic degrees of freedom 
in the case of a real vector gauge field $A_M$ in $D$ dimensions is $D-2$, while 
the (real) number of on-shell fermionic degrees of freedom in the case of a Dirac
spinor $\l$ is $2^{[D/2]}$. Either the Weyl or the Majorana condition on $\l$ 
reduces the
last number by a factor of $1/2$. Therefore, the maximal dimension where the 
numbers of bosonic and fermionic degrees of freedom match for a minimal vector 
supermultiplet comprising $(A_M,\l)$ is $D=10$ {\it provided} that $\l$ is 
Majorana and Weyl simultaneously, which is allowed in ten 
dimensions.~\footnote{Similarly, the $N=2$~ SYM theory can be obtained by 
dimensional reduction from the super- \newline ${~~~~~}$ symmetric gauge theory 
in $D=6$ provided that the superpartner of the Yang-Mills field is a
\newline ${~~~~~}$ Weyl spinor in the adjoint representation of the gauge
group~\cite{bss}.}

The action of the supersymmetric Yang-Mills theory in ten dimensions reads
$$ I_{10}=\int d^{10} x\,\tr \left[ -\fracm{1}{4}F_{MN}F^{MN}-\fracm{1}{2}
\bar{\l}\G^M(D_M\l)\right]~,\eqno(4.1)$$
where both fields $A^a_M$ and $\l^a$ are in the adjoint of the gauge group, and
$$ (1-\G_{11})\l=0~,\qquad \bar{\l}=\l^{\rm T}C_{10}~.\eqno(4.2)$$
We use the standard notation:
$$F^a_{MN}=\pa_MA^a_N-\pa_NA_M^a-ef^{abc}A^b_MA^c_N~,\quad
(D_M\l)^a=\pa_M\l^a-ef^{abc}A^b_M\l^c~,\eqno(4.3)$$
as usual. In eq.~(4.2) one has $\G_{11}=\G_0\G_1\G_2\cdots\G_9$, while $C_{10}$ 
is the charge conjugation matrix in ten dimensions, 
$C_{10}\G_M C_{10}^{-1}=-\G_M^{\rm T}$. The early lower-case Latin letters are 
still used for the gauge group indices, while the capital Latin letters, 
$M,N,\ldots=0,1,\ldots,9$, are used to denote the Lorentz 
indices in ten dimensions. It is straightforward to verify that the action (4.1) 
is invariant under the supersymmetry transformations 
$$\d A^a_M=\bar{\ve}\G_M\l^a~,\qquad \d\l^a=-\s^{MN}F^a_{MN}\ve~,\eqno(4.4)$$
where the infinitesimal supersymmetry parameter $\ve$ is also a Majorana-Weyl 
spinor, and $\s^{MN}=\fracmm{1}{4}\[\G^M,\G^N\]$.

The dimensional reduction essentially amounts to requiring all the 
fields be only dependent on the four-dimensional space-sime coordinates 
$x^{\m}$, while $x^M=(x^{m},x^i)$ and $\m=0,1,2,3$. From the group-theoretical
viewpoint, it reduces the Lorentz group $SO(1,9)$ to $SO(1,3)\otimes SO(6)$. As
a result, the fermionic field $\l$ decomposes off-shell as 
$16=(2_+,4_+)+(2_-,4_-)$, where the
subscripts denote the space-time chirality. The ten-dimensional Dirac matrices 
can also be represented in terms of the four-dimensional Dirac matrices and some
internal $4\times 4$ matrices. Similarly, the gauge fields are decomposed 
off-shell as $10=(4,1)+(1,6)$, which leads to a gauge field, three scalars and 
three pseudo-scalars, all in the adjoint, in four dimensions. Because of the 
isomorphism $Spin(6)\equiv SU(4)$, the resulting four-dimensional Lagrangian can 
be written in various forms. For instance, the six scalar fields can be united 
into an antisymmetric complex matrix $\f_{ij}$ subject to the $SU(4)$ 
self-duality condition 
$$ \f_{ij}^{\dg}=\f^{ij}=\ha\ve^{ijkl}\f_{kl}~,\eqno(4.5)$$
where $i,j,\ldots=1,2,3,4$. As a result, the Lagrangian of the $N=4$~ SYM theory,
which follows from eq.~(4.1) after the dimensional reduction, is given by
$$\eqalign{
\Lag_{N=4~SYM}~=~&~\tr \left( -\fracmm{1}{4}F_{\m\n}F^{\m\n} +i\l_i\s^{\m}
D_{\m}\bar{\l}^i +\ha D_{\m}\f_{ij}D^{\m}\f^{ij}\right. \cr
~&~\left. +i\l_i\[ \l_j,\f^{ij}\] +i\bar{\l}^i\[ \bar{\l}^j,\f_{ij}\] + 
\fracmm{1}{4} \[\f_{ij},\f_{kl}\] \[\f^{ij},\f^{kl}\]\right)~.\cr}\eqno(4.6)$$
  
The $N=4$~ SYM theory also has monopole and dyon solutions, similar to the $N=2$~
SYM theory~\cite{os2}. In the $N=4$~ theory, it is actually possible to have 
monopoles carrying spin 1, which overcomes one of the obstacles mentioned in
Part I. Indeed, since there is a {\it unique} $N=4$ multiplet with the highest 
spin 1, the monopole $N=4$ supermultiplet must be isomorphic to the $N=4$
gauge supermultiplet, have 16 states, and one state of spin 1, in 
particular.~\footnote{In the $N=2$~ SYM theory, the monopole solution belongs to
a hypermultiplet~\cite{os2}, which does \newline ${~~~~~}$ not contain a spin-1
 state.} Moreover, the $N=4$~ SYM theory is known to be {\it 
UV-finite}~\cite{sow,sm,bln}, i.e. it has vanishing beta-function and it is 
exactly scale invariant. Altogether, it selects the $N=4$~ SYM theory as a good
candidate which may support the {\it exact} Montonen-Olive duality. In the $N=2$~
SYM theory, the S-duality can only be {\it effective}, not exact, being a subgroup
of $SL(2,{\bf Z})$ (see Part III for details).
\vglue.2in

\section{Moduli space of the ~$N=2$~ SYM~ theory}

The $N=2$~ SYM scalar potential has flat directions to be determined as solutions
to eq.~(3.7). All the vacuum field configurations define the vacuum `manifold'
(Part I) which is parametrized by the vacuum expectation values of the scalar
(Higgs) field. Since the vacua related by a gauge transformation describe the
same physics, we are interested in the gauge-inequivalent vacua forming the
{\it moduli space} ${\cm}$ and corresponding to the physically inequivalent
configurations. The moduli space ${\cm}$ generically has the structure of an
orbifold, i.e. it possesses singularities. The singularities of $\cm$ appear at
the points where the vacuum symmetry group is enhanced or, equivalently, its
dimension jumps.

The moduli space  ${\cm}$ of the $N=2$~ SYM theory has the natural
gauge-invariant vacuum `order' parameter, given by the quadratic Casimir 
eigenvalue,
$$ u\equiv \VEV{\tr\f^2}~.\eqno(5.1)$$
Eq.~(5.1) equally applies to the quantum moduli space, and any gauge group too.

In the $SU(2)$ case, the Higgs field is given by $\f=\f^a(x)t^a$, where the
$SU(2)$ generators $t^a$ have been introduced, $\tr(t^at^b)=2\d^{ab}$. The
classical vacuum configurations satisfying eq.~(3.7) can always be put by a
gauge transformation into the form $\VEV{\f}=\fracmm{1}{2}a t^3$ or, equivalently,
$$\VEV{\f}=\fracmm{1}{2}a \s_3~,\eqno(5.2)$$
where a complex constant $a$ has been introduced.  Hence, semiclassically. one
has $u=\fracmm{1}{2}a^2$ (see sect.~III.5 also).

Given a non-vanishing $\VEV{\f}$ or $a\neq 0$ semiclassically, the $SU(2)$ gauge 
symmetry is spontaneously broken to $U(1)$ by the Higgs mechanism. The gauge 
bosons $W_{\m}^{\pm}=\fracmm{1}{\sqrt{2}}(A^1_{\m}\pm iA^2_{\m})$ get mass
$m=\sqrt{2}a$ from the scalar kinetic term $\abs{\de_{\m}\f}^2$,~\footnote{The
corresponding gauginos also get the same mass by supersymmetry, thus forming a
 massive \newline ${~~~~~}$ $N=2$ vector multiplet.} whereas the rest of the
fields, comprising an abelian $N=2$ vector multiplet and a scalar one in the
$t^3$-direction, remain massless. The situation is different when $a=0$, where
the $SU(2)$ symmetry is unbroken, and all the fields are massless. Note that the
$SU(2)$ rotations by $\p$, forming the so-called (discrete) {\it Weyl subgroup} 
of $SU(2)$, change $a$ to $-a$, so that the corresponding vacuum states are 
gauge-equivalent. The classical moduli space is therefore given by the upper half
of a complex plane punctured at the  origin. The semiclassical (weak coupling) 
region corresponds to the area far away from the origin, while the strong coupling
region appears in the vicinity of the origin. 

It should be noticed that, after all quantum fluctuations are taken into account,
the {\it quantum} moduli space $\cm_{\rm q}$ may be very different from the
classical one. On the one hand, one should expect on physical grounds that a
classical singularity may disappear if the associated massless particle is not
stable under quantum corrections. On the other hand, new singularities in the
quantum moduli space may appear when a charged particle in the full quantum
spectrum of the theory becomes massless which results in the enhanced symmetry of
the physical vacuum. Although it is not known how to determine the structure of
the quantum moduli space from first principles, it can nevertheless be fixed
from a consistency of the full quantum theory (see Part III).

The existence of the quantum moduli space is guaranteed by the non-vanishing
{\it Witten index}~\cite{w1} and the {\it non-renormalization theorem} in $N=2$
supersymmetry~\cite{sei} (see also ref.~\cite{grsie} and the 
books~\cite{wb,ggrs,west,psbook} for more about the non-renormalization in 
supersymmetry). As was noticed in sect.~3, the $N=2$ supersymmetry
does not allow a superpotential for the $N=1$ chiral matter superfields in the
$N=1$ superspace formulation of the $N=2$~ SYM theory. Therefore, the classical
flat direction (5.2) remains in the full quantum theory provided that
the $N=2$ supersymmetry is not dynamically broken. A restriction on possible
dynamical supersymmetry breaking can be obtained from a calculation of the
Witten index $\tr(-1)^F$ which is essentially a topological index  counting a
difference between the zero-energy bosonic and fermionic states~\cite{w1}. The
supersymmetry is spontaneously broken if the vacuum energy is non-vanishing,
which implies the vanishing Witten index. A calculation shows that the Witten
index for the $N=2$~ SYM theory is different from zero~\cite{w1}, which means
that the $N=2$ supersymmetry is this theory is not going to be dynamically
broken and, hence, the existence of the quantum moduli space is justified.

Though the $SU(2)$ gauge symmetry is spontaneously broken to $U(1)$ in a
generic point of the moduli space, the $N=2$~ SYM low-energy effective action
is still $N=2$ supersymmetric. The low-energy effective action is therefore given
by an abelian $N=2$ gauge theory, whose $N=1$ superspace form is essentially 
described by eq.~(3.16), namely 
$$I_{\cf}^{\rm abelian} =\fracmm{1}{16\p}\,{\rm Im}\,\int d^4x\,
\left[ \int d^2\q\,\cf''(\F)W^{\a}W_{\a} +\int d^2\q d^2\bar{\q}\,\F^{\dg}
\cf'(\F)\right]~.\eqno(5.3)$$
After being written in components, eq.~(5.3) yields the kinetic terms
$$\eqalign{
I_{\cf}^{\rm abelian,~kin.}=
\fracmm{1}{4\p}\,{\rm Im}\,\int d^4x & \left[  -\fracm{1}{4}\cf''(\f) F_{\m\n}
(F^{\m\n}-i{}^*F^{\m\n}) + \cf''(\f) \abs{\pa_{\m}\f}^2 \right. \cr
&\left. -i\cf''(\f)(\l\s^{\m}\pa_{\m}\bar{\l}-\j\s^{\m}\pa_{\m}\bar{\j})
\right]~.\cr}\eqno(5.4)$$

A scalar field theory whose scalar fields are the coordinates of an (internal)
manifold is called the {\it non-linear sigma-model} (NLSM). The NLSM metric
$G$ is defined by the NLSM kinetic terms. In particular, as far as eq.~(5.4)
is concerned, one has $G_{\f\f^{\dg}}\sim\,{\rm Im}\,\cf''(\f)$. If the field
$\f$ is replaced by its vacuum expectation value $a$ parametrizing the modular
space of the $N=2$~ SYM theory, the NLSM metric reduces to the so-called
{\it Zamolodchikov metric} on the moduli space~\cite{zam},
$$ds^2={\rm Im}\,\cf''(a)dad\bar{a}={\rm Im}\,\t(a)dad\bar{a}~,\eqno(5.5)$$
where the effective (complexified) coupling constant $\t(a)$,
$$ \t(a)\equiv \cf''(a)~,\eqno(5.6)$$
has been introduced ({\it cf.} sect.~3). Unitarity requires the kinetic terms
to be positive definite, which implies that
$$ {\rm Im}\,\t(a) >0~.\eqno(5.7)$$
Since $\cf$ is a holomorphic function, ${\rm Im}\,\t$ is a harmonic function
and, therefore, it cannot have a minimum on the compactified complex plane.
This means that eq.~(5.7) {\it cannot} be satisfied in quantum theory unless
the $N=2$ prepotential $\cf$ is not globally defined throughout the moduli
space.~\footnote{The only exception is the classical formula (3.15) where
$\t$ is a constant.} Therefore, to ensure the kinetic terms in the effective
action be non-singular, the function $\cf$ can only be {\it locally} defined.
It means that we should use different $u$-coordinates to cover the whole quantum
moduli space $\cm_{\rm q}$, each of them being appropriate only in a certain
region of $\cm_{\rm q}$. It is the structure of singularities on $\cm_{\rm q}$
that tells us how many different local coordinates we really need (Part III).
\vglue.2in

\section{$N=2$~ SYM~ low-energy effective action \\ and renormalization group}

The Zamolodchikov metric is related to the renormalization group and the effective
action~\cite{zam}.~\footnote{See Chapter VIII of ref.~\cite{book} for a review.}
The effective action $\G[\vf]$ in quantum field theory is defined as the 
generating functional of {\it one-particle-irreducible} (1PI) Feynman 
diagrams. The functional $\G[\vf]$ is formally given by a Legendre transform of 
the generating functional $W[\vf]$ of connected Feynman diagrams. Since the latter
has to be renormalized, it introduces a dependence upon the renormalization scale
$\m$ into $W[\vf]$ and $\G[\vf]$. In spontaneously broken gauge theories, the 
scale $\m$ is usually identified with the mass scale to be determined by the 
Higgs mechanism, i.e. the vacuum expectation value of the Higgs scalar. The 
effective coupling constant $e_{\rm eff}(\m)$ is defined as the coefficient at the
corresponding 1PI vertex function, with its external momenta squared being equal 
to $\m^2$. If a quantum field theory has massless particles, as it usually 
happens in the gauge theories, on should introduce both an {\it ultra-violet}
(UV) cutoff and an {\it infra-red} (IR) one, in order to fully regularize the 
theory. It then becomes important whether momentum integrations in loop 
diagrams are performed from the UV-cutoff (to be taken to infinity after 
divergence subtractions) down to zero, or they are only performed down to 
${\m}$ which usually serves as the IR-cutoff. In the latter case, the 
corresponding effective action $S_W[\vf;\m]$ is called the {\it Wilsonian} 
effective action~\cite{russian}. In supersymmetric gauge theories, one should 
also distinguish between the two definitions of effective action, because of the 
so-called {\it Konishi anomaly}~\cite{kon}, which implies that the physical 
beta-functions to be defined with respect to the two effective actions are also
different.~\footnote{The Konishi anomaly is the field theory analogue of the 
two-dimensional holomorphic anomaly \newline ${~~~~~}$ which is well-known 
in string theory~\cite{dph}.} The Wilsonian effective coupling $e_{\rm eff}(\m)$ 
of a supersymmetric gauge theory is {\it holomorphically} dependent upon the scale
$\m$, which is not the case for the standard effective action $\G$. It is the
property that makes the Wilsonian effective action to be preferable in the case of
the quantum $N=2$~ SYM theory, whose low-energy effective action has the 
holomorphic structure due to $N=2$ supersymmetry. Eqs.~(I.5.9), (3.10) and (5.6)
imply the following relation between the Zamolodchikov metric and the 
renormalized (Wilsonian) coupling constants:
$$ {\rm Re}\,\t(\m)=\fracmm{\q(\m)}{2\p}~,\qquad
{\rm Im}\,\t(\m)=\fracmm{4\p}{e^2(\m)}~,\eqno(6.1)$$
where the effective vacuum angle ($\q$-parameter) $\q(\m)$ has been introduced. 
Though being unrenormalized in perturbation theory, the vacuum angle is expected 
to receive non-perturbative corrections from multi-instanton processes.

Because of the renormalization, the question arises is it the renormalized or the 
unrenormalized coupling that enters the Dirac quantization condition (I.2.20) and
its DZS generalization (I.2.23)~?  It does not matter for the $N=4$~ SYM theory
which is UV-finite, but it matters for the $N=2$~ SYM theory which is not 
UV-finite, and, therefore, whose duality properties need to be elaborated further.

The pure (without extra matter) $N=2$~ SYM theory with the gauge group $SU(2)$
is an {\it asymptotically free} theory. The running of its coupling constant 
$e(\m)$ is governed by the beta-function which receives both perturbative
{\it and} non-perturbative (due to instanton corrections) contributions. The 
perturbative one-loop beta-function can be calculated by standard perturbation
theory, with the result
$$ \b(e)\equiv \m \fracmm{de}{d\m} =-\,\fracmm{e^3}{4\p^2}~.\eqno(6.2)$$
It is remarkable that the higher-loop orders of perturbation theory do not
contribute to that (Wilsonian) beta-function. It can be argued by using either 
instanton methods~\cite{russian}, or superfield perturbation theory in the 
ordinary $(N=1)$ covariant superspace~\cite{grisi}, in the $N=2$ extended 
covariant superspace~\cite{hsw2}, or in the light-cone $N=2$ 
superspace~\cite{ket1}. The extended supersymmetry is crucial in all that 
approaches. As far as an $N=1$ supersymmetric gauge theory (with matter) is 
concerned, the general criterion of perturbative UV-finiteness, based on the 
knowledge of one-loop beta-function, was given in ref.~\cite{lpsi} (see also 
the book~\cite{psbook}). It should be noticed that all known finite $N=1$ 
supersymmetric gauge theories are based on a {\it simple} gauge group, i.e. 
they have a single gauge coupling, and their Yukawa couplings are functions of 
the gauge coupling. Both features are automatic in the extended supersymmetric 
gauge theories under consideration --- see e.g., eq.~(3.16).

A simple argument for the absence of all higher loop corrections to the $N=2$~
SYM beta-function (6.2) was given by Seiberg~\cite{sei2}. He noticed that the 
classical $N=~2$ SYM theory has the global symmetry $SU(2)\otimes U(1)$, where 
the $SU(2)$ rotates the two spinor superspace coordinates whereas the $U(1)$ 
(also called {\it R-symmetry}) multiplies them by a phase: $\q\to e^{-i\a}\q$, 
$\bar{\q}\to e^{-i\a}\bar{\q}$ and $\J\to e^{2i\a}\J$. The $R$-symmetry is 
anomalous, while the anomlay is given by the index theorem in the presence of an
instanton~\cite{sei2},
$$ \pa_{\m}j_R^{\m}=\fracmm{e^2}{8\p^2}\ve^{\m\n\r\l}F_{\m\n}F_{\r\l}~,
\eqno(6.3)$$
which is a non-perturbative phenomenon. The invariance of the perturbative 
effective action under the $U(1)_R$ symmetry restricts, however, the $N=2$
prepotential to the form
$$ \cf_{\rm per} (\J)=\J^2\left[ b_1 + b_2 \log\fracmm{\J^2}{\L^2}
\right]~,\eqno(6.4)$$
where $b_1$ and $b_2$ are two parameters to be determined from eqs.~(3.15) and
(6.2), respectively, and $\L$ is the renormalization-invariant scale at which the
gauge coupling becomes strong (see below). Some care should be excercised here,
since, though the perturbative effective action is $U(1)_R$ invariant, the 
effective Lagrangian is actually not. In fact, under an $U(1)_R$ rotation, the 
perturbative effective Lagrangian, 
$\Lag^{\rm eff}_{\rm per}=\int d^4\q  \cf_{\rm per} +h.c.$, transforms as
$$ \d \Lag^{\rm eff}_{\rm per}=\fracmm{\a}{4\p}
\ve^{\m\n\r\l}\tr(F_{\m\n}F_{\r\l})~,\eqno(6.5)$$
in agreement with eq.~(6.3).

It is clear from eq.~(6.4) that the first term represents the classical 
contribution whereas the second one is a one-loop effect,
$$ \cf_{\rm per}=\cf_{\rm cl} + \cf_{\rm 1-loop}~,\eqno(6.6)$$
where $\cf_{\rm cl}=\frac{1}{2}\t_{\rm cl}\J^2$ and 
$$\cf_{\rm 1-loop}(\J)=\fracmm{i}{2\p}\J^2\log\fracmm{\J^2}{\L^2}~.\eqno(6.7)$$
Therefore, after differentiating eq.~(6.6) twice. one finds 
$$ \fracmm{4\p}{e^2(\m)}+\fracmm{1}{\p}\log\fracmm{a^2}{\m^2}
=\fracmm{4\p}{e^2(a)}\equiv\fracmm{1}{\p}\log\fracmm{a^2}{\L^2}~,\eqno(6.8)$$
where the renormalization-invariant scale $\L$ is given by
$$\L^2=\m^2\exp\left\{-\,\fracmm{4\p^2}{e^2(\m)}\right\}~.\eqno(6.9)$$
In particular, one easily gets back eq.~(6.2).

The effective {\it field-dependent} coupling constant arises by setting the
renormalization scale ${\m}$ equal to the characteristic scale of the theory given
by the vacuum expectation value of the Higgs field: $e_{\rm eff}(\m)\to
 e_{\rm eff}(a)$. Eqs.~(6.6) and (6.7) imply at $a\to \infty$ that
$$ \t(a)=\fracmm{\pa^2\cf_{\rm per}(a)}{\pa a^2}\sim \fracmm{i}{\p}\left(
\log\fracmm{a^2}{\L^2}+3\right)~.\eqno(6.10)$$
The Zamolodchikov metric
${\rm Im}\,\t(a)\sim\fracmm{1}{\p}\log\fracmm{\abs{a}^2}{\L^2}~$
is therefore single-valued and positive in the semiclassical region 
$u\sim\ha a^2\to\infty$, as it should because of unitarity. 

Some useful information about multi-valued functions $f(u)$ can be obtained by
analyzing their behaviour as $u$ is taken around a closed contour. If there are no
special (singular) points inside the contour, the function $f(u)$ will return
to its initial value once $u$ has completed the loop. However, if there is a 
singularity, the multi-valued function $f(u)$ does not usually return to its 
initial value, which is known as a non-trivial {\it monodromy}. For example, it 
follows from eq.(6.10) that the loop around $u\sim\infty$ in the classical moduli
space produces a shift  $\t\to\t-2$ because of the branch cut of the logarithm. 
In its turn, it results in an irrelevant shift of
the vacuum angle ($\t$ like $\cf$ is also a multi-valued function~!). The full
story requires knowing the full set of singularities in the quantum moduli space
and the monodromy properties of $\cf$ (or $\t$), which are going to be discussed
in Part III.

In the IR-region (below $\L$), the positivity of ${\rm Im}\,\t$ is no longer 
secured by perturbation theory, and the instanton corrections become important.
One is left with an effective abelian gauge theory having vanishing beta-function.
In terms of the effective $\t$-parameter, one has~\cite{sei2}
$$ \fracmm{\q(a)}{2\p}+\fracmm{4\p i}{e^2(a)}=\fracmm{4\p i}{e^2_0}+\fracmm{i}{\p}
\log \fracmm{a^2}{\L^2}-\fracmm{i}{2\p}\sum^{\infty}_{l=1}c_l\left(
\fracmm{\L^2}{a^2}\right)^{2l}~,\eqno(6.11)$$
where the infinite sum over the instanton configurations with topological charge 
$l$ has been introduced. The unknown coefficients $c_l$ can, in principle, be 
calculated from zero-momentum correlators of the Higgs and gaugino's fields in 
multi-instanton backgrounds but, in practice, it was only done for a small number
of instantons. It is the recent achievement due to Seiberg and 
Witten~\cite{sw1} who determined the {\it exact} function $\cf$ and,
hence, the coefficients $c_l$ altogether (Part III). 

According to eq.~(6.11), one should expect the full $N=2$ prepotential to be of
the form
$$\cf(\J)=\ha\t_{\rm cl}\J^2 +\fracmm{i}{2\p}\J^2\log\fracmm{\J^2}{\L^2}+
\fracmm{1}{4\p i}\J^2\sum^{\infty}_{l=1}c_l\left( \fracmm{\L^2}{\J^2}
\right)^{2l}~,
\eqno(6.12)$$
which reproduces eq.~(6.11) after differentiating $\cf$ twice at 
$\left. a=\VEV{\J}\right|_{\q=0}$~.

To conclude this section, as well as the Part II, let me summarize some of the 
general features, which are apparent in the case of the $N=2$~ SYM~ theory.
Namely,
\begin{itemize}
\item the structure of the quantum moduli space does not need to be the same
as that of the classical moduli space,
\item one should use the Wilsonian effective action to compute the 
beta-function of renormalization group,
\item as far as the (Wilsonian) exact low-energy effective action is concerned,
it is the one-loop perturbative effects and non-perturbative instanton 
contributions that are only relevant, while the perturbation theory beyond one 
loop is irrelevant. 
\end{itemize}
\vglue.2in

\newpage

{\Large\bf PART III: Seiberg--Witten theory}
\vglue.2in

In the last Part III of our review, the exact solution to the low-energy effective
action in the $SU(2)$ pure (i.e. without $N=2$ matter) ~$N=2$~ SYM theory will be 
described, along the lines of the original work of Seiberg and Witten~\cite{sw1}.
Some generalizations to other gauge groups, as well as adding ~$N=2$~ matter,
will also be considered. We conclude with a very short discussion of the impact 
of that results on confinement and string theory.
\vglue.2in

\setcounter{section}{0}

\section{Quantum moduli space in the $SU(2)$ pure $N=2$ \\ SYM theory}

Unlike the $N=4$~ SYM theory which is supposed to be exactly self-dual in the
sense of Montonen-Olive, the $N=2$~ SYM theory cannot be self-dual. It is enough
to notice that the `fundamental' fields belong to an $N=2$ vector multiplet
whereas the magnetic monopoles belong to an $N=2$ scalar multiplet, i.e. an $N=2$~
hypermultiplet (Part II).  Nevertheless, the $N=2$~ theory still possesses the 
{\it effective} duality, which is now going to be explained.

First of all, one should understand the exact global structure of the quantum 
moduli space $\cm_{\rm q}$ of vacua. It is entirely determined by 
{\it singularities} of $\cm_{\rm q}$, which should be associated with
certain massless physical excitations. Therefore, the
global structure of  $\cm_{\rm q}$ can be physically motivated. 
The classical singularity at $u=0$ is due to extra massless gauge bosons 
$W^{\pm}$, and it results in the gauge symmetry enhancement from
$U(1)$ to $SU(2)$. The other singularity at $u=\infty$~\footnote{The moduli space
 is supposed to be compactified by adding the point at infinity.} is due to a
branch cut of the logarithm in eq.~(II.6.4) which is the one-loop renormalization
effect, and it is going to survive in the semiclassical region near $u=\infty$ in
the full quantum theory because of asymptotic freedom.

It was postulated by Seiberg and Witten~\cite{sw1} that $\cm_{\rm q}$ has just
{\it two} extra singularities at finite $u=\VEV{\tr\f^2}=\pm \L^2$, where $\L$ is
the dynamically generated quantum scale, while the classical singularity at $u=0$
in $\cm_{\rm cl}$ is absent in $\cm_{\rm q}$~. The absence of a singularity in the
origin of $\cm_{\rm q}$ means the absence of massless $W^{\pm}$ bosons in the full
quantum theory. Their presence would otherwise imply a superconformal invariance 
in the IR-limit, which is not compatible with any scale. Hence, the gauge 
symmetry is abelian over the whole quantum moduli space, at it never becomes 
restored to the full $SU(2)$ symmetry. The appearance of just two strong coupling
 singularities, where
certain t'Hooft-Polyakov monopoles (or dyons) become massless, is consistent with
earlier calculations of the Witten index, $\tr(-1)^F=2$, and they can be further
justified by the ultimate consistency of the solution (see the end of this
 section). If there were no quantum singularities at all, the coordinate $a$
 would be defined globally and unitarity would be lost --- see eq.~(II.5.7) and 
the discussion after that.~\footnote{The global ${\bf Z}_2$ symmetry $u\to -u$ 
implies that the number of strong coupling singularities must \newline
${~~~~~}$  be even. The only
fixed points of the ${\bf Z}_2$ symmetry are $u=\infty$ and $u=0$.}

Since the semiclassical masses of the BPS states are protected against quantum
corrections (Part II), the BPS mass formula (I.5.17) is valid in the full quantum
theory. In terms of the $N=2$~ SYM low-energy effective action, the dual variable
 $a_{\rm D}$ is simply given by
$$ a_{\rm D} =\fracmm{\pa\cf(a)}{\pa a}~,\eqno(1.1)$$
while $\pa a_{\rm D}/\pa a=\pa^2\cf/\pa a^2=\t(a)$.

In physical terms, the $a_{\rm D}$ is the `magnetic dual' of the `electric' Higgs
field $a$. By $N=2$ supersymmetry, the $a_{\rm D}$  has to be a part of the $N=2$ 
abelian vector multiplet containing the `magnetic dual' photon $A^{\rm D}_{\m}$. 
The electro-magnetic duality~\footnote{An explicit duality transformation will be
given in the next section~2.} relates $A^{\rm D}_{\m}$ to the `fundamental'
gauge potential $A_{\m}$~. Hence, the magnetic monopoles/dyons couple 
{\rm locally} to the dual photon, just like the `fundamental' $N=2$~
hypermultiplets, if present, locally couple to the electro-magnetic gauge 
potential $A_{\m}$~.
 The dual theory looks like the $N=2$ quantum electrodynamics which is not 
asymptotically free, and whose `magnetic' beta-function is positive ({\it cf}. 
eq.~(II.6.2)),
$$ \b_{\rm D}(e_{\rm D})\equiv \m \fracmm{de_{\rm D}}{d\m}=\,+\,
\fracmm{e^3_{\rm D}}{8\p^2}~.\eqno(1.2)$$
The $U(1)$ gauge theory does not contribute to the beta-function (1.2) whose
appearance is entirely due to the dual $N=2$ matter with unit charge coupling 
to the dual $N=2$ abelian vector multiplet. 

The BPS formula (I.5.17) is also consistent with the appearance of the quantum
singularity at $u=+\L^2$ where one should expect $a_{\rm D}=0$ but $a\neq 0$.
Indeed, a monopole hypermultiplet with charges $n_{\rm e}=0$ and $n_{\rm m}=1$ 
would
then be massless indeed, in agreement with eq.~(I.5.17). Also, since $\cm_{\rm q}$
is supposed to have no singularity at $u=0$, the semiclassical relation $u\simeq
\ha a^2$ cannot be globally valid in the full quantum moduli space.

The effective duality means that the variable $a_{\rm D}(u)$ should be considered
on equal footing with $a(u)$.~\footnote{We thus confine ourselves to the 
low-energy effective action, the duality is absent for the full \newline
${~~~~~}$ S-matrix~!} In
other words, it does not matter which variable is used to describe the theory ---
it only depends upon the region (in $\cm_{\rm q}$) to be described. It is the
semiclassical (`electric') region (near $u=\infty$) where the preferred local
variable is $a(u)$, whereas it is $a_{\rm D}(u)$ that is the preferred variable 
near the (`magnetic') strong coupling singularity at $u=\L^2$. Also, 
as was already noticed above,  the $a_{\rm D}$ belongs to the dual gauge 
multiplet that couples locally to magnetically charged excitations, in the same 
way that the $a(u)$ locally
couples to `electric' excitations. The full theory is of course non-local, which
manifests itself in the multi-valuedness of the prepotential $\cf$. In
the semiclassical region, the instanton sum in eq.~(II.6.12) converges well as
long as $a\simeq \sqrt{2u}\to\infty$. However, the same sum does not make sense 
outside 
the convergence domain. Since $\cf$ is not an analytic function, the instanton
terms in the strong coupling region have to be resummed in terms of some other
variables. In particular, near $u=\L^2$, one should expect another (dual) form of 
the effective Lagrangian,
$$ \cf_{\rm D}(\J_{\rm D})=\ha\t^{\rm D}_{\rm cl}\J^2_{\rm D}\,-\,\fracmm{i}{4\p}
\J^2_{\rm D}\log\left[ \fracmm{\J^2_{\rm D}}{\L^2}\right] + \fracmm{i}{2\p}\L^2
\sum^{\infty}_{l=1} c^{\rm D}_l \left(\fracmm{i\J_{\rm D}}{\L}\right)^l~,
\eqno(1.3)$$
which converges as $\J_{\rm D}\to 0$. In terms of the original variables, 
eq.~(1.3) describes a strong coupling. The coefficient in front of the
logarithm in eq.~(1.3) follows from eq.~(1.2), and it will be calculated below.

The other singularity at $u=-\L^2$ can be treated in a similar way, after 
replacing $a_{\rm D}$ in $\cf_{\rm D}(a_{\rm D})$ by $a_{\rm D}-2a$ (see below).
Hence, three patches are enough to cover the whole moduli space $\cm_{\rm q}$~. 
Inside of each patch (or {\it phase}), the theory is weakly coupled in
proper variables, and a 
local effective Lagrangian exists. The relation between the Lagrangians in 
different phases is however non-local. It is the patching together of
the local data about $\cm_{\rm q}$ in a globally consistent way that
will completely fix the theory. In other words, it is the absence of a `global' 
anomaly in the full quantum theory that is important.
    
Under an $SL(2,{\bf Z})$ duality transformation, the section $\left( 
\begin{array}{c} a_{\rm D}(u) \\ a(u) \end{array}\right)$ on 
$\cm_{\rm q}$ gets  transformed as
$$\left( \begin{array}{c} a_{\rm D}(u) \\ a(u) \end{array}\right) \longrightarrow
M \left( \begin{array}{c} a_{\rm D}(u) \\ a(u) \end{array}\right)~,\eqno(1.4)$$
where $M\in SL(2,{\bf Z})$ is nothing but a monodromy matrix, which is entirely
determined by the logarithmic terms in eqs.~(II.6.12) and (1.3). In particular,
in the semiclassical region near $u=\infty$, one has $u\simeq \ha a^2$ and
$$ a_{\rm D} =\fracmm{\pa\cf(a)}{\pa a} \simeq \fracmm{i}{\p}a\left( \log
\fracmm{a^2}{\L^2} +1\right)~,\eqno(1.5)$$
because of asymptotic freedom. Hence, taking the argument $u$ around a loop
encircling the point at infinity in $\cm_{\rm q}$ (which looks like 
$\cm_{\rm cl}$ near $u=\infty$) in a clockwise direction $(u\to e^{2\p i}u)$,
one finds that $a\simeq \sqrt{2u}\to-a$ and~\footnote{Eq.~(1.6) implies that
the mass of the magnetic monopole becomes infinite in the semiclassical 
\newline ${~~~~~}$ limit $a\to\infty$, as it should (Part I).}
$$ a_{\rm D}\to \fracmm{i}{\p}(-a)\left[ \log \fracmm{e^{2\p i}a^2}{\L^2} +1
\right]=-a_{\rm D}+2a~,\eqno(1.6)$$
because $u=\infty$ is a branch point of the logarithmic function in
eq.~(1.5), i.e.
$$\left( \begin{array}{c} a_{\rm D}(u) \\ a(u) \end{array}\right) \longrightarrow
M_{\infty} \left( \begin{array}{c} a_{\rm D}(u) \\ a(u) 
\end{array}\right)~,\eqno(1.7)$$
where
$$M_{\infty}=\left( \begin{array}{cc} -1 & 2 \\ 0 & -1 \end{array}\right)~.
\eqno(1.8)$$

Near the quantum singularity $u=+\L^2$, the renormalization  scale is proportional
to $a_{\rm D}\simeq \VEV{\F_{\rm D}}\sim 0$, which is the only scale there. In
the abelian gauge theory one has $\q_{\rm D}=0$ and, hence, $\t_{\rm D}=\fracmm{4
\p i}{e^2_{\rm D}(a_{\rm D})}$. We can now rewrite eq.~(1.2)  to the form
$$ a_{\rm D}\fracmm{d}{da_{\rm D}}\t_{\rm D}
=\,-\,\fracmm{i}{\p}~,\quad {\rm or}\quad
\t_{\rm D}=\,-\,\fracmm{i}{\p}\ln a_{\rm D}~,\eqno(1.9)$$
and integrate it further $(\t_{\rm D}=\,-\,da/da_{\rm D})$. Hence, near
 $a_{\rm D}\sim 0$, one finds in the leading order that 
$$ a \approx \fracmm{i}{\p}a_{\rm D}\ln a_{\rm D}~.\eqno(1.10)$$
It is enough to fix the coefficient in front of the logarithm in eq.~(1.3), 
as well as the monodromy as $u$ goes around the loop encircling $+\L^2$:
$$\left( \begin{array}{c} a_{\rm D}(u) \\ a(u) \end{array}\right) \longrightarrow
\left( \begin{array}{c} a_{\rm D}(u) \\ a(u)-2a_{\rm D}(u) \end{array}\right) =
M_{+\L^2} \left( \begin{array}{c} a_{\rm D}(u) \\ a(u) 
\end{array}\right)~,\eqno(1.11)$$
where
$$M_{+\L^2}=\left( \begin{array}{cc} 1 & 0 \\ -2 & 1 \end{array}\right)~.
\eqno(1.12)$$

The remaning monodromy matrix at $u=-\L^2$ can be calculated from the 
factorization condition
$$ M_{\infty}=M_{+\L^2}M_{-\L^2}~,\eqno(1.13)$$
which, in its turn, follows from the fact that a contour around $u=\infty$ can be
deformed into two contours, one encircling $\L^2$ and another encircling $-\L^2$.
One finds
$$M_{-\L^2}=\left( \begin{array}{cc} -1 & 2 \\ -2 & 3 \end{array}\right)~.
\eqno(1.14)$$

As was already noticed in sect.~I.5, a monodromy transformation can also be 
interpreted
as changing the magnetic and electric numbers $q_m=(n_{\rm m},n_{\rm e})$ by the
right multiplication with $M^{-1}$. The BPS state with vanishing mass,
which is responsible for a quantum singularity, should be {\it
invariant} under  the monodromy $M$, i.e. $q_m$ has to be the
eigenvector of $M^{-1}$ (or $M$) with {\it unit} eigenvalue. It is
obviously the case for
the magnetic monopole, with $q_m=(1,0)$ and the monodromy matrix (1.12).
Similarly, the eigenvector of $M_{-\L^2}$ in eq.~(1.14) with unit eigenvalue is
$(n_{\rm m},n_{\rm e})=(1,-1)$ which is {\it a dyon}~!~\footnote{An explicit
dyonic solution was constructed by Sen~\cite{sen}.}

In general, $(n_{\rm m},n_{\rm e})$ is the eigenvector of
$$M_{(n_{\rm m},n_{\rm e})}=\left( \begin{array}{cc} 1+2n_{\rm m}n_{\rm e} & 
+2n^2_{\rm e}  \\ -2n^2_{\rm m}  & 1-2n_{\rm m}n_{\rm n} \end{array}\right)~,
\eqno(1.15)$$
with unit eigenvalue. The matrix (1.15) would appear as the monodromy matrix for
the singularity due to a massless dyon with charges  
$q_m=(n_{\rm m},n_{\rm e})$.~\footnote{The monodromy matrix $M_{\infty}$   is not
of the form (1.15) since it does not correspond to a massless \newline
${~~~~~}$ physical state.} Again, one
finds a consistency with the initial proposal about the existence of only two
quantum singularities at $u=\pm\L^2$. Remarkably, no solution to the monodromy 
factorization condition exist in the case of more (finite number of) strong 
coupling singularities~\cite{dublin}. 

For comparison, it should be noticed that the monodromy group generated by the
singularities of the {\it classical} moduli space $\cm_{\rm q}$ is
{\it abelian}, and it reduces to irrelevant shifts of the vacuum angle,
$\q\to \q+2\p n$, $n\in {\bf Z}$. 

In conclusion, the general lessons from this section are:
\begin{itemize}
\item the classical vacuum degeneracy is not lifted by quantum corrections,
even after the non-perturbative instanton contributions are fully taken 
into account,
\item the monodromies around singularities in $\cm_{\rm q}$ represent
the duality transformations which either shift
the vacuum angle or connect weak and strong coupling,
\item the duality is not a symmetry of the theory, though the charges of the
{\it massless states} to be responsible for quantum singularities are invariant 
under the duality,
\item a consistency of the quantum theory severely restricts the
global structure of the quantum moduli space $\cm_{\rm q}\,$.
\end{itemize}

\section{Duality transformations}

The low-energy effective action is given by the $N=2$ supersymmetric {\it abelian}
 gauge theory whose form in $N=1$ superspace was written down in eq.~(II.5.3). 
Its dual can be explicitly constructed by the Legendre transform,
${\cf}_{\rm D}(\F_{\rm D})=\cf(\F)-\F\F_{\rm D}$, where 
$ \F_{\rm D}\equiv\cf'(\F)$, which implies
$${\cf'}_{\rm D}(\F_{\rm D})=\,-\,\F~.\eqno(2.1)$$
The  Legendre transform is known to be very similar to a canonical transformation,
with $\cf'(\F)$ playing the role of a canonical momentum. Since the canonical
transformations preserve the phase-space measure, it should not be surprising 
that the Jacobian of the duality transformation is also equal to one.

The second term in eq.~(II.5.3) is obviously invariant under the duality 
transformation,
$$\eqalign{
{\rm Im}\,\int d^4xd^2\q d^2\bar{\q}\,\F^{\dg}\cf' (\F)~&~=~
{\rm Im} \,\int d^4xd^2\q d^2 \bar{\q}\, \left( -{ \cf'}_{\rm D} (\F_{\rm D})
\right)^{\dg}\F_{\rm D}\cr
~&~=~{\rm Im}\,\int d^4xd^2\q d^2\bar{\q}\,\F^{\dg}_{\rm D}{ \cf'}_{\rm D}
(\F_{\rm D})~.\cr}\eqno(2.2)$$

As far as the first term in eq.~(II.5.3) is concerned, we need a dual  
$W^{\a}_{\rm D}$ to the abelian superfield strength $W^{\a}$. Unlike
the duality relation bewteen $\F_{\rm D}$ and $\F$, the relation between the  
$W^{\a}_{\rm D}$ and $W^{\a}$ cannot be local since it includes, in particular,
the duality relation between the component (abelian) field strengths 
$F^{\m\n}_{\rm D}$ and $F^{\m\n}$ (see Part I). The component Bianchi identity 
for the  $F^{\m\n}$ is a part of the superspace constraint (II.2.17),
which is equivalent to 
$${\rm Im}\,(D_{\a}W^{\a})=0~,\eqno(2.3)$$ 
and it follows from the abelian version of eq.~(II.2.13). Hence, the integration 
over the unconstrained superfield $V$ in the functional integral defining the 
quantum theory can be exchanged for the integration over $W^{\a}$ subject to the 
the constraint (2.3). The latter can be enforced by using a real Lagrange
multiplier $V_{\rm D}$ as follows:
$$\int {\cal D}V \exp \left\{ \fracmm{i}{16\p}\,{\rm Im}\,\int d^4x d^2\q\,
\cf''(\F)W^{\a}W_{\a}\right\} \simeq \eqno(2.4)$$
$$ \int {\cal D}W{\cal D}V_{\rm D}
\exp \left\{ \fracmm{i}{16\p}\,{\rm Im}\,\int d^4x \left(\int d^2\q\,
\cf''(\F)W^{\a}W_{\a} +\ha \int  d^2\q d^2\bar{\q}\,V_{\rm D}D_{\a}W^{\a}
\right)\right\}~.$$
One finds
$$\int  d^4x d^2\q d^2\bar{\q}\,V_{\rm D}D_{\a}W^{\a}=
\int d^4xd^2\q(\bar{D}^2D_{\a}V_{\rm D})W^{\a}=-4\int d^4xd^2\q\,(W_{\rm D})_{\a}
W^{\a}~,\eqno(2.5)$$
after integrating by parts, and using the relations $\bar{D}_{\dt{\b}}W^{\a}=0$
and $W_{{\rm D}\a}\equiv-\fracmm{1}{4}\bar{D}^2D_{\a}V_{\rm D}$. The remaining
functional integral over $W$ is Gaussian, and it yields  the dual action
$$ \int {\cal D} V_{\rm D} \exp \left\{ \fracmm{i}{16\p}\,{\rm Im}\,\int d^4x 
d^2\q\, \left(-\,\fracmm{1}{\cf''(\F)} W^{\a}_{\rm D}W_{{\rm D}\a}\right)
\right\}~.\eqno(2.6)$$
Note that the effective coupling $\t(a)=\cf''(a)$ has been replaced by the dual 
one, $-1/\t(a)$, which is nothing but the S-duality (I.5.11). Since
$$ \cf''_{\rm D}(\F_{\rm D})=-\,\fracmm{d\F}{d\F_{\rm D}}
=-\,\fracmm{1}{\cf''(\F)}~,\eqno(2.7)$$
one finds
$$ -\,\fracmm{1}{\t(a)}=\t_{\rm D}(a_{\rm D})~.\eqno(2.8)$$
The dual to the whole action (II.5.3) now takes the same form,
$$ \fracmm{1}{16\p}\,{\rm Im}\,\int d^4x\left\{
\int d^2\q\,{\cf''}_{\rm D}(\F_{\rm D})W^{\a}_{\rm D}W_{{\rm D}\a}+
\int d^2\q d^2\bar{\q}\F^{\dg}_{\rm D}{\cf'}_{\rm D}(\F_{\rm D})\right\}~,
\eqno(2.9a)$$
and it can be rewritten as
$$ \fracmm{1}{16\p}\,{\rm Im}\,\int d^4x d^2\q\,\fracmm{d\F_{\rm D}}{d\F}
W^{\a}W_{\a}+ \fracmm{1}{32\p i} \int d^4x d^2\q d^2\bar{\q} \left(
\F^{\dg}\F_{\rm D}-\F_{\rm D}^{\dg}\F\right)~.\eqno(2.9b)$$

The S-duality (I.5.11) is only a part of the the full duality group (sect.~I.5), 
and it corresponds to the transformation ({\it cf.} eq.~(I.2.6)) 
$$ \left( \begin{array}{c} \F_{\rm D} \\ \F \end{array} \right)
\longrightarrow  
\left( \begin{array}{cc} 0 & 1 \\ -1 & 0 \end{array} \right)
\left( \begin{array}{c} \F_{\rm D} \\ \F \end{array} \right)~.\eqno(2.10)$$
The transformation (2.10) is not a symmetry of the theory, but it relates its two
different parametrizations, one being more suitable for weak coupling while the
other for strong coupling. It follows from the form (2.9b) of the dual action 
that there is a symmetry
$$ \left( \begin{array}{c} \F_{\rm D} \\ \F \end{array} \right)
\longrightarrow
\left( \begin{array}{cc} 1 & b \\ 0 & 1 \end{array} \right)
\left( \begin{array}{c} \F_{\rm D} \\ \F \end{array} \right)~,\quad
{\rm where}\quad b\in {\bf Z}~,\eqno(2.11)$$ 
which only results in an irrelevant shift of the first term in eq.~(2.9b) by
$$\fracmm{b}{16\p}\,{\rm Im}\,\int d^4x d^2\q\,W^{\a}W_{\a}=\,-\,
\fracmm{b}{16\p}\int d^4x F_{\m\n}{}^*F^{\m\n}=\,-\,2\p b n~,\eqno(2.12)$$
where $n$ is the instanton number (sect.~I.3). The transformations (2.10) and 
(2.11) together generate the full S-duality group $SL(2,{\bf Z})$.

Since $a_{\rm D}(u)=\pa\cf(a)/\pa a$, the Zamolodchikov metric (II.5.5) can be 
rewritten in the explicitly $SL(2,{\bf Z})$-invariant form as
$$ds^2={\rm Im}\,(da_{\rm D}d\bar{a})=\fracmm{i}{2}\left(dad\bar{a}_{\rm D}
-da_{\rm D}d\bar{a}\right)=\,
-\,\fracmm{i}{2}\ve_{mn}\fracmm{dv^m}{du}\fracmm{d\bar{v}^n}{d\bar{u}}dud\bar{u}~,
\eqno(2.13)$$
where the two-dimensional vector
$$v^m\equiv \left( \begin{array}{c} a_{\rm D} \\ a \end{array} 
\right) \eqno(2.14)$$
is considered as a function of $u$.
\vglue.2in

\section{Seiberg-Witten elliptic curve}

A solution to the low-energy effective action or, equivalently, a calculation of
multi-valued functions $a_{\rm D}(u)$ and $a(u)$, was reduced in sect.~1 to the
standard {\it Riemann-Hilbert} (RH) problem of finding the functions with a given
monodromy around the singularities. A solution to the RH problem is known to be 
unique up to a multiplication by an entire function. The last ambiguity can be 
resolved in our case by the known asymptotical behaviour.

The monodromy matrices (1.12) and (1.14) generate the monodromy group $\G(2)$ 
which is a subgroup of the modular group $SL(2,{\bf Z})$,
$$ \G(2) =\left\{ \left(\begin{array}{cc} a & b \\ c & d\end{array}\right) \in
SL(2,{\bf Z})~,\quad b=0\,{\rm mod}\,2 \right\}~.\eqno(3.1)$$
The fact that the $N=2$ theory is
not self-dual becomes transparent by noticing that the S-duality (I.5.11) having
$b=1$ does not belong to the $\G(2)$. Still, there are other transformations in
eq.~(3.1) which relate weak and strong coupling, and it is the precise definition
of the effective duality in the $N=2$~ theory under consideration. 
The quantum moduli space is therefore given by
$$ \cm_{\rm q}\cong {\bf H}^+/\G(2)~,\eqno(3.2)$$
where ${\bf H}^+$ is the upper half-plane. 
 
It was the Seiberg-Witten idea~\cite{sw1} to introduce an auxiliary genus-one 
{\it Riemann surface} (elliptic curve) whose moduli space is precisely given by
$\cm_{\rm q}$ of eq.~(3.2), and whose period `matrix' (or elliptic modulus) is 
presicely the gauge coupling $\t(u)$. That auxiliary construction automatically 
guarantees positivity of the Zamolodchikov  metric $(\,{\rm Im}\,\t>0)$ because of
the well known `Riemann second relation' in the theory of Riemann 
surfaces~\cite{fkra}. In addition, it secures integer monodromy (see below). 

The relevant Riemann surface is defined by an algebraic equation 
$$ y^2(x,u)=(x^2-u)^2 -\L^4 \equiv \prod^4_{i=1}(x-e_i(u,\L))~,\eqno(3.3)$$
where
$$\eqalign{
e_1=-\sqrt{u+\L^2}~,\quad ~&~ e_2=-\sqrt{u-\L^2}~,\cr
e_3=+\sqrt{u-\L^2}~,\quad ~&~ e_4=+\sqrt{u+\L^2}~,\cr}\eqno(3.4)$$
and it can be represented in terms of two sheets (complex planes) connected 
through the cuts $\[e_1,e_2\]$ and $\[e_3,e_4\]$. The point at infinity is 
supposed to be added to each sheet, so that one gets the topology of a torus.

The period `matrix' ~$\t(u)$ of the torus is defined by a ratio of 
its {\it period} integrals,
$$ \t(u)=\fracmm{\o_{\rm D}(u)}{\o(u)}~,\eqno(3.5)$$
where 
$$ \o_{\rm D}(u)=\oint_{\b} \tilde{\o}~,\qquad \o(u)=\oint_{\a} \tilde{\o}~,\quad
{\rm with}\quad \tilde{\o}\equiv \fracmm{dx}{y(x,u)}~,\eqno(3.6)$$
and $(\a,\b)$ is a canonical homology basis of the torus.~\footnote{The cycle $\a$
can be chosen as a loop around $e_1$ and $e_2$, while the cycle $\b$ goes through
the cuts \newline ${~~~~~}$ and encircles $e_2$ and $e_3$.}

Since $\t=\pa a_{\rm D}/\pa a$, eq.~(3.5) suggests to identify
$$ \o_{\rm D}(u)=\fracmm{d a_{\rm D}(u)}{d u}~,\qquad  
\o(u)=\fracmm{d a(u)}{d u}~.\eqno(3.7)$$
Hence, both functions $a_{\rm D}(u)$ and $a(u)$, as well as the prepotential, 
$\cf=\int da\, a_{\rm D}(a)$, can be obtained by integration of the 
torus periods. One finds
$$ a_{\rm D}(u)=\int_{\b} \l~,\qquad a(u)=\int_{\a} \l ~,\eqno(3.8)$$
where the meromorphic one-form $\l$ is given by
$$ \l=x^2\tilde{\o}=x^2\fracmm{dx}{y(x,u)}~.\eqno(3.9)$$

The monodromy properties of the periods in eqs.~(3.6) and (3.8)
around the singularities in $\cm_{\rm q}$ fix them completely. Hence, it remains
to identify the singularities, and find the monodromy properties in the case of
basis cycles $\a$ and $\b$ of the Riemann surface (3.3).

The singularities arise when the torus degenerates, which happens if any two of
the branch points $e_i$ coincide, i.e. when the {\it discriminant}
$$ \prod^4_{i<j} (e_i-e_i)^2=(2\L)^8 (u^2-\L^4) \eqno(3.10)$$
vanishes. It results in the three possibilities:\\
(i)~~ $e_2\to e_3$ or $u\to +\L^2$, the cycle $\n_{+\L^2}\equiv \b$ degenerates,\\
(ii)~ $e_1\to e_4$ or $u\to -\L^2$, the cycle $\n_{-\L^2}\equiv \b-2\a$ 
degenerates,\\
(iii) $e_1\to e_2$ and  $e_3\to e_4$, or $\L^2/u\to 0$.

Going around a singularity in $\cm_{\rm q}$ results in an exchange of the branch 
points $e_i(u)$ along certain paths (called {\it vanishing cycles}) $\n$ 
shrinking to zero when one of the branch points approaches another one. 
For example, looping around the singularity $u=+\L^2$ results in the rotation of 
$e_2$ and $e_3$ around each other, so that the cycle  $\a$ gets transformed to 
$\a-2\b$, while the cycle $\b$ remains intact. This means that the monodromy
action is 
$$\left( \begin{array}{c} \b \\ \a \end{array}\right) \longrightarrow
M_{+\L^2} \left( \begin{array}{c} \b \\ \a \end{array}\right)~,\eqno(3.11)$$
where the monodromy matrix $M_{+\L^2}$ is exactly the one as in eq.~(1.12). 
Similarly, one finds that the monodromy matrix to be derived from the vanishing
cycle in the case (ii), near the singularity $u=-\L^2$, is precisely given by the
matrix $M_{-\L^2}$ of eq.~(1.14). The monodromy  $M_{\infty}$ has to be given by 
eq.~(1.8), just because of the consistency relation (1.13). The approach based on
the vanishing cycles is therefore justified. An explicit solution will be given 
in the next section 4.

The vanishing cycles are closely related to massless BPS states. Given a vanishing
cycle $\n$, it can always be decomposed with respect to the homology basis, 
$$\n=n_{\rm m}\b+n_{\rm e}\a~,\eqno(3.12)$$
where $n_{\rm m}$ and 
 $n_{\rm e}$ are integers. One finds at a given singularity that
$$0=\oint_{\n} \l =n_{\rm m}\int_{\b} \l + n_{\rm e}
\int_{\a} \l =n_{\rm m}a_{\rm D} +n_{\rm e}a\equiv Z~,\eqno(3.13)$$
which corresponds to a massless BPS state with the magnetic and electric charges 
$(n_{\rm m},n_{\rm e})$ at the singularity~! Therefore, the dyon
charges are just the coordinates of the corresponding vanishing cycle in the
homology basis~\cite{lerche}. Under a canonical change of the homology basis 
(a duality transformation~!), the intersection number
$$ \#(\n^i,\n^j)=n^i_{\rm m}n^j_{\rm e}-n^j_{\rm m}n^i_{\rm e} \in 
{\bf Z}~,\eqno(3.14)$$
has to be invariant. Note that eq.~(3.14) is nothing but the DZS quantization 
condition (I.2.23). Two BPS states are mutually local with respect to each other 
if eq.~(3.14) vanishes, and they are non-local otherwise. There exists the 
general (Picard-Lefshetz) formula~\cite{agv} that determines the 
monodromy for any vanishing cycle (3.12), and it just gives rise to eq.~(1.15).
\vglue.2in

\section{Solution to the low-energy effective action}

It is not difficult to write down the differential equation for a multi-valued
section $(a_{\rm D}(u),a(u))$ having a given monodromy around known singularities
 in the moduli space parametrized by a local coordinate $u$. Consider the 
{\it second-order} Schr\"odinger-type equation in the complex plane $u$, 
$$ \left[ \,-\,\fracmm{d^2}{du^2} +V(u)\right]\j(u)=0~,\eqno(4.1)$$
whose potential $V(u)$ is a meromorphic (single-valued) function with a finite 
number of poles at some points $u_i$  where, for example, $u_1=1$, $u_2=-1$ and
$u_3=\infty$ as in sect.~1.~\footnote{We take $\L^2=1$ for simplicity.} Eq.~(4.1)
is known to have only two linearly independent solutions, let's call them 
$a_{\rm D}(u)$ and $a(u)$. As $u$ goes around any of the poles, there can be a 
non-trivial monodromy, as in eq.~(1.4). As is well known in the theory of
differential equations~\cite{agv}, the non-trivial constant monodromies
correspond to those poles of the potential that are of second order 
{\it at most}.~\footnote{That singularities are called {\it regular}~\cite{agv}.}
 The general form of the potential in our case is therefore fixed up to a few 
coefficients, 
$$V(u)=\,\fracmm{d_1}{(u+1)^2}+\fracmm{d_2}{(u-1)^2}+\fracmm{d_3}{(u+1)(u-1)}~~.
\eqno(4.2)$$
Eq.~(4.1) with the potential (4.2) can be transformed into the standard 
hypergeometric differential equation, whose explicit solutions are known. It 
remains to compare its general solution, in terms of a hypergeometric function 
to be parametrized by the potential residues $d_i$, with the known asymptotics 
(sect.~1) at each singularity, in order to identify the coefficients $d_i$, and
hence, fix the particular solutions both for $a_{\rm D}(u)$ an $a(u)$ in terms of
hypergeometric functions~\cite{bilal}. The information contained in the 
asymptotics is equivalent to that contained in the monodromies (sect.~1).

Having obtained the representation (3.8) for the solution in terms of the 
auxiliary elliptic curve, one can make a `short cut' by verifying that the
right-hand sides of eqs.~(3.8) are annihilated by the the second-order 
differential operator 
$$ \tilde{\Lag}(w,\q_w)=\q_w\left( \q_w-\ha\right)-w\left(\q_w-\fracmm{1}{4}
\right)^2~,\eqno(4.3)$$
where the new variables $w=u^2$ and $\q_w\equiv w\pa_w$ have been introduced.
Eq.~(4.3) defines the hypergeometric system $F(-\frac{1}{4},-\frac{1}{4};
\frac{1}{2},w)$. It is easy to check that 
$$\pa_u\tilde{\Lag}=\tilde{\Lag}_{\rm PF}\pa_u~,\eqno(4.4)$$
where another operator 
$$\tilde{\Lag}_{\rm PF}(w,\q_w)
=\q_w\left( \q_w-\ha\right)-w\left(\q_w+\fracmm{1}{4}\right)^2 \eqno(4.5)$$
has been introduced. In terms of the original variable $u$, the operator (4.5)
takes the form 
$$\tilde{\Lag}_{\rm PF}=(1-u^2)\pa_u^2-2u\pa_u-\fracmm{1}{4}~,\eqno(4.6)$$
while the corresponding differential equation, $\tilde{\Lag}_{\rm PF}\j(u)=0$, is
known as the {\it Picard-Fuchs} (PF) equation, and it plays the role of 
eq.~(4.1) here. All the periods of the Seiberg-Witten elliptic curve are known
to satisfy the PF equation~\cite{fkra,agv}. For those of them, which are given by
eqs.~(3.6) and (3.7), it was just argued. In our case, 
matching the asymptotic expansions of the period integrals in accordance with the
results of sect.~1 yields the particular combination of hypergeometric 
functions~\cite{bilal}, 
$$\eqalign{
a_{\rm D}(u)=~&~\fracmm{i}{2}(u-1)F\left(\frac{1}{2},\frac{1}{2};
2,\fracmm{1-u}{2}\right)~,\cr
a(w)=~&~\sqrt{2(u+1)}F\left(-\frac{1}{2},\frac{1}{2};1,\fracmm{2}{u+1}\right)~.
\cr}\eqno(4.7a)$$
Using standard integral representations of the hypergeometric 
functions~\cite{erd}, one can rewrite eq.~(4.7a) to the very explicit
form~\cite{sw1},
$$\eqalign{
a_{\rm D}(u)=~&~\fracmm{\sqrt{2}}{\p}\int^u_1 \fracmm{dx\sqrt{x-u}}{\sqrt{x^2-1}}
~,\cr
a(u)=~&~\fracmm{\sqrt{2}}{\p}\int^1_{-1} \fracmm{dx\sqrt{x-u}}{\sqrt{x^2-1}}~.\cr}
\eqno(4.7b)$$

It is straightforward to calculate the prepotential $\cf(a)$ from the explicit 
expressions given above. For example, one can invert the second equation in
eq.~(4.7) and insert the result into the first one, in order to obtain 
$a_{\rm D}$ as a function of $a$. Integrating the latter once with respect to $a$
yields $\cf(a)$. For example, actual calculations in the case of large $a$  (the 
semiclassical region) produce eq.~(II.6.12) as expected, now with all concrete 
values for the instanton coefficients $c_l$, namely~\cite{lerche}
$$\begin{array}{c|cccccc} l & 1 & 2 & 3 & 4 & 5 & \cdots \\ \hline 
                         c_l & \fracmm{1}{2^5} & \fracmm{5}{2^{14}}
& \fracmm{3}{2^{18}} & \fracmm{1469}{2^{31}} & \fracmm{4471}{5\cdot 2^{34}} &
\cdots \end{array} \eqno(4.8)$$
Similarly, one can treat the dual magnetic region near the singularity $u=+\L^2$,
where the monopole becomes massless. One finds eq.~(1.3) indeed, whose lowest
threshold correction coefficients read ~\cite{lerche}  
$$\begin{array}{c|cccccc} l & 1 & 2 & 3 & 4 & 5 & \cdots \\ \hline
                         c^D_l & 4 & -\fracmm{3}{4} & \fracmm{1}{2^{4}} & 
\fracmm{5}{2^{9}} & \fracmm{11}{2^{12}} & \cdots \end{array} \eqno(4.8)$$

The numbers above were confirmed by multi-instanton calculations~\cite{dkma}. 
The modular-invariant (uniformizing) coordinate $u$ of $\cm_{\rm q}$ is given 
by~\cite{padova}
$$ u(a)=\p i\left( \cf(a) -\frac{1}{2}a\pa_a\cf (a) \right)~.\eqno(4.9)$$
\begin{itemize}
\item It is the power of {\it holomorphicity} together with {\it duality} that 
determine the whole function $\cf$ from its known asymptotics near the 
singularities. 
\end{itemize} 

\section{Other groups, and adding ~$N=2$~ matter}

Once the exact low-energy effective action of the $SU(2)$ pure $N=2$~ SYM theory 
is understood, it is straightforward to generalize the Seiberg-Witten results to 
other gauge groups~\cite{dous,klt,af,dsu}. Let us take $G=SU(n)$ for 
definiteness, where $n=N_{\rm c}$ is the number of `colors'.

The classical moduli space $\cm_{\rm cl}$ of the inequivalent vacua is the space
of all solutions to eq.~(II.3.7) modulo gauge transformations. The vacuum 
expectation value of the Higgs field can be chosen in the {\it Cartan subalgebra}
 (CSA) of $G$,~\footnote{The brackets indicating vacuum expectation values are 
often omitted in what follows, in order \newline ${~~~~~}$ to simplify the 
formulas.}
$$ \f =\sum^r_{k=1}a_kH_k~,\quad {\rm where}\quad r={\rm rank}\,G~.\eqno(5.1)$$
In the case of $G=SU(n)$, one has $r=n-1$ and $H_k=E_{k,k}-E_{k+1,k+1}$, where 
$(E_{k,l})_{ij}=\d_{ik}\d_{jl}$. In generic point of $\cm_{\rm cl}$, the gauge 
group $G$ is spontaneously broken to $U(1)^r$. When some eigenvalues coincide, a
(non-abelian) subgroup $H_P\subset G$ remains unbroken.

The electric charge of the $SU(2)$ theory is replaced by the charge vector
$\vec{q}$ belonging to the root lattice $\L_R(G)$ in Dynkin basis of $G$. The BPS
mass formula (without magnetic charges) $m^2(q)=2\abs{Z_q(a)}^2$~, where 
$Z_q(a)=\vec{q}\cdot\vec{a}$, determines which gauge bosons remain massless for a
given background $\vec{a}=\{ a_k \}$.

The SCA variables $\vec{a}$ are, however, not invariant under the gauge 
transformations. They do not even have the residual gauge invariance under the 
discrete transformations from the {\it Weyl group} $S(n)$.~\footnote{The Weyl 
group $S(n)$ acts on the weights $\l_i$ of $G$ by permitation.} The 
gauge-invariant description is provided in terms of the Weyl-invariant Casimir
eigenvalues $u_k(a)$ belonging to ${\bf C}^{n-1}/S(n)$. The polynomials $u_k(a)$ 
parametrizing the CSA modulo the Weyl group can be easily obtained by looking at 
the characteristic equation
$$ \det ( x{\bf 1}-\f)=0~,\eqno(5.2)$$
whose coefficients are Weyl-invariant. In the case of $SU(n)$, one has
$$ \f={\rm diag}(a_1,a_2,\ldots,a_n)~,\quad {\rm and}\quad \sum_ia_i=0~.
\eqno(5.3)$$
Hence, eq.~(5.2) yields
$$ x^n+x^{n-2}\sum_{i<j}a_ia_j -x^{n-3}\sum_{i<j<k}a_ia_ja_k+\ldots +(-1)^n
\prod_ia_i=0~.\eqno(5.4)$$
Taking $n=2$ gives $\f=\ha a\s_3$ and $u\equiv\VEV{\tr\f^2}=\ha a^2$, as 
expected (sect.~II.5). In the case of $SU(3)$, one easily finds
$$ x^3-x\ha\tr\f^2 -\fracmm{1}{3}\tr\f^3=0~,\eqno(5.5)$$
where
$$\eqalign{
u\equiv +\ha\VEV{\tr\f^2}=~&~-\sum_{i<j}a_ia_j=a_1^2+a_2^2 +a_1a_2~,\cr
v\equiv -\fracmm{1}{3}\VEV{\tr\f^3}=~&~-a_1a_2a_3=a_1a_2(a_1+a_2)~.\cr}
\eqno(5.6)$$
Similarly, in the case of $SU(n)$, one finds the {\it symmetric} polynomials
$$\det(x{\bf 1}-\f)=x^n+x^{n-2}c_2(\f)+\ldots + (-1)^j x^{n-j}c_j
(\f) +\ldots =0~,\eqno(5.7)$$
where
$$ c_j(\f)= \sum_{n_1<n_2< \ldots <n_j}a_{n_1}a_{n_2}\cdots a_{n_j}~.\eqno(5.8)$$

It is more convenient to introduce linear combinations $Z_{\l_i}(a)\equiv 
\vec{\l}_i\cdot\vec{a}$, where $\{\l_i\}$ are the weights of the $n$-dimensional
fundamental representation of $SU(n)$. It is the $Z_{\l_i}(a)$ that have direct 
group-theoretical meaning, and that enter the BPS mass formula. The corresponding
 characteristic equation reads
$$ \prod^n_{i=1} \left(x-Z_{\l_i}(a)\right)=x^n-\sum^{n-2}_{l=0}u_{l+2}(a)
x^{n-2-l}\equiv W_{A_{n-1}}(x,u_k)~.\eqno(5.9)$$
The non-linear transition from $Z_{\l_i}(a)$ to $u_k(a)$ is known as a classical
{\it Miura transformation},
$$ u_k(a) =(-1)^{k+1}\sum_{j_1< j_2< \ldots < j_k}Z_{\l_{j_1}}(a)
Z_{\l_{j_2}}(a)\cdots Z_{\l_{j_k}}(a)~.\eqno(5.10)$$
The polynomial $W_{A_{n-1}}(x,u_k)$ is called the {\it simple singularity} 
associated with $A_{n-1}$ (or with $SU(n)$) in the theory of partial differential
equations~\cite{agv}, or as the {\it Landau-Ginzburg} (LG) potential in conformal
field theory~\cite{book}. In the cases of $SU(2)$ and $SU(3)$, one finds
$$ W_{A_1}=x^2-u~,\qquad W_{A_2}=x^3-xu-v~.\eqno(5.11)$$

Extra massless non-abelian gauge bosons appear in the classical theory whenever
$$ Z_{\l_i}(a)=Z_{\l_j}(a)\eqno(5.12)$$
for some $i\neq j$. Eq.~(5.12) describes classical singularities 
which are the fixed points of the Weyl transformations. Hence,
$$ \cm_{\rm cl}=\{u_k\}/\S_0~,\eqno(5.13)$$
where $\S_0=\{u_k:~~\D_0(u_k)=0\}$, and the discriminant~\cite{agv}
$$ \D_0(u)=\prod^{n}_{i<j}\left(Z_{\l_i}(u)-Z_{\l_j}(u)\right)^2
=\prod_{\rm positive~roots} Z^2_{\a}(u)~,\eqno(5.14)$$ 
has been introduced. The discriminant of the simple singularity therefore encodes
all information about the classical symmetry breaking patterns in the 
gauge-invariant way.

The $N=2$ supersymmetry restricts the form of the low-energy effective action to 
an $N=2$ abelian gauge theory with the prepotental $\cf$. The theory contains
$r={\rm rank}\,G$ abelian $N=2$ vector multiplets which can be decomposed into 
$r$~ $N=1$ chiral 
multiplets $A_i$ and $r$~ $N=1$ abelian vector multiplets $W^{\a}_i$. The $N=1$ 
superspace Lagrangian is given by
$$ \Lag=\fracmm{1}{4\p}{\rm Im}\left[ \int d^4\q\left(\sum_i\fracmm{\pa\cf}{\pa
A_i}\bar{A}_i\right)+\int d^2\q\ha\left(\sum_{i,j}\fracmm{\pa^2 \cf}{\pa A_i\pa
A_j}W^{\a}_iW_{\a j}\right)\right]~.\eqno(5.15)$$
Accordingly, the $N=1$ K\"ahler potential reads
$$ K(A,\bar{A})=\,{\rm Im}\, \sum_i \fracmm{\pa\cf(A)}{\pa A_i}\bar{A}_i~,
\eqno(5.16)$$
the effective couplings are
$$ \t_{ij}(A)=\fracmm{\pa^2\cf(A)}{\pa A^i\pa A^j}~,\eqno(5.17)$$
and the dual fields are defined by
$$ A_{\rm D}^i=\fracmm{\pa\cf(A)}{\pa A_i}~.\eqno(5.18)$$
As usual, the leading component of the superfield $A_i$ is called $a_i$, and
similarly for $A^i_{\rm D}$: $\left.A^i_{\rm D}\right|_{\q=0}=a^i_{\rm D}$. 

Zamolodchikov's metric is defined by
$$ ds^2={\rm Im}\,\fracmm{\pa^2\cf(a)}{\pa a_i\pa a_j}da_id\bar{a}_j~,\eqno(5.19)
$$
where $i,j,\ldots =1,2,\ldots,r$. The metric has to be positively definite,
$$ {\rm Im}\,\t_{ij} >0~.\eqno(5.20)$$
The dual coordinates $a^i_{\rm D}\equiv \fracmm{\pa\cf}{\pa a_i}$ together
with the initial coordinates $a_i$ parametrize a $2r$-dimensional vector space
$X\cong {\bf C}^{2r}$. Hence, one arrives at a {\it vector bundle} which locally 
looks
like $\cm{\rm q}\otimes X$. The $X$ can be endowed with the {\it symplectic} form
$$ \o =\fracmm{i}{2}\sum_i\left( da_i\wedge d\bar{a}^i_{\rm D}-da^i_{\rm D}
\wedge d\bar{a}_i\right)~,\eqno(5.21)$$
and the {\it holomorphic form}
$$ \o_{\rm hol}=\sum_i da_i\wedge da^i_{\rm D}~.\eqno(5.22)$$
We are interested in the sections, $f:\cm_{\rm q}\to X$, which take the form
$$ \left( \begin{array}{cc} a_{\rm D}^i (u) \\ a_i(u) \end{array}\right)~,
\eqno(5.23)$$
and are restricted by the condition that the pullback of 
$\o_{\rm hol}$ vanishes: $f^*(\o_{\rm hol})=0$.
 
The Zamolodchikov metric 
$$ds^2=\,{\rm Im}\,\fracmm{\pa a_{\rm D}^i}{\pa u_n}
\fracmm{\pa \bar{a}_i}{\pa\bar{u}_m}du_nd\bar{u}_m \eqno(5.24)$$
is invariant under the symplectic transformations $Sp(2r,{\bf R})$. In accordance
 with sect.~4, we should expect that only a subgroup $\G_{\rm M}$ of the 
{\it discrete} group $Sp(2r,{\bf Z})$ is going to survive in the quantum theory, 
the  $\G_{\rm M}$ being generated by actual monodromies in $\cm_{\rm q}\,$. It is
also known that the same group $Sp(2r,{\bf Z})$ is the modular group of a 
genus--$r$ Riemann surface, whose generators can be visualized 
in terms of Dehn twists around homology cycles~\cite{fkra}. Therefore, it is a 
good idea to look for an auxiliary Seiberg-Witten (SW) curve (a Riemann surface)
whose moduli space is precisely given by $\cm_{\rm q}\,$. Given the SW curve, 
the positivity of Zamolodchikov's metric would then be guaranteed. In order to 
identify the right Riemann surface, one notices that it should have something to 
do with the simple 
singularity $W_{A_{n-1}}$ playing the key role in determining the structure of 
the classical moduli space $\cm_{\rm cl}$. For instance, as is well-known in the 
two-dimensional $N=2$~ supersymmetric conformal field theory, the classical LG 
potential is still relevant in determining the structure of the quantum 
theory~\cite{book}. Hence, it is not very surprising that the SW curve exists, 
and it is given by an algebraic curve~\cite{klty}
$$ y^2=\left(W_{A_{n-1}}(x,u_k)\right)^2-\L^{2n}~.\eqno(5.25)$$
Since eq.~(5.25) can be rewritten as
$$  y^2=\left(W_{A_{n-1}}-\L^n\right)\left(W_{A_{n-1}}+\L^n\right)~,\eqno(5.26)$$
it happens that each classical singularity {\it splits} into two quantum
singularities to be associated with massless dyons, with the distance between 
them being governed by the quantum scale $\L$. Accordingly, every single isolated
 branch of $\S_0$ splits into two barnches of $\S_{\L}$. The points $Z_{\l_i}$ 
also split,
$$ Z_{\l_i}(u)\Longrightarrow Z^{\pm}_{\l_i}(u,\L)~,\eqno(5.27)$$
and become $2n$ {\it branch points}. The (SW) Riemann surface itself can
be represented as a two-sheeted covering of the Riemann sphere branched at $2n$
points, $Z^+_{\l_i}$ and $Z^{-}_{\l_i}\,$, with cuts running between them. Hence,
the SW curve appears to be {\it hyperelliptic}.

By definitition, a Riemann surface is called hyperelliptic, if it admits a 
meromorphic function with exactly two poles~\cite{fkra}. Then, the ramification 
(branch) points have branch number $1$ and, by the {\it Riemann-Hurwitz} theorem,
the number of branch points is related to the genus $h$ by $2n=2h+2$, so that
$h=n-1=r$.~\footnote{In fact, any elliptic curve of genus $h\leq 2$ is
hyperelliptic~\cite{fkra}.}

A generalization to the other {\it simply-laced}~\footnote{A simply-laced Lie 
group has all roots of the same length.} Lie groups is now obvious: one should
simply replace the simple singularity $W_{A_{n-1}}$ with the proper one,
$W_{D_n}$ or $W_{E_m}$, associated with $SO(2n)$ and $E_{6,7,8}$, respectively. 

Given a Riemann surface of genus $h$, there exists $h$ holomorphic abelian 
differentials $\o_k$ (of the first kind)~\cite{fkra}. As far as the SW curve 
(5.25) is concerned, they are given by
$$ \o_{k}=\fracmm{x^{n-k-1}dx}{y}~,\qquad k=1,2,\ldots,n-1~.\eqno(5.28)$$
The period integrals are defined by
$$ A_{ij}=\oint_{\a_j}\o_i~,\qquad B_{ij}=\oint_{\b_j}\o_i~,\eqno(5.29)$$
while the period matrix is $\t\equiv A^{-1}B$. Hence, one can identify 
$$A_{ij}(u)=\fracmm{\pa}{\pa u_{i+1}}a_j(u)~,\qquad B_{ij}(u)=
\fracmm{\pa}{\pa u_{i+1}}a^j_{\rm D}(u)~,\eqno(5.30)$$ 
similarly to that in eq.~(3.7). One finds by integration that ~\cite{klt}
$$ a^i_{\rm D} =\oint_{\b_i}\l~,\qquad  a_i =\oint_{\a_i}\l~,\eqno(5.31)$$
where ({\it cf.} eq.~(3.9))
$$ \l =\fracmm{\rm const.}{2\p i}\left(\fracmm{\pa}{\pa x}W_{A_{n-1}}(x,u_k)
\right)\fracmm{xdx}{y} \eqno(5.32)$$
is an abelian differential of the second kind (with vanishing residues). The
constant in eq.~(5.32) can be fixed from the known asymptotics of 
$(\vec{a}_{\rm D},\vec{a})$.

The quantum charges of the massless dyons associated with quantum singularities
are determined by the vanishing cycles (see sect.~3). Indeed, any vanishing cycle
$\n$ can be decomposed with respect to a homology basis $(\vec{\a},\vec{\b})$ 
on the SW curve,
$$ \n=\vec{q}\cdot\vec{\a} +\vec{g}\cdot\vec{\b}~,\eqno(5.33)$$
where the charge vector $\vec{q}$ has integer components and belongs to the root
lattice $\L_{\bf R}$, while the charge vector $\vec{g}$ also has integer 
components but belongs to the dual (simple root) lattice $\L_{\bf R}^{\bf D}\,$.
One has ({\it cf}. eq.~(3.13))
$$ 0=\oint_{\n}\l=\left(\vec{q}\cdot \oint_{\vec{\a}}+\vec{g}\cdot
\oint_{\vec{\b}}\right)\l=\vec{q}\cdot\vec{a}+\vec{g}\cdot\vec{a}_{\rm D}\equiv
Z_{(q,g)}~,\eqno(5.34)$$ 
where the central charge $Z_{(q,g)}$, entering the BPS mass formula $m^2(q,g)=
2\abs{Z_{(q,g)}}^2$, appears. Hence, similarly to the $SU(2)$ solution 
(sect.~3), the quantum numbers can be read off from the vanishing cycles. Since
the section (5.23) non-trivially transforms under the duality transformations,
the charges $\vec{\n}=(\vec{g},\vec{q})$ have to transform accordingly, so that
the central charge and the BPS mass remain invariant. The 
{\it intersection number},
$$ \n_i\cap\n_j\equiv\vec{\n}_i^{\rm T}\left(\begin{array}{cc} 0 & {\bf 1} \\
-{\bf 1} & 0 \end{array} \right)\vec{\n}_j=\vec{g}_i\cdot\vec{q}_j-
\vec{g}_j\cdot\vec{q}_i \in {\bf Z}~,\eqno(5.35)$$
is also invariant under a change of homology basis (a duality transformation~!),
and it yields the generalized DZS quantization condition ({\it cf}. eq.~(I.2.23)).
Two BPS states are, therefore, local with respect to each other (i.e. a local 
Lagrangian containing both particles exists), if and only if the intersection 
number vanishes. 

The rest of calculations is quite similar to the $SU(2)$ case considered in 
sect.~4. The Picard-Lefshetz formula,
$$ M_{(g,q)}=\left( \begin{array}{cc} {\bf 1}+\vec{q}\otimes \vec{g} &
+\vec{q}\otimes\vec{q} \\ -\vec{g}\otimes \vec{g} &  {\bf 1}-\vec{g}\otimes 
\vec{q} \end{array} \right) \in Sp(2r,{\bf Z})~,\eqno(5.36)$$
determines the monodromies from the known charges of a given quantum singularity
and vice versa.
The period integrals of the SW curve satisfy the (second-order) system of 
$h=r$ Picard-Fuchs differential equations, and they determine the section (5.23)
by eq.~(5.30). The information from the semiclassical region provided by the 
perturbative one-loop beta-function (asymptotic freedom~!) fixes the monodromy 
around infinity or, equivalently, determines the perturbative contribution to the
 $N=2$ prepotential ({\it cf}. eq.~(II.6.12)) as
$$ \cf_{\rm 1-loop}(a)=\fracmm{i}{4\p}\sum_{\rm positive~roots}Z^2_{\a}\log
\left[ \fracmm{Z^2_{\a}}{\L^2}\right]~,\eqno(5.37)$$
where $Z_{\a}(a)=\vec{\a}\cdot\vec{a}$ for simply-laced Lie groups. 
The weakly coupled
dual prepotential (in proper dual variables) near a quantum singularity looks 
like that in eq.~(1.3), and it is also fixed by the beta-function of the 
corresponding abelian $N=2$ supersymmetric gauge theory (no asymptotic freedom).
Putting all together, one arrives at the well-defined Riemann-Hilbert problem,
whose unique solution can be calculated by solving the Picard-Fuchs equations
subject to the known asymptotics near the singularities. It is then 
straightforward to calculate the $N=2$ prepotential $\cf$. For example, in the 
case of $SU(3)$, the solution can be expressed in terms of the so-called 
{\it Appel} functions which generalize the hypergeometric functions to the case 
of two variables~\cite{klty,klt}.

Let us now briefly discuss what happens when an $N=2$ matter to be represented
by some number $(N_{\rm f})$ of $N=2$ hypermultiplets in the fundamental 
representation of the gauge group $SU(N_{\rm c})$ is added.~\footnote{See e.g., 
the second paper in ref.~\cite{sw1} and refs.~\cite{dsu,aas} for details.}  Each
$N=2$ hypermultiplet comprises two $N=1$ chiral superfields $Q(q,\j_q)$ and
$\tilde{Q}(\tilde{q},\j_{\tilde{q}})$. Under the internal $SU(2)$ symmetry
associated to $N=2$ supersymmetry, the `squarks' $(q,\tilde{q}^{\dg})$ form a
doublet, whereas their `quark' superpartners $\j_q$ and $\j_{\tilde{q}}$ are
singlets.~\footnote{A `mirror' particle $\j_{\tilde{q}}$ for each quark $\j_q$ 
makes $N=2$ supersymmetry to be phenomenologically \newline ${~~~~~}$ 
unacceptable. $N=2$ supersymmetry has to be softly broken to $N=1$ supersymmetry 
which, \newline ${~~~~~}$ in its turn, is spontaneously broken in realistic 
models (see the next sect.~6 for an example).} The $N=1$ superpotential in the 
$N=2$ abelian gauge theory with matter has some additional terms,
$$ V_{\rm matter}=
\sum^{N_{\rm f}}_{i=1}\left( \sqrt{2}\tilde{Q}_i\F Q_i + m_i\tilde{Q}_i
Q_i\right) +{\rm h.c.}~,\eqno(5.38)$$
where $\F$ is the chiral $N=1$ superfield in the $N=2$ vector multiplet, and
$\{m_i\}$ are mass parameters. Because of eq.~(5.38), one should expect both the 
supercurrents (to be derived from the full action), and the central charges
in the supersymmetry algebra (to be derived from the supercurrents) to receive 
contributions from the matter terms too. Accordingly, the BPS mass formula 
(I.5.17) gets modified. One finds~\cite{sw1}
$$ Z=n_{\rm e}a+n_{\rm m}a_{\rm D}+\sum_kS_km_k/\sqrt{2}~,\eqno(5.39)$$
where $S_k$ are the $U(1)$ charges of the matter hypermultiplets. Eq.~(5.39)
implies that the masses $\{m_i\}$ will enter as the additional parameters in the 
Seiberg-Witten approach to the low-energy effective action. In particular,
the positions of the quantum singularities, as well as the SW curve itself, are
all going to be deformed by them.

The R-symmetry anomaly in eq.~(II.6.3) is replaced by
$$\pa_{\m} j^{\m}_{\rm R}=\left(2N_{\rm c}-N_{\rm f}\right)\fracmm{F{}^*F}{32\p^2}
~.\eqno(5.40)$$
The perturbative (to all loop-orders) beta-function (II.6.2) is also modified as
$$ \b(e)=\m\fracmm{de(\m)}{d\m}=\,-\,(2N_{\rm c}-N_{\rm f})\fracmm{e^3}{16\p^2}~,
\eqno(5.41a)$$
or, equivalently $(\a\equiv e^2/4\p)$,
$$ \fracmm{1}{\a_{N_{\rm f}}(\m)}=\fracmm{2N_{\rm c}-N_{\rm f}}{4\p}
\ln\fracmm{\m^2}{\L_{N_{\rm f}}^2}~.\eqno(5.41b)$$
In the $SU(2)$ case, eqs.~(5.40) and (5.41) tell us that one should take 
$N_f<4$, in order to keep the asymptotic freedom. If $N_f=4$ and there 
are no `quark' masses, the particular $N=2$ gauge theory with the $SU(2)$ gauge 
group and four $N=2$ matter hypermultiplets is finite to all orders of 
perturbation theory, and it is expected to be conformally invariant even 
non-perturbatively. That is obviously  consistent with the vanishing $R$-anomaly
(5.40) and the vanishing beta-function (5.41), and it presumably gives yet 
another example of an exactly self-dual theory in the sense of Montonen-Olive 
with respect to the S-duality, like the $N=4$~ SYM theory though the details are 
quite different~\cite{sw1}. There is the `flavor' $SO(8)$ global symmetry in
the self-dual $N=2$ theory with matter, while the related $SO(8)$ triality 
symmetry is non-trivially mixed with the S-duality ({\it cf.} the U-duality in 
a compactified type-II superstring theory, sect.~7).

The global structure of the quantum moduli space and the low-energy effective
action crucially depend on the number of `flavors' $N_{\rm f}$. In the $SU(2)$
case, if $N_{\rm f}=1$, only a (strong coupling) Coulomb phase appears where 
$\VEV{\f}\neq 0$ and $SU(2)$ is broken to $U(1)$, like in the pure ($N_{\rm f}=0$)
theory considered in the previous sections. If $1<N_{\rm f}<4$, one can have 
(strong coupling) Higgs phases also, where the gauge symmetry in completely 
broken while the light scalars parametrize a unique hyper-K\"ahler manifold (the 
existence of a hyper-K\"ahler structure is dictated by $N=2$ 
supersymmetry~\cite{bagw}). In the case of general gauge groups with matter, one 
finds a rich spectrum of vacua having non-abelian Coulomb phases and mixed 
Coulomb-Higgs phases as well. Many examples, including a construction of the SW
curves in the presence of $N=2$ matter, can be found in the 
literature~\cite{sw1,dsu,aas}.

It is remarkable that the choice of an auxiliary manifold (SW curve) is not 
unique~! In fact, it could be any manifold $\cg$ whose moduli space
is $\cm_{\rm q}\,$, and whose period integrals (to be obtained by integration of 
proper meromorphic forms over $\cg$) coincide with that of the SW curve. 
For example, a six-dimensional {\it Calabi-Yau} (CY) manifold is 
known~\cite{kachru} which is equally good for describing the low-energy effective
action of the $SU(3)$ pure $N=2$~ SYM~ theory like the SW hyperelliptic curve 
considered above. When the ten-dimensional 
type-IIB superstring theory is compactified on that CY space $\cg$ down to four 
dimensions, the resulting four-dimensional $N=2$ supersymmetric string theory 
contains the $SU(3)$ pure $N=2$~ SYM theory in the point-particle limit 
$\a'\to 0$. Hence, one should expect generalizations of the Seiberg-Witten 
duality to string theory, which is another big story (see sect.~7 also).
\vglue.2in

\section{Seiberg-Witten version of confinement}

The Seiberg-Witten results about the exact low-energy effective action in the
$N=2$ supersymmetric gauge theories provide some non-perturbative information 
about the $N=1$ supersymmetric gauge theories, including the $N=1$ super-QCD. One
should expect, for example, that the quantum moduli in the $SU(2)$ pure $N=1$ 
gauge theory are also given by two points $\pm\L^2$ related by a ${\bf Z}_2$ 
transformation (R-symmetry), because the Witten index is the same for both 
theories. In that $N=1$ theory, it is possible to add a mass term $W=m\,\tr\F^2$ 
to the potential, where $\F$ is the chiral $N=1$ superfield (sect.~II.3). 
The mass term lifts the 
flat direction of the $N=2$ potential, and it can be considered as a {\it soft}~
$N=2$ supersymmetry breaking term which allows one to define the $N=1$~ SYM 
theory as the low-energy effective field theory of the $N=2$ theory. It is 
believed that the $N=1$ theory has a mass gap and, hence, a non-vanishing gaugino
condensation vacuum expectation value, $\VEV{\bar{\l}\l}\neq 0$~\cite{russian}.
The existence of the mass gap in the $N=1$ theory also implies that the dual
 `magnetic' photon becomes massive by some Higgs mechanism in the vacua
corresponding to the two singularities $\pm\L^2$ in the quantum moduli space. 
The only obvious candidate for the role of the Higgs field is given by the
t'Hooft-Polyakov monopole or dyon of the initial $N=2$ theory. Such `dual' Higgs
 effect can be interpreted as the dual mechanism to the well-known Meissner
effect in the theory of superconductivity, and it can explain quark confinement 
as the phenomenon arising from the condensation of the magnetic monopoles carrying
 global quantum numbers. 
 
The relevant terms in the $N=1$ supersymmetric action with the dual photon and
the monopole field read
$$ W = m\,\tr\F^2 + a_{\rm D}M\tilde{M}~,\eqno(6.1)$$
where $M$ and $\tilde{M}$ are the $N=1$ chiral superfields representing the 
monopole, and the second term gives the coupling of the monopole to the 
dual photon as required by $N=2$ supersymmetry. Since $a_{\rm D}=a_{\rm D}(u)$
and $u=\tr\F^2$, one can rewrite eq.~(6.1) to the form
$$ W(M)=mu(a_{\rm D}) + a_{\rm D}M\tilde{M}~.\eqno(6.2)$$
Vacua correspond to solutions of $dW=0$, and satisfy $\abs{M}=\abs{\tilde{M}}$,
since the latter is necessary for the vanishing of the $D$-term. Assuming that
$du/da_{\rm D}\neq 0$, one easily finds from eq.~(6.2) the equations of motion,
$$ m\fracmm{du}{da_{\rm D}}+M\tilde{M}=0~,\qquad a_{\rm D}M=a_{\rm D}\tilde{M}
=0~.\eqno(6.3)$$
Eq.~(6.3) has a non-trivial solution: $a_{\rm D}=0$ and
$$ \VEV{M}=\VEV{\tilde{M}}=\sqrt{-m\fracmm{du}{da_{\rm D}}}~\neq 0~.\eqno(6.4)$$
The non-vanishing magnetic order parameter $\VEV{M}$ implies the mass gap in the
$N=1$ theory by the dual Higgs mechanism, and the confinement of abelian charge
as well.
\vglue.2in

\section{Conclusion}

In string theory,  the Yang-Mills coupling constant is determined by the vacuum
expectation value of the {\it dilaton} field $d$, while the Yang-Mills vacuum
angle is similarly related to the {\it axion} field $\x$ in four space-time
dimensions,
$$ \fracmm{e^2}{4\p}=\VEV{e^d}=\fracmm{8G}{\a'}~,\qquad
\fracmm{\q}{2\p}=\VEV{\x}~,\eqno(7.1)$$
where the string constant $\a'$ and the gravitational (Newton's) constant $G$ are
both dimensionful. Hence, the S-duality in string theory acts on the complex 
(dilaton-axion) field $S\equiv \x+ie^{-d}$, and it is supposed to relate strong
and weak couplings. It gives a reason to expect that a strongly coupled string
theory may well be represented by yet another weakly-coupled string 
theory.~\footnote{Perhaps, it may be something else than a string theory 
(M-theory)~\cite{witt}.} The compactified superstrings have another 
well-established target space duality called T{\it -duality}~\cite{gpra}, 
which is usually represented by a non-compact discrete group $G_{\rm T}$. The 
T-duality group $G_{\rm T}$ together with the S-duality group $SL(2,{\bf Z})$ are
actually the subgroups of an even larger non-compact discrete group $G_{\rm U}$ 
known as U{\it -duality}~\cite{hallt}. 
The group $G_{\rm U}$ appears to be a discrete
subgroup of the hidden non-compact continuous symmetry known to be present in the
extended supergravity theory arising from the compactified superstring theory in 
the point-particle limit $\a\to 0$~\cite{filq}. For example, the $N=8$ maximally
extended supergravity in four spacetime dimensions has a non-compact global 
symmetry $E_7$, while there is an evidence for the existence of a discrete $E_7$ 
as the U-duality group $G_{\rm U}$ in the corresponding (compactified) type-II 
superstring theory~\cite{hallt}.

Also, in the spirit of Seiberg and Witten, it is quite natural to interpret the 
so-called {\it conifold} singularities in the moduli space $\cm(\cg)$ of
complex structures of a Calabi-Yau manifold $\cg$ (in the type-II superstring
compactified on that $\cg$) as that coming from the BPS (stable) massless charged
hypermultiplets. The latter are usually interpreted as charged massless {\it
black holes} in string theory~\cite{stromi}. The known dual pairs of string 
theories provide some examples in which the classical moduli in one theory appear
as the quantum moduli in the dual one, thus relating $\cm_{\rm c}$ and 
$\cm_{\rm q}$ in the string theory context. The Seiberg-Witten approach to the 
extended supersymmetric gauge theories can therefore be further promoted to the
level of superstrings in the very natural way. A thorough discussion of the 
string dualities is however beyond the scope of this paper.

There exists a deep relation between the Seiberg-Witten low-energy effective 
theory and {\it integrability}~\cite{kmmm,itep}. In particular, the SW solution 
can be reformulated in terms of certain integrable systems on the moduli space of
instantons. The effective dynamics in the space of coupling constants $(\t)$ is 
governed by the equations belonging to the generalized {\it KP-Toda hierarchy} 
whose solutions are known to be naturally parametrized in terms of auxiliary 
special surfaces, like the SW curves. For instance, the key relations (III.3.8) 
can be understood as just the action-integrals (in proper parametrization) in the
sine-Gordon model~\cite{kmmm}~! 

One can verify that the known prepotentials $\cf(a_i)$, $i=1,2,\ldots,N-1$, of 
the N=2 supersymmetric Yang-Mills theory with the $SU(N)$ gauge group satisfy the 
WDVV-type~\cite{w,dvv} equations 
$$\cf_i\cf^{-1}_k\cf_j=\cf_j\cf^{-1}_k\cf_i~,\quad {\rm where}\quad
(\cf_i)_{jk}\equiv\fracmm{\pa^3\cf}{\pa a_i\pa a_j\pa a_k}~~.\eqno(7.2)$$
The WDVV-equations are known to express the associativity of the algebra of 
primary fields in (conformal) topological field theory~\cite{w,dvv}. That 
observation provides yet another unexpected link between the four-dimensional 
N=2 SYM low-energy effective action and the two-dimensional topological field 
theories. The full story is thus far from being closed~! 
\vglue.2in

\section*{Acknowledgements}

I would like to thank J. Fuchs, J. Gomiz, K. Konishi, D. Korotkin, W. Lerche, 
A. Morozov, H. Nicolai, K. Sibold, S. Theisen and C. Vafa for discussions, and
A. Bilal, S. J. Gates Jr., M. Matone and P. West for correspondence. 

\vglue.2in

\end{document}